\documentclass[aps,twocolumn,reprint,showpacs, groupedaddress]{revtex4-2}
\usepackage{amsmath}
\usepackage[latin9]{inputenc}
\usepackage{graphicx}
\usepackage{hyperref}
\usepackage{graphicx}
\usepackage[color= blue!20]{todonotes}
\usepackage[normalem]{ulem}
\usepackage{color}
\usepackage{ulem}
\usepackage{xcolor}
\usepackage{comment}
\hypersetup{
	colorlinks,
	linkcolor={red!70!blue},
	citecolor={red!70!blue},
	urlcolor={red!70!blue}
}
\DeclareMathOperator{\atan2}{atan2}

\begin{document}

\title{Extracting the current-phase-relation of a monolithic three-dimensional nano-constriction using a DC-current-tunable superconducting microwave cavity}

\author{Kevin~Uhl}\email{kevin.uhl@pit.uni-tuebingen.de}
\affiliation{Physikalisches Institut, Center for Quantum Science (CQ) and LISA$^+$, Universit\"at T\"ubingen, 72076 T\"ubingen, Germany}
\author{Daniel~Hackenbeck}
\affiliation{Physikalisches Institut, Center for Quantum Science (CQ) and LISA$^+$, Universit\"at T\"ubingen, 72076 T\"ubingen, Germany}
\author{Dieter~Koelle}
\affiliation{Physikalisches Institut, Center for Quantum Science (CQ) and LISA$^+$, Universit\"at T\"ubingen, 72076 T\"ubingen, Germany}
\author{Reinhold~Kleiner}
\affiliation{Physikalisches Institut, Center for Quantum Science (CQ) and LISA$^+$, Universit\"at T\"ubingen, 72076 T\"ubingen, Germany}
\author{Daniel~Bothner}\email{daniel.bothner@uni-tuebingen.de}
\affiliation{Physikalisches Institut, Center for Quantum Science (CQ) and LISA$^+$, Universit\"at T\"ubingen, 72076 T\"ubingen, Germany}

\begin{abstract}
Superconducting circuits with nonlinear elements such as Josephson tunnel junctions or kinetic inductance nanowires are the workhorse for microwave quantum and superconducting sensing technologies.
For devices, which can be operated at high temperatures and large magnetic fields, nano-constrictions as nonlinear elements are recently under intense investigation.
Constrictions, however, are far less understood than conventional Josephson tunnel junctions, and their current-phase-relationships (CPRs) -- although highly important for device design -- are hard to predict.
Here, we present a niobium microwave cavity with a monolithically integrated, neon-ion-beam patterned three-dimensional (3D) nano-constriction.
By design, we obtain a DC-current-tunable microwave circuit and characterize how the bias-current-dependent constriction properties impact the cavity resonance.
Based on the results of these experiments, we reconstruct the CPR of the nano-constriction.
Finally, we discuss the Kerr nonlinearity of the device, a parameter important for many high-dynamic-range applications and an experimental probe for the second and third derivatives of the CPR.
Our platform provides a useful method to comprehensively characterize nonlinear elements integrated in microwave circuits and could be of interest for current sensors, hybrid quantum systems and parametric amplifiers.
Our findings furthermore contribute to a better understanding of nano-fabricated 3D constrictions.

\end{abstract}

\maketitle
\let\oldaddcontentsline\addcontentsline
\renewcommand{\addcontentsline}[3]{}

\section*{Introduction}
\vspace{-3mm}
Josephson junctions (JJs) and nonlinear weak links between two superconducting electrodes form an essential ingredient for a wide variety of groundbreaking technologies, such as superconducting quantum interference devices (SQUIDs) \cite{Kleiner04, Clarke04}, voltage standards \cite{Hamilton00} or superconducting microwave quantum circuits \cite{Nakamura99, Clarke08, You11}.
The main characteristics of a superconducting nonlinear element are its critical current $I_0$ and its potentially non-trivial current-phase-relation (CPR) $I(\delta)$, which relates the phase difference $\delta$ across it to the current $I$ flowing through it \cite{Josephson62, Golubov04}.
These two quantities also determine both the inductance and the anharmonicity of a Josephson microwave circuit, extremely important parameters for engineering high-quality superconducting qubits \cite{Krantz19, Willsch23} or Josephson parametric amplifiers \cite{CastellanosBeltran08, Bergeal10, Macklin15}.
For standard superconductor-insulator-superconductor (SIS) Josephson junctions, the CPR usually has the ideal sinusoidal form $I = I_0\sin\delta$ and the only remaining design parameter for particular applications is the critical current.
Other types of JJs, for example trilayer junctions with normal-conducting (SNS) or ferromagnetic (SFS) barriers as well as constriction-type junctions (cJJs) in many cases exhibit a significant deviation of their CPR from the ideal sinusoidal shape \cite{Golubov04}.
To properly design devices and technologies based on these non-sinusoidal Josephson elements, it is therefore of utmost importance to gather knowledge about $I_0$ and the CPR.
Lately, there has been growing interest in constriction type Josephson junctions.
Those cJJs have been already investigated in the early days of superconducting weak links \cite{Likharev79}, they have been implemented into DC-SQUID magnetometers and nanoSQUIDs \cite{Bouchiat01, Hasselbach02, Mitchell12, Chen16, Wang17, MartinezPerez17, Wyss22} and more recently into superconducting field-effect transistors \cite{DeSimoni18, Paolucci18} and superconducting microwave circuits \cite{Vijay10, Kennedy19, Uhl23}.
The latter -- nano-constriction microwave circuits -- are used for dispersive magnetometry \cite{Hatridge11, LevensonFalk16}, microwave optomechanics \cite{Rodrigues19, Bothner22}, photon-pressure devices \cite{Bothner21, Rodrigues21}, parametric amplifiers \cite{Xu23}, current detectors \cite{Schmidt20}, and quantum bits \cite{Rieger23}.
Constrictions are interesting for these applications, since they combine a small junction area with high critical current densities but without adding large capacitances or lossy materials.
Hence, they are ideal for experiments in large magnetic fields and for microwave circuits with small anharmonicity.
On the other hand, the exact constriction CPR depends quite strongly on the material properties, the constriction dimensions as well as the fabrication method \cite{Likharev79, Golubov04}.
Therefore, a platform for investigating simultaneously the transport characteristics, the constriction CPR and the cJJ impact on the microwave circuit properties such as frequency tunability, Kerr anharmonicity and change of decay rates would be ideal.
Some experiments in that direction have already been realized in the past \cite{Rifkin76, Schmidt20a, Dou21, Haller22}, none of which checking all the boxes on the wishlist though.

\begin{figure*}
	\centerline{\includegraphics[width=0.9\textwidth]{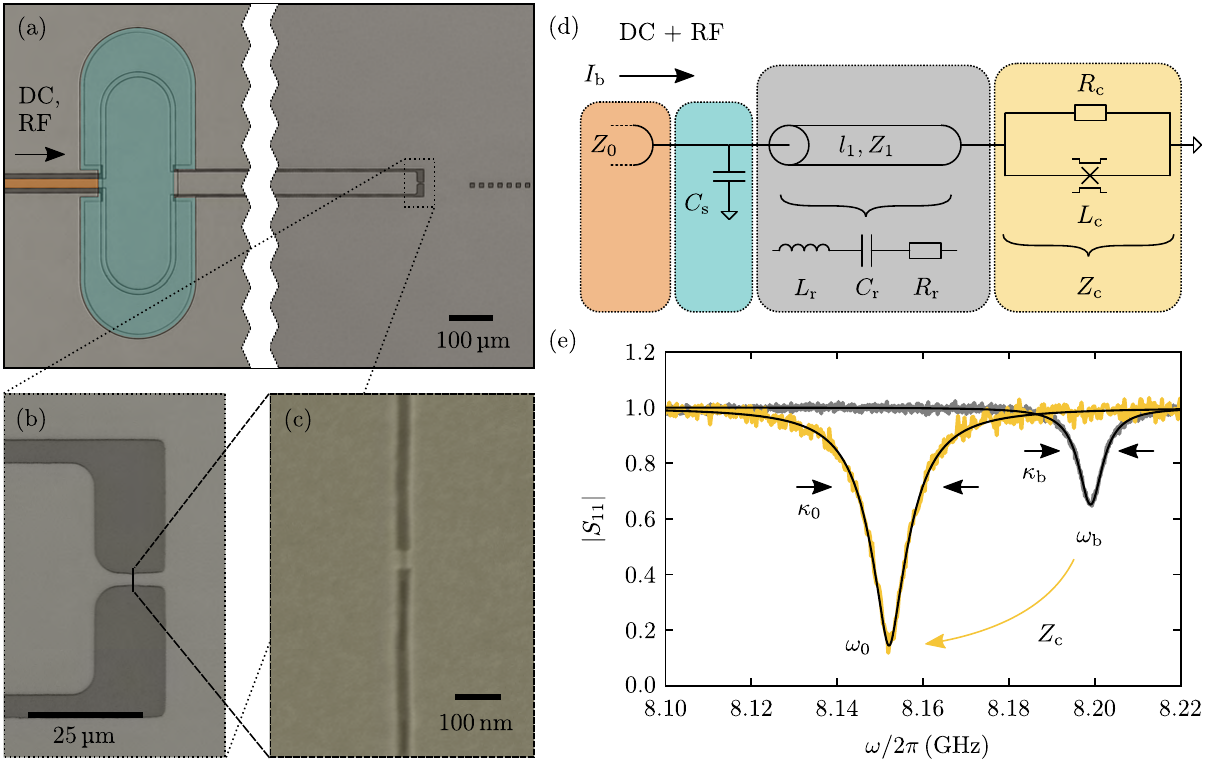}}
	\caption{\textsf{\textbf{A superconducting microwave cavity with integrated niobium nano-constriction and direct current access.} (a) False-color optical micrograph of the $\lambda/2$ transmission line cavity. At the input port, the cavity is shunt-coupled by a large parallel-plate shunt capacitance $C_\mathrm{s}$ (Nb-Si$_3$N$_4$-Nb, colored in teal) to a $Z_0 = 50\,\Omega$ feedline (center conductor colored in orange) for simultaneous microwave (RF) and direct current (DC) probing of cavity and integrated nano-constriction. The cavity itself has a characteristic impedance $Z_1 = 32\,\Omega$ and a length of $l_1 \approx 7200\,$\textmu m (the zigzag lines indicate that most of the cavity length is omitted here). At the far end (far from the input), close to the connection to ground, a monolithic 3D niobium nano-constriction-type Josephson junction (cJJ) is patterned into the center conductor of the coplanar waveguide cavity by means of a focused neon ion beam. Bright and colored parts are niobium, dark parts are silicon substrate. (b) Zoom to the end part of the resonator; (c) False-color scanning electron microscopy image of the 3D cJJ. (d) Circuit equivalent of the cavity with the cJJ at the end, which is modelled as a parallel combination of a loss channel (resistor $R_\mathrm{c}$) and a reactive channel (inductance $L_\mathrm{c}$), in total described by the impedance $Z_\mathrm{c}$; the coplanar waveguide part of the cavity can be described by a lumped element series equivalent with an inductor $L_\mathrm{r}$, a capacitor $C_\mathrm{r}$, and a resistor $R_\mathrm{r}$, respectively. (e) Reflection response $S_{11}$ of the cavity before (gray) and after (yellow) cutting the constriction, measured at $T_\mathrm{s} = 3.9\,$K. From the change in resonance frequency $\omega_\mathrm{b} \rightarrow \omega_0$ and linewidth $\kappa_\mathrm{b} \rightarrow \kappa_0$ induced by the added constriction impedance $Z_\mathrm{c}$, the cJJ elements $R_\mathrm{c}$ and $L_\mathrm{c}$ can be determined. Gray and yellow noisy lines are data, black smooth lines are fits.}}
	\label{fig:Figure1}
\end{figure*}

Here, we present a superconducting microwave cavity with an integrated 3D niobium nano-constriction, which has been monolithically cut into the cavity by using the focused ion beam (FIB) of a neon (Ne) ion microscope (NIM).
Our particular choice of cavity layout allows simultaneous DC current-voltage access to the cJJ and microwave characterization of the DC-current-biased cavity using microwave reflectometry \cite{Bosman15, Schmidt20}.
Compared to similar experiments with different types of JJs \cite{Schmidt20, Schmidt20a, Haller22}, the monolithic approach -- in which the cJJ is cut into the cavity at the very end of the fabrication -- has the advantage that one and the same cavity can be characterized without and with the cJJ.
From the change of cavity resonance frequency and linewidth by the current-biased cJJ, we reconstruct the constriction CPR and observe that it considerably deviates from the simple sinusoidal shape and that it gets more linear with decreasing temperature.
Finally, we measure the cavity Kerr anharmonicity, and demonstrate that it is both in good agreement with calculations based on the forward-skewed CPR found beforehand and sufficiently small for high-dynamic range applications.
Our analysis also reveals, that very small deviations in the second and third derivatives of the CPR can considerably impact the value of the Kerr constant, even at zero bias current.
Overall, our results show that the cavity-integrated characterization of nonlinear superconducting inductances can be used to reconstruct the CPR of familiar and novel weak links, to deepen the understanding of Ne-FIB patterned cJJs and that 3D niobium cJJs are promising for applications in superconducting microwave circuits, radiation-pressure systems and current sensors.

\vspace{-3mm}
\section*{Results}
\vspace{-3mm}
\subsection*{The Device}
\vspace{-3mm}
Our device is presented in Fig.~\ref{fig:Figure1} and it is based on a $\lambda/2$ (half wavelength) coplanar waveguide cavity, near-shorted to ground at both ends.
The cavity is patterned from a $90\,$-nm-thick DC-magnetron-sputtered niobium film on high-resistivity silicon substrate, it has a characteristic impedance $Z_1 \approx 32\,\Omega$ and a total length $l_1 \approx 7200\,$\textmu m.
The niobium film has a critical temperature $T_\mathrm{c} \sim 9.0\,$K and a residual resistivity of $\rho\sim 7.3\,$\textmu$\Omega\cdot$cm at $10\,$K.
At its input port, the cavity is shunt-coupled to a $Z_0 = 50\,\Omega$ coplanar waveguide feedline for microwave driving and readout via a three-layer parallel plate capacitance $C_\mathrm{s} \approx 17.5\,$pF to ground.
Due to this particular coupling scheme, the center conductor remains uninterrupted and we can pass a DC bias current along the waveguide from the feedline all the way through the cavity \cite{Bosman15}. 
At the far end of the cavity, the center conductor has a narrower part, into which the cJJ is cut with a Ne-FIB after a first round of basic cavity characterization experiments.
The cJJ has a length of $\sim25\,$nm, a width of $\sim40\,$nm and a thickness of $\sim 40\,$nm, cf. Fig.~\ref{fig:Figure1}(c).
For more details on the sample fabrication, cf. Supplementary Note~I.
All experiments, before and after the junction cutting, are performed in the vacuum chamber of a liquid helium cryostat.
For the measurements, the $10\times10\,$mm$^2$ large chip containing the constriction-cavity is mounted on and wirebonded to a microwave printed circuit board (PCB), which is connected to a coaxial microwave cable.
Close to the sample, a bias-tee is connected to the coaxial cable to combine the DC and microwave inputs, and a $10\,$dB directional coupler is added to split microwave input and output signals.
The microwave input line is strongly attenuated to thermalize the input noise to the sample temperature, and the output line is connected to a cryogenic $\sim 35\,$dB HEMT (high-electron-mobility transistor) amplifier.
The DC cables are two cryogenically low-pass-filtered twisted pair copper wires.
For a stable temperature control between $T_\mathrm{min} \approx 2.5\,$K and $T_\mathrm{max} \approx 6\,$K (range of the experiments presented here), a vacuum pump is connected to the liquid helium container of the cryostat and a feedback-loop-controlled heating resistor is included in the vacuum sample chamber.
More details on the experimental setup and a corresponding schematic can be found in Supplementary Note~II.
As any Josephson element at temperatures of few kelvin typically has a reactive and a resistive part to its impedance, we model the cJJ as a parallel combination of an inductor $L_\mathrm{c}$ and a resistor $R_\mathrm{c}$ similar to the two-fluid model, cf. Fig.~\ref{fig:Figure1}(d).
We omit the additional junction capacitance here as well as a possible quasiparticle inductance in the resistive branch, since according to our estimates both are negligible at the frequencies relevant for this work.
In order to model the cJJ-shorted transmission line resonator with analytic expressions, we use a lumped element equivalent of the resonator with resistor $R_\mathrm{r}$, inductor $L_\mathrm{r}$ and capacitor $C_\mathrm{r}$, cf. Fig.~\ref{fig:Figure1}(d), which is a very good approximation near the resonance frequency (cf. Supplementary Notes~III and IV) and simplifies the calculations.
For sufficiently low excitation powers to be safely in the linear response regime, the resonance frequency of the cavity after junction cutting is in good approximation given by
\begin{equation}
	\omega_0 = \frac{\omega_\mathrm{b}}{1 + \frac{L_\mathrm{c}^*}{2L_\mathrm{r}}}
	\label{eqn:w0}
\end{equation}
with the resonance frequency before the cutting $\omega_\mathrm{b} = 1/\sqrt{L_\mathrm{r}C_\mathrm{tot}}$ and $C_\mathrm{tot} = C_\mathrm{r}C_\mathrm{s}/\left(C_\mathrm{r} + C_\mathrm{s}\right)$.
The resonator inductance is given by
\begin{equation}
	L_\mathrm{r} = \frac{\pi Z_1}{2\omega_1}
	\label{eqn:Lr}
\end{equation}
with the shunt-capacitor-less resonance frequency $\omega_1$ (cf. Supplementary Note~III).
From Eqs.~(\ref{eqn:w0}) and (\ref{eqn:Lr}) it also becomes clear why we chose $Z_1$ to be smaller than the usual $50\,\Omega$, since a smaller $L_\mathrm{r}$ increases the device sensitivity to changes in $L_\mathrm{c}^*$.
The resonance linewidth after constriction cutting is 
\begin{equation}
	\kappa_0 \approx \kappa_\mathrm{b} + \omega_0^2R_\mathrm{c}^*C_\mathrm{tot}
\end{equation}
with the total linewidth before cutting $\kappa_\mathrm{b}$.
The two effective lumped elements $R_\mathrm{c}^*$ and $L_\mathrm{c}^*$ are related to the actual junction resistance and inductance $R_\mathrm{c}$ and $L_\mathrm{c}$, respectively, via 
\begin{equation}
	R_\mathrm{c}^* = \frac{R_\mathrm{c}\omega_0^2L_\mathrm{c}^2}{R_\mathrm{c}^2 + \omega_0^2 L_\mathrm{c}^2}, ~~~~~ L_\mathrm{c}^* = \frac{L_\mathrm{c}R_\mathrm{c}^2}{R_\mathrm{c}^2 + \omega_0^2 L_\mathrm{c}^2}
	\label{eqn:RJLJ}
\end{equation}
and after finding the values for $R_\mathrm{c}^*$ and $L_\mathrm{c}^*$ from the properties of the cavity, we invert these relations and obtain the values for $R_\mathrm{c}$ and $L_\mathrm{c}$.
To demonstrate the effect of cutting the cJJ into the resonator and to analyze the unbiased constriction properties, we show in Fig.~\ref{fig:Figure1}(e) the reflection response $S_{11}$ around resonance of the cavity without and with the junction, i.e., before and after constriction cutting.
As always in this work, the response was measured by means of a vector network analyzer (VNA).
Before cutting the junction, the cavity has a resonance frequency $\omega_\mathrm{b} \approx 2\pi\cdot8.199\,$GHz and a total linewidth of $\kappa_\mathrm{b} = 2\pi\cdot7.6\,$MHz with internal and external contributions $\kappa_\mathrm{i, b} = 2\pi\cdot 1.3\,$MHz and $\kappa_\mathrm{e, b} = 2\pi\cdot6.3\,$MHz, respectively.
After the junction is cut into the cavity, the resonance frequency has shifted to $\omega_\mathrm{0} \approx 2\pi\cdot\,8.152\,$GHz and the linewidth to $\kappa_\mathrm{0} \approx 2\pi\cdot 15.6 \,$MHz.
Here, the junction-induced decay rate is $\kappa_\mathrm{c} = \omega_0^2R_\mathrm{c}^*C_\mathrm{tot} \approx 2\pi\cdot 8.0\,$MHz.
Using relations Eq.~(\ref{eqn:RJLJ}) we obtain $L_\mathrm{c} = 11.9\,$pH and $R_\mathrm{c} = 7.4 \,\Omega$.
Note that we only show the resonance lines for a single sample temperature $T_\mathrm{s} = 3.9\,$K here, but in Supplementary Notes III and VII more data on the temperature dependence of the junction-less cavity and the cavity with constriction can be found.
\begin{figure*}
	\centerline{\includegraphics[width=0.9\textwidth]{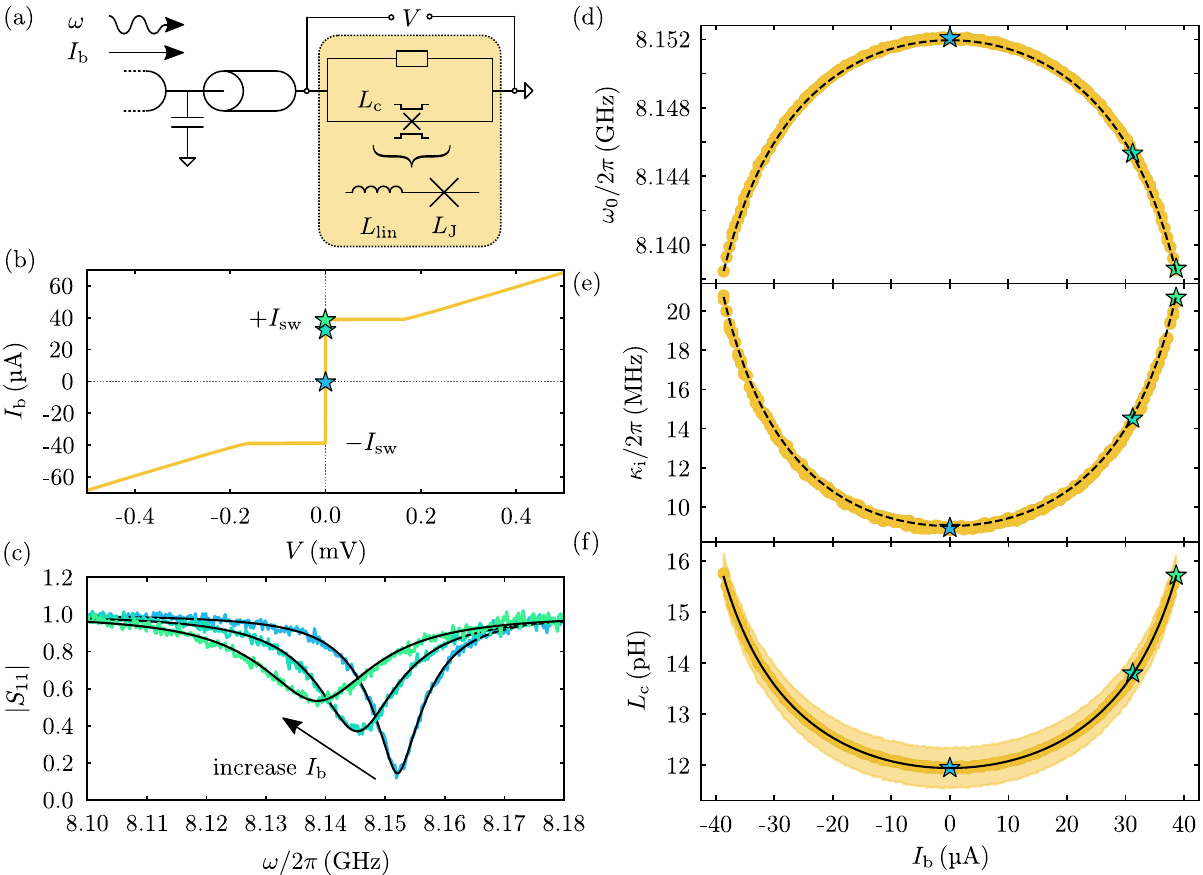}}
	\caption{\textsf{\textbf{Tuning microwave cavity and constriction inductance with a sub-critical DC current.} (a) Schematic of the experiment. A DC bias current $I_\mathrm{b}$ is sent through the feedline and the cavity center conductor to the cJJ and the corresponding (time-averaged) voltage $V$ is detected. For various DC currents, the resonator reflection $S_{11}$ is measured with a vector network analyzer. For the analysis, we split the constriction inductance $L_\mathrm{c}$ into a linear part $L_\mathrm{lin}$ and a Josephson part $L_\mathrm{J}$. All data in this figure were taken at $T_\mathrm{s} = 3.9\,$K. (b) shows a typical non-hysteretic current-voltage characteristic (IVC) of the constriction with a switching current of $I_\mathrm{sw} \sim \pm 39\,$\textmu A. The three stars mark the bias current values $I_\mathrm{b} = 0$, $I_\mathrm{b} \sim 0.8I_\mathrm{sw}$ and $I_\mathrm{b} \sim 0.99I_\mathrm{sw}$, for which we show the corresponding cavity reflection response $S_{11}$ in panel (c). With increasing DC current, the resonance frequency shifts to lower values and the cavity linewidth increases. Colored noisy lines are data, black smooth lines are fits. From the fits to each reflection measurement, the resonance frequency $\omega_0(I_\mathrm{b})$ and the internal decay rate $\kappa_\mathrm{i}(I_\mathrm{b})$ are extracted, the results are shown in panels (d) and (e), respectively. We attribute the frequency shift to a change of the nonlinear constriction inductance $L_\mathrm{c}(I_\mathrm{b})$ and the change in linewidth to a current-dependent loss channel $R_\mathrm{c}(I_\mathrm{b})$. The values we extract for the current-dependent junction inductance $L_\mathrm{c}$ are shown in panel (f). In (d)-(f), symbols are data, solid lines are fits, dashed lines are calculated from theory expressions and fit values, and star symbols correspond to the resonances shown in (c). The color-shaded area around the data points in (f) corresponds to the estimated error due to uncertainties of the constriction-free cavity parameters as discussed in Supplementary Note VII.}}
	\label{fig:Figure2}
\end{figure*}

\vspace{-3mm}
\subsection*{Tuning the cavity with a DC current}
\vspace{-3mm}

As a next step, we investigate the impact of a DC bias current $I_\mathrm{b}$ through the cJJ on the cavity properties $\omega_0(I_\mathrm{b})$ and $\kappa_\mathrm{i}(I_\mathrm{b})$, respectively.
The experiment and the results are presented in Fig.~\ref{fig:Figure2}.
Again, we show exemplarily the data for a single sample temperature $T_\mathrm{s} = 3.9\,$K, but more analogous data for different temperatures are presented and discussed in Supplementary Note~VII.
First, we measure the current-voltage characteristic (IVC) of the cJJ without any microwave tone by sending a DC current through the junction and tracking the corresponding DC voltage.
We observe switching from the zero-voltage- to a finite-voltage-state at a current of $I_\mathrm{sw}\approx 39\,\mu$A in both current directions.
We also find a non-hysteretic IVC indicating that the constriction is in the overdamped regime, and we observe a considerable excess current of $I_\mathrm{ex} \sim 23.7\,$\textmu A, possibly related to self-heating or to Andreev reflections \cite{Blonder82, Tinkham03}.
Overall, the IVCs behave very much as expected from earlier experiments and observations in niobium constrictions \cite{Troeman08, Chen16}.
The (differential) DC resistance that we determine from the slope of the linear part of the voltage-branch is $R_\mathrm{lin} = 11\,\Omega$, so slightly larger than the $R_\mathrm{c} \approx 7.4\,\Omega$ we got from the microwave experiment.
They are also not expected to coincide, however, since heating effects for instance can significantly impact the measured DC resistance.
To investigate the microwave properties of the current-biased cavity in the sub-critical $I_\mathrm{b}$ regime, we then measure the reflection response $S_{11}$ for varying bias currents  $I_\mathrm{b} < I_\mathrm{sw}$ and extract $\omega_0(I_\mathrm{b})$ and $\kappa_\mathrm{i}(I_\mathrm{b})$ from fits to the reflection data.
The equations and fitting routines used are detailed in Supplementary Notes V and VI.
Here, we use sufficiently low microwave probe powers to keep the cavity in the linear response regime, which we ensure by staying far below the input powers needed to observe nonlinearities in the resonance lineshape.
With increasing bias current, the resonance frequency is shifting to lower values as observed also in earlier experiments \cite{Schmidt20, Schmidt20a} and expected from the nature of a superconducting nonlinear inductance, which usually increases with increasing bias current.
The total frequency range that we can cover with the bias-current-tuning is strongly temperature-dependent (see Supplementary Note~VII), but for $T_\mathrm{s} = 3.9\,$K it is $\sim 13.5\,$MHz.
The internal linewidth on the other hand is increasing with increasing bias current, from $\kappa_\mathrm{i} \sim 2\pi \cdot 8.9\,$MHz at zero bias current to $\kappa_\mathrm{i} \sim 2\pi \cdot 20.7\,$MHz at $I_\mathrm{b} \sim I_\mathrm{sw}$.
For currents larger than the switching current the resonance vanishes abruptly, as the losses in the constriction and the internal linewidth get so high that the resonance cannot be discriminated from the background anymore.
Both effects the decrease in resonance frequency and the increase in linewidth can ultimately be attributed to a bias-current-dependent decrease of Cooper pair density and an increase of quasiparticle density inside the cJJ.
A reduced Cooper pair density leads to an increase of the kinetic supercurrent inductance $L_\mathrm{c}$.
Simultaneously, it leads to a reduced value for $R_\mathrm{c}$, i.e., more and more of the microwave current is passing through the resistor (quasiparticle current) instead of the inductor (Cooper pair current).
By applying the same data extraction routine as dicussed in the context of Fig.~\ref{fig:Figure1}, we can determine for all bias currents the value of $L_\mathrm{c}$, the result is plotted in Fig.~\ref{fig:Figure2}(f).
The constriction inductance increases from around $12\,$pH at zero bias current to around $15.8\,$pH at $I_\mathrm{b}\sim I_\mathrm{sw}$.
We perform a fit of the data by assuming that the inductance of the constriction $L_\mathrm{c}$ can be described by a series combination of a linear inductance $L_\mathrm{lin}$ and an ideal Josephson inductance $L_\mathrm{J}$, an often surprisingly accurate model for nano-constrictions, that has already been discussed in Ref.~\cite{Likharev79} and more recently in Ref.~\cite{Wang23}.
The dependence of the total inductance on the bias current in this case is given by
\begin{eqnarray}
	L_\mathrm{c}(I_\mathrm{b}) & = & L_\mathrm{lin} + L_\mathrm{J}(I_\mathrm{b}) \\
	& = & L_\mathrm{lin} + \frac{L_\mathrm{J0}}{\sqrt{1 - \frac{I_\mathrm{b}^2}{I_0^2}}}
	\label{eqn:six}
\end{eqnarray}
with the zero-bias-current Josephson inductance $L_\mathrm{J0} = \Phi_0/(2\pi I_0)$ and the theoretical critical current $I_0$.
The latter can be different from the experimental switching current $I_\mathrm{sw}$, for instance due to thermal or quantum activated escape, phase diffusion or electronic noise coupling into the device.
And indeed, as indicated by the black fit line in Fig.~\ref{fig:Figure2}(f), this approach works very well with the fit parameters $L_\mathrm{lin} = 5.55\,$pH and $I_0 = 50.5\,$\textmu A, which corresponds to $L_\mathrm{J0} = 6.5\,$pH.
From this fit it indeed seems that the critical current $I_0$ is significantly larger than the experimental switching current $I_\mathrm{sw}$ and we will discuss this effect further for various sample temperatures in the next sections and in Supplementary Note~VII.
Regarding the extracted resistance $R_\mathrm{c}$ we do not have a clear microscopic or intuitive physical model for the functional bias-current dependence, but we can fit it with an even fourth order polynomial for all temperatures, cf. also Supplementary Note~VII.
Re-inserting the obtained fit curves for $L_\mathrm{c}(I_\mathrm{b})$ and $R_\mathrm{c}(I_\mathrm{b})$ back into the equations for the resonance frequency and linewidths leads to the black dashed lines in panels (d) and (e), showing excellent agreement with the data.

\vspace{-3mm}
\subsection*{Extracting the constriction current-phase relation}
\vspace{-3mm}
\begin{figure*}
	\centerline{\includegraphics[width=0.87\textwidth]{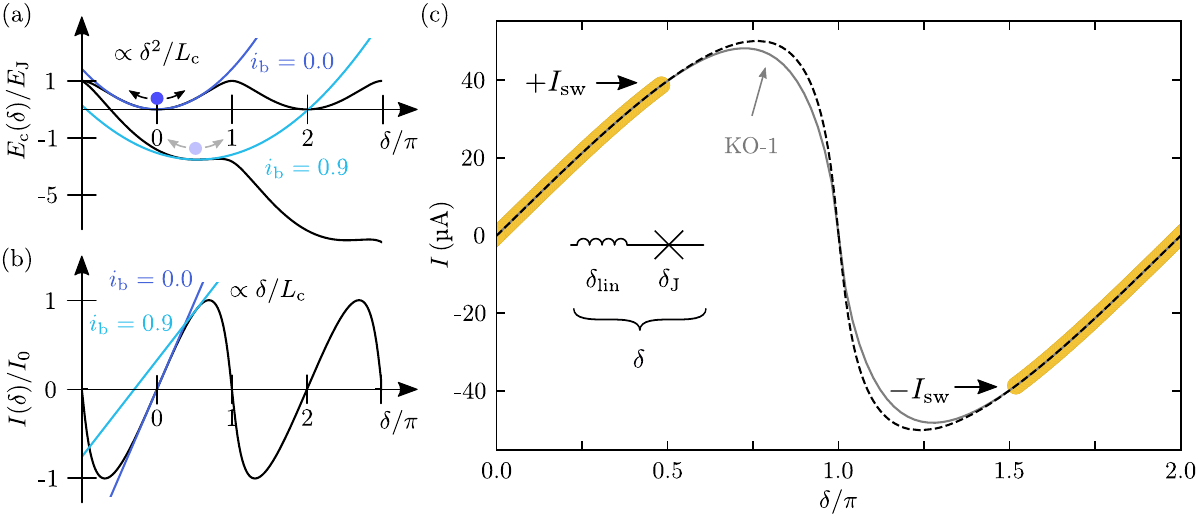}}
	\caption{\textsf{\textbf{Reconstructing the current-phase-relation of the nano-constriction from microwave data.} (a) Schematic drawing of a $2\pi$-phase-periodic constriction potential $E_\mathrm{c}(\delta)$, in which the superconducting phase particle can oscillate around its equilibrium position $\delta_0$. We show the two exemplary cases $\delta_0 = 0$ for $i_\mathrm{b} = I_\mathrm{b}/I_0 = 0$ and $\delta_0 > 0$ for $i_\mathrm{b} = 0.9$, where $I_0$ is the critical current. For small oscillation amplitudes, the potential can be approximated by a parabola $\propto \delta^2/L_\mathrm{c}(I_\mathrm{b})$. A bias current $i_\mathrm{b} > 0$ is equivalent to a potential tilt, which reduces the curvature in the minimum. (b) The schematic current-phase-relation (CPR) can be obtained from the potential by differentiation and the slope of the CPR at the equilibrium position of the particle is $\propto1/L_\mathrm{c}$. Hence, from knowing the inductance $L_\mathrm{c}$ for many bias currents $I_\mathrm{b}$ the CPR can be reconstructed by integration (see main text). We perform this integration here for the dataset discussed in Fig.~\ref{fig:Figure2}(f) and the result is plotted in (c) as yellow circles. Temperature is $T_\mathrm{s} = 3.9\,$K and data points extend to $\pm I_\mathrm{sw}$. Data points for $I < 0$ have been manually shifted by $+2\pi$ in phase. By describing $L_\mathrm{c}$ as a linear-plus-Josephson inductance, the CPR $I(\delta)$ can be modeled with a total phase $\delta = \delta_\mathrm{lin} + \delta_\mathrm{J}$ with the phase across the linear part $\delta_\mathrm{lin}$, the result is shown as black dashed line and resembles a forward-skewed sinusoidal shape. Here $\delta_\mathrm{lin}$ and $\delta_\mathrm{J}$ describe the individual phases across each component, linear and sinusoidal part, respectively. For comparison an exemplary CPR based on the KO-1 model is shown as gray fit line.}}
	\label{fig:Figure3}
\end{figure*}
Once we know the cJJ inductance for each bias current, we can reconstruct the current-phase-relation (CPR) of the constriction and in what follows we will describe the relation between $L_\mathrm{c}$ and $I(\delta)$ in more detail.
We have treated the cJJ so far as a bias-current-dependent but microwave-linear inductance, which was justified by the low powers used in the experiment.
But what exactly does that mean and how is it related to the CPR of the cJJ?
In circuit theory, a linear inductance $L$ is directly related to a harmonic potential, where the potential energy is given by $E_\mathrm{pot} = \frac{\varPhi^2}{2L}$ with the generalized flux through the inductor $\varPhi$.
A Josephson-like element now usually has an anharmonic and $2\pi$-periodic potential, e.g., a cosine-shaped potential $E_\mathrm{pot} = E_\mathrm{J}\left(1 - \cos\delta \right)$ in the case of a standard junction, which is expressed in terms of the phase difference across the junction $\delta = 2\pi\frac{\varPhi}{\Phi_0}$ instead of the flux and where $E_\mathrm{J} = \frac{\Phi_0 I_0}{2\pi}$ is the Josephson energy.
For a linear inductance, the potential can therefore be expressed in terms of the phase as $E_\mathrm{pot} = \frac{\Phi_0^2}{4\pi^2}\frac{\delta^2}{2L}$.
At the same time, such a quadratic term is also the first dynamically relevant term in a Taylor expansion of a nonlinear $2\pi$-periodic potential around the equilibrium phase $\delta_0$, which explains why for small phase oscillations (=low microwave excitation powers) we can also treat the nano-constriction as a linear inductance.
In both cases, the current-phase relation is given by
\begin{equation}
	I(\delta) = \frac{2\pi}{\Phi_0}\frac{\partial E_\mathrm{pot}}{\partial \delta}
\end{equation}
and the linear inductance of the element is expressed as
\begin{equation}
	\frac{1}{L} = \frac{4\pi^2}{\Phi_0^2}\frac{\partial^2 E_\mathrm{pot}}{\partial \delta^2}\bigg|_{\delta_0}
\end{equation}
where the equilibrium phase is defined via $I(\delta_0) = I_\mathrm{b}$.
Hence, there is a direct relation between the CPR and $L$, which reads in terms of our device variables
\begin{equation}
	L_\mathrm{c} = \frac{\Phi_0}{2\pi}\left( \frac{\partial I}{\partial \delta}\bigg|_{\delta_0} \right)^{-1}.
\end{equation}
In other words, the constriction inductance is identical to the reciprocal of the slope of the CPR at any bias point and we can obtain the phase for any given bias current by
\begin{equation}
	\delta = \frac{2\pi}{\Phi_0} \int_0^{I_\mathrm{b}} L_\mathrm{c} dI_\mathrm{b}',
\end{equation}
given that we have tracked the inductance for all currents up to $I_\mathrm{b}$.
Since of course we only have a finite number $N$ of discrete current and inductance values, we have to replace the integration by summation
\begin{equation}
	\delta_j = \frac{2\pi}{\Phi_0}\sum_j^N L_{\mathrm{c}, j}\Delta I_j
\end{equation}
with $j>0$ and $\Delta I_j = I_{\mathrm{b}, j} - I_{\mathrm{b}, j-1}$.
Figure~\ref{fig:Figure3} summarizes and illustrates these ideas and the result of our CPR reconstruction using this method for the data presented and discussed already in Fig.~\ref{fig:Figure2}.
What we find when performing this discrete integration is shown in Fig.~\ref{fig:Figure3}(c): we obtain a rather linear dependence $I(\delta)$ for $\delta \leq \pi/4$, which starts to bend towards smaller slopes for higher currents and then suddenly stops at around $\delta \approx \pi/2$, when $I_\mathrm{b} = I_\mathrm{sw}$.
Very clearly, this CPR deviates significantly from the sinusoidal shape of an ideal Josephson tunnel junction, which would have a maximum and zero slope at $\delta = \pi/2$.
In our data, we do not reach a point where the slope approaches zero, which is typically the case when $I_\mathrm{b} \sim I_0$.
When we add the theoretical curve for the CPR of a series combination of $L_\mathrm{lin}$ and an ideal Josephson element $L_\mathrm{J}$ though, shown as dashed line, the picture seems to match quite perfectly.
The theoretical CPR is piecewise calculated using
\begin{equation}
	\delta = (-1)^n\arcsin{\left( \frac{I}{I_0} \right)} + \frac{2\pi}{\Phi_0}L_\mathrm{lin}I + n\pi
\end{equation}
and inversely plotted for $n\in\{-1, 0, 1\}$.
Note that we used exactly the values for $L_\mathrm{lin}$ and $I_\mathrm{0}$, that we obtained from the fit in Fig.~\ref{fig:Figure2}(f), and so the agreement is not a complete surprise.
\begin{figure*}
	\centerline{\includegraphics[width=0.9\textwidth]{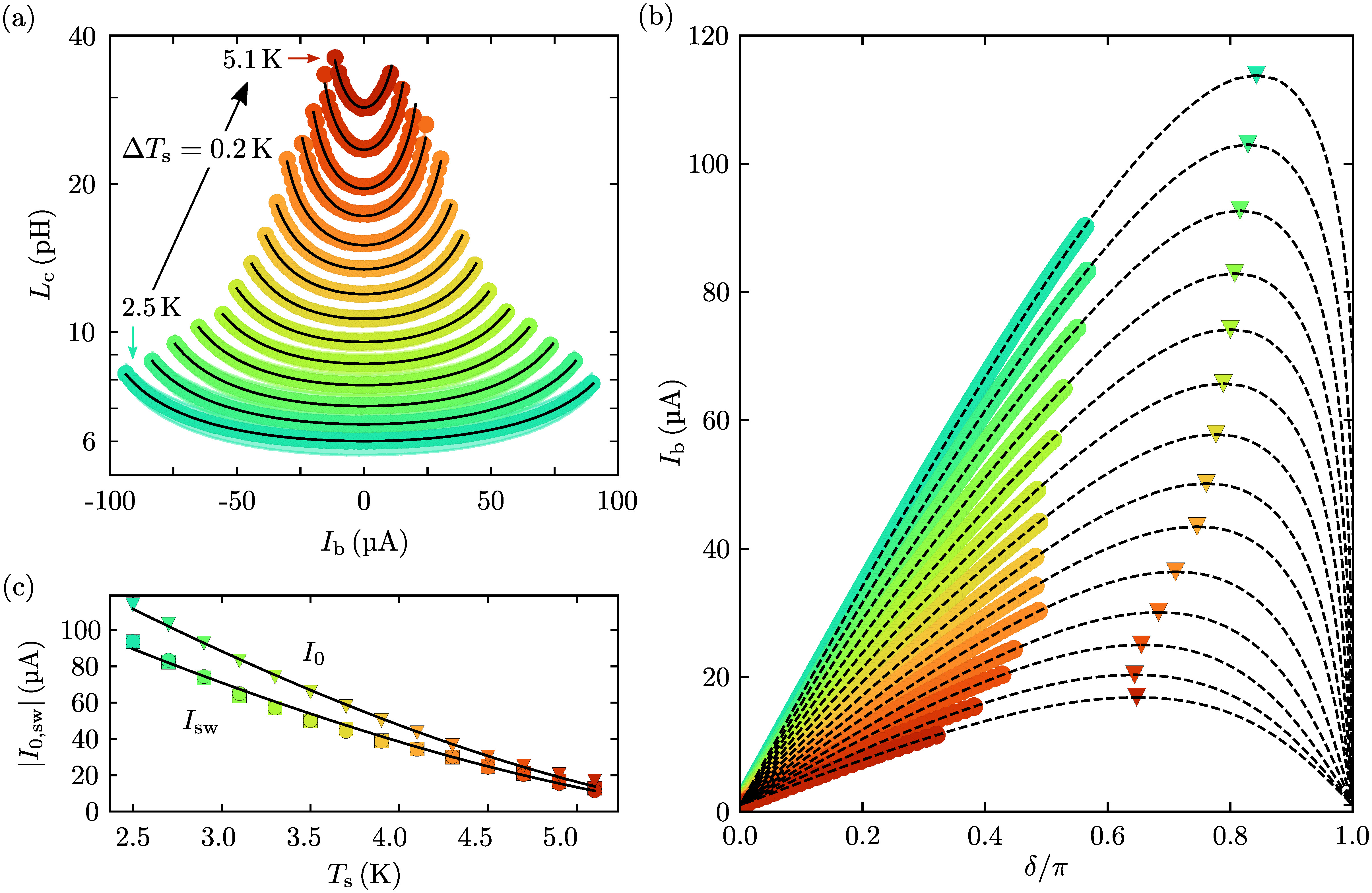}}
	\caption{\textsf{\textbf{Change of the constriction properties with sample temperature.} (a) The constriction inductance $L_\mathrm{c}(I_\mathrm{b})$ for several different sample temperatures $T_\mathrm{s}$. Lowest temperature is $T_\mathrm{s}^\mathrm{min} = 2.5\,$K (cyan data, lowest $L_\mathrm{c}$), highest temperature $T_\mathrm{s}^\mathrm{max} = 5.1\,$K (dark red data, highest $L_\mathrm{c}$), increased in steps of $\Delta T_\mathrm{s} = 0.2\,$K. The inductance increases with both increasing bias current $I_\mathrm{b}$ and with increasing temperature. Circles are data, lines are fits using Eq.~(\ref{eqn:six}), and colored shades around the data points correspond to the error range analogous to the one in Fig.~\ref{fig:Figure3}(f). By integrating the inductance $L_\mathrm{c}$, the corresponding constriction CPR can be reconstructed for each $T_\mathrm{s}$. The result is plotted in (b), circles are data, dashed lines are theoretical CPRs with the two-inductor-model and based on the fits of (a). With increasing temperature, the CPRs get less skewed and less linear in the regime $\delta \leq \pi/2$. For all curves (except for the highest few temperatures) $I_\mathrm{sw} \sim 0.8 I_0$, but the phase for which either of them is reached gets smaller with increasing $T_\mathrm{s}$. The CPR critical currents $I_0$ are marked with triangles at the maximum of each dashed line. (c) Critical current $I_0$ and switching current $I_\mathrm{sw}$ vs temperature in the range $T_\mathrm{s} = 2.5\,$K$-5.1\,$K. Circles and squares are $I_\mathrm{sw}$ data obtained by microwave and IV-measurements, respectively. Triangles are the theoretical critical currents $I_0$, which we get from the fits of $L_\mathrm{c}$. Solid lines in (c) are fits with Eq.~(\ref{eqn:I0_Ts}).}}
	\label{fig:Figure4}
\end{figure*}
Such a forward-skewed CPR has has been found in many papers for constriction junctions and generalized weak links before \cite{Likharev79, Golubov04, Hasselbach02, Vijay10, Troeman08, Wang23}, in particular for niobium constrictions, but also for aluminum and other materials.
Similar CPRs are also predicted by different theoretical models, such as for instance the Kulik-Omelyanchuk (KO-1) model \cite{Likharev79}, Ginzburg-Landau models \cite{Wang23} and others, for a review cf. Refs.~\cite{Likharev79, Golubov04}.
Of course, we cannot be completely certain how the experimental CPR will continue beyond $\pi/2$, where we do not have further experimental data points.
However, it has been demonstrated in previous studies that the linear-plus-Josephson-inductance model matches numerically calculated CPRs with high accuracy, see e.g. the very recent work by Wang \textit{et al}. on 3D niobium constrictions \cite{Wang23}.
The forward-skewed CPRs of analytical models are furthermore indeed quite similar in shape, and for a qualitative comparison we show a KO-1 fit of our data, although strictly speaking we are most likely not in the regime of validity of this theory with our device properties.
According to Ref.~\cite{Golubov04}, the KO-1 model is valid in the regime where the length of the constriction $l_\mathrm{c} \ll \sqrt{\xi_0 l_\mathrm{e}}$ and the transverse size $w_\mathrm{c} \ll l_\mathrm{c}$.
Both conditions are not fulfilled in our device since the BCS coherence length is $\xi_0 = 38\,$nm for niobium, the electron mean free path in our device is $l_\mathrm{e} \approx 5.7\,$nm (cf. Supplementary Note~VII) and $l_\mathrm{c} = 25\,$nm, $w_\mathrm{c} = 40\,$nm.
Nevertheless does the KO-1 fit resemble both, the skewedness and the critical current $I_0$ in good approximation and at the same time demonstrates that small deviations in the CPR are well possible in the regime $I\sim I_0$ without compromising the behaviour for $I<I_\mathrm{sw}$.
Since we have the possibility to vary the sample temperature between $T_\mathrm{s}^\mathrm{min} \sim 2.5\,$K and $T_\mathrm{s}^\mathrm{max} > T_\mathrm{c} \approx 9\,$K, we will use this opportunity to study the temperature-dependence of cavity and constriction properties and of the resulting CPR in the next part.
\vspace{-3mm}
\subsection*{Change of CPR with sample temperature}
\vspace{-3mm}
To this end, we repeat all the measurements and data analyses discussed above for $T_\mathrm{s} = 3.9\,$K for all temperatures $2.5\,$K$\,\leq T_\mathrm{s}\leq5.1\,$K in steps of $\Delta T_\mathrm{s} = 0.2\,$K.
In Supplementary Note~VII a collection of all cavity resonance frequencies and internal linewidths as function of $I_\mathrm{b}$ and $T_\mathrm{s}$ is presented.
Here, we want to focus on the resulting constriction inductances $L_\mathrm{c}$, the re-constructed current-phase-relationships and the constriction currents $I_\mathrm{sw}$ and $I_0$.
Our main findings are collected and presented in Fig.~\ref{fig:Figure4}.
The constriction inductance $L_\mathrm{c}$ is increasing with both, temperature and bias current, and over all temperatures the minimum and maximum values are $\sim 6\,$pH and $36\,$pH, respectively.
With decreasing temperature, the inductance tunes faster with increasing $I_\mathrm{b}$, a signature for a decreasing $I_0$.
The relative tuning range however, i.e., the ratio of maximum to minimum inductance for a fixed $T_\mathrm{s}$ seems to be rather constant, except for the highest temperatures, where the range decreases somewhat.
For a further quantitative analysis and the extraction of the CPRs, we again fit the measured inductances with our linear-plus-sinusoidal inductance model Eq.~(\ref{eqn:six}) and it gives excellent agreement for all temperatures.
Then, we integrate the data points of $L_\mathrm{c}$ over the bias current $I_\mathrm{b}$ for each $T_\mathrm{s}$ and plot the resulting CPRs in panel (b) in direct comparison with the curves calculated through the two-inductance model and the corresponding fit.
There are several important observations we can make in this representation.
Firstly, the integrated CPRs are always in excellent agreement with the calculated ones, there is no indication of deviations in the experimentally accessible regime.
This shows that the inductance model we applied to obtain the fits is well-applicable over the complete temperature and bias-current range.
Secondly, the CPR skewedness is getting reduced with increasing temperature, an effect that has already been observed earlier for superconducting junctions with skewed CPRs \cite{Troeman08} and which is also an intrinsic property of theories like the KO-1 model for instance \cite{Likharev79, Golubov04}.
One way of intuitively interpreting this is that the constriction inductance becomes more linear with decreasing temperature and deviates stronger from that of an ideal tunnel junction.
Ultimately, it is defined by a competition of different length scales such as the constriction dimensions, the electron mean free path and the superconducting coherence length \cite{Golubov04}, some of which are temperature-dependent.
Importantly, however, we do not observe any multi-valued CPRs as in many earlier studies on niobium constrictions \cite{Hasselbach02, Troeman08}, which is a signature of the high quality of our devices \cite{Vijay10}.
Since for the lowest temperatures used here the CPR slope around $\delta = \pi$ is approaching very large values, it could, however, be that for even lower temperatures the single-valuedness vanishes and that the critical current is maybe even shifted to $\delta_{I_0} > \pi$.
Finally, we observe that the switching current for all temperatures except for the very highest ones is reached at $\delta \sim \pi/2$ and that the ratio $I_\mathrm{sw}/I_0$ does not show a strong temperature dependence, cf. also panel (c) and Supplementary Note VII.
In fact, the ratio $I_\mathrm{sw}/I_0$ stays nearly constant for all temperatures at around $0.8 \pm 0.05$, except for the highest $T_\mathrm{s}= 5.1\,$K, where the switching happens even earlier, cf. also Supplementary Note VII.
Although we do not know the exact origin of the premature switching, we believe we can exclude several possible suspects due to the constant value of $I_\mathrm{sw}/I_0$.
If the switching was triggered by an external noise source with a constant noise amplitude or by thermal noise, the suppression of the switching current would increase with increasing $T_\mathrm{s}$ or remain constant in units of \textmu A.
Also, the switching is not induced by the microwave probe signals, except maybe for the highest $T_\mathrm{s}$.
In panel (c) we show both, the switching currents determined from the IV characteristics in absence of any microwave signals (circles) and the switching currents observed in the experiment, in which we stepwise increase $I_\mathrm{b}$ and probe the cavity for each value (squares).
The results do not show any remarkable difference.
It is however, not unusual that in experiments with superconducting nanowires and constrictions the switching current is considerably lower than the critical current.
It has been attributed to phase slips and phase diffusion in the past \cite{Pekker09, Li11, Baumans17, Friedrich19}, which can be thermally activated or by quantum tunneling, and the level of suppression can depend on the bias current sweep rate (here $22.5\,$nA/ms), on the thermalization and heating details of the sample and on the current-phase-relation of the system under consideration.
To illuminate and analyze this premature switching in more detail, further experiments and systematic investigations will be necessary, for instance through measuring switching statistics as function of bias-current sweep rate and temperature.
To gain further insights from the data we do have, we fit the temperature dependence of both $I_\mathrm{sw}$ and $I_0$ with a function
\begin{equation}
	I_\mathrm{x}(T_\mathrm{s}) = I_\mathrm{c, \mathrm{x}}\left( 1 - \frac{T_\mathrm{s}}{T_\mathrm{cc, x}} \right)^{\frac{3}{2}}
	\label{eqn:I0_Ts}
\end{equation}
where $\mathrm{x} = \{0, \mathrm{sw}\}$.
Although not based on a specific theory for constrictions, such a temperature-dependence reproduces very well the experimentally observed behaviour, cf. panel (c) in Fig.~\ref{fig:Figure4}.
Here, the fit parameters are the critical/switching current at zero temperature  $I_\mathrm{c, x}$, and the constriction critical temperature $T_\mathrm{cc, x}$.
We find $T_\mathrm{cc, 0} = 5.96\,$K and $T_\mathrm{cc, sw} = 5.98\,$K, so the two values are very close together, but quite different from $T_\mathrm{c} \approx 9.0\,$K of the superconducting film.
This demonstrates that our device corresponds to a so-called SS'S constriction, where the superconductor in the leads (S) is different from the superconductor forming the constriction itself (S').
The exact reason for both the reduction of $T_\mathrm{cc}$ compared to $T_\mathrm{c}$ and the difference between similar samples are currently not completely clear, but we suspect neon ion implantation or the creation of a thin normal-conducting surface layer to be the origin \cite{Troeman07}.
\vspace{-3mm}
\subsection*{Characterizing the nonlinear CPR corrections}
\vspace{-3mm}
\begin{figure*}
	\centerline{\includegraphics[width=0.9\textwidth]{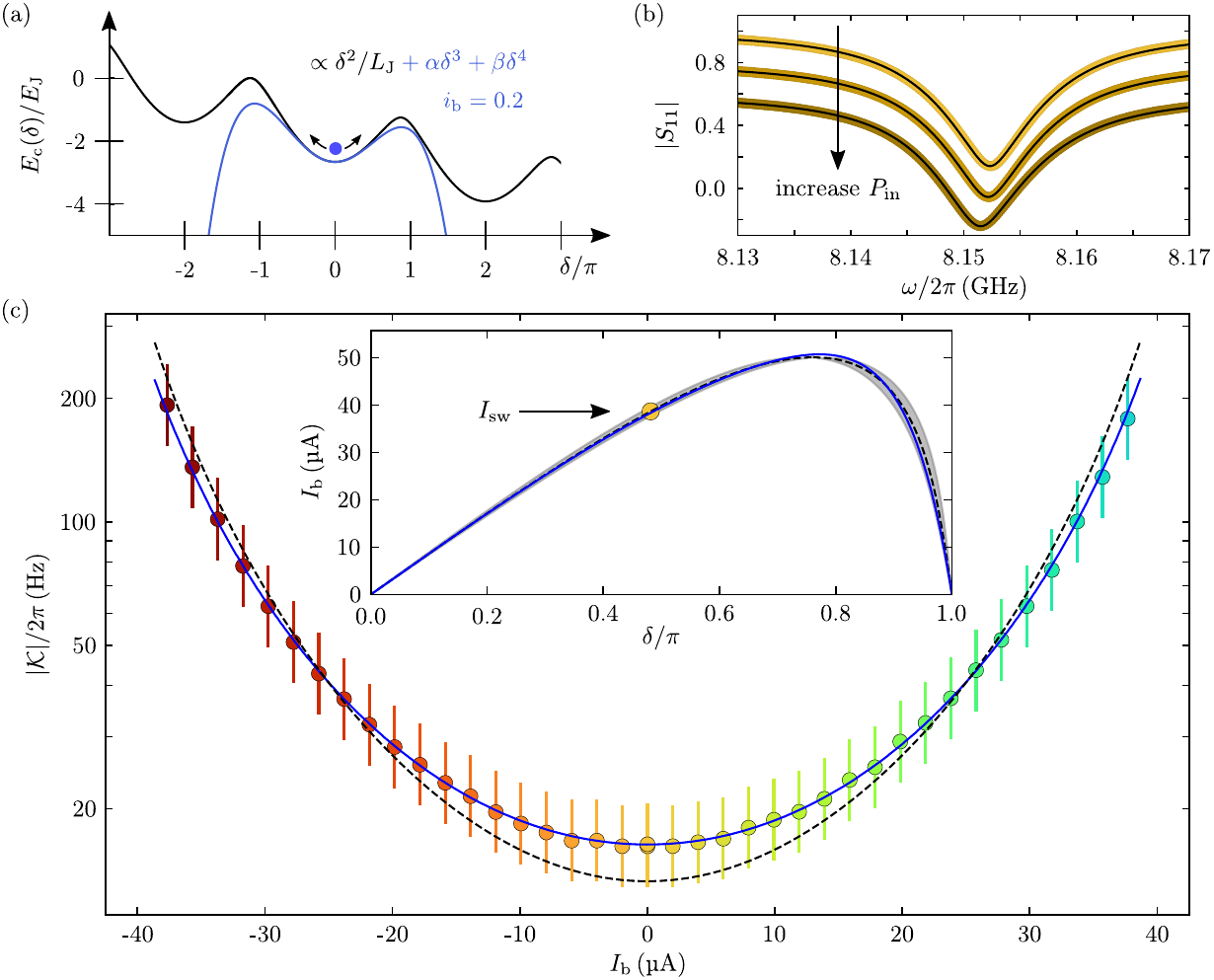}}
	\caption{\textsf{\textbf{Higher order derivatives of the CPR shape the Kerr anharmonicity.} (a) Schematic of a particle in a $2\pi$-periodic potential, which is oscillating with large amplitudes. For larger oscillation amplitudes than considered so far, the particle will experience a deviation from the harmonic approximation and the next relevant contribution in a Taylor-expansion of the potential is $\propto \delta^4$ for the case of $I_\mathrm{b} = 0$ or both third order and forth order corrections $\propto \delta^3$ and $\propto \delta^4$ for $I_\mathrm{b}>0$, respectively. Shown is the case for $i_\mathrm{b} = 0.2$. The Taylor-coefficients $\alpha$ and $\beta$ are proportional to the second and third derivatives of the CPR, respectively. (b) Reflection data $S_{11}$ of the cavity for three different microwave probe tone powers $P_\mathrm{in}$. Subsequent curves are offset by $-0.2$ each for better visibility. For the larger powers, the resonance shifts to lower frequencies and deviates from the Lorentzian lineshape. We fit the data using a nonlinear model (black lines), from which we obtain the Kerr constant $\mathcal{K}$, which has contributions from both the third order and the forth order correction terms, see Supplementary Note IV for details. The values of $|\mathcal{K}| = -\mathcal{K}$ for multiple bias currents are shown in panel (c), they increase in magnitude with increasing bias current from around $\mathcal{K} \approx 2\pi\cdot 18\,$Hz for $I_\mathrm{b} = 0$ to $\mathcal{K} \approx 2\pi\cdot 190\,$Hz for $I_\mathrm{b} \sim \pm I_\mathrm{sw}$. Error bars consider a $\pm1\,$dB uncertainty in on-chip power of the microwave tone. For comparison we plot two different theory lines based on nonlinear current-conservation calculations; the black dashed one takes as input the second and third derivatives of the experimentally determined CPRs (linear-plus-sinusoidal model), the blue solid one is based on an artificially created CPR, that resembles very closely the one based on the linear-plus-sinusoidal-inductance model for $I_\mathrm{b}<I_0$. Both CPRs are shown in direct comparison in the inset. The switching current in the experiment is marked by a yellow circle, the uncertainty in the linear-plus-sinusoidal CPR is shown as gray shade. Sample temperature for experimental data was $T_\mathrm{s} = 3.9\,$K.}}
	\label{fig:Figure5}
\end{figure*}
So far, we have operated in all experiments and analyses with low microwave powers and assuming the constriction to be a linear microwave inductance.
With higher microwave powers, however, we can also gather information about higher-order corrections to the periodic and non-parabolic constriction potential, and compare these higher-order corrections with the ones obtained from our CPR model.
The picture is simple: when we apply a microwave signal to the cavity, the phase particle in the anharmonic potential is oscillating around the minimum, and the oscillation amplitude depends on microwave excitation power and frequency.
For very small oscillations, the particle only feels a parabolic term, for higher powers the particle will experience the deviations, cf. also Fig.~\ref{fig:Figure5}(a).
Experimentally, these deviations will be observable by a dependence of the cavity resonance frequency on intracavity microwave photon number, which originates from a change of the time-averaged constriction inductance with higher powers.
It has been demonstrated, that both third order and fourth order nonlinearities in the potential lead to a shift of the resonance frequency proportional to intracavity photon number and that the shift of both terms can be condensed into a single effective Kerr constant $\mathcal{K}$ \cite{Frattini18}.
The total cavity frequency shift is then $\mathcal{K}n_\mathrm{c}$ with the intracavity photon number $n_\mathrm{c}$.
We determine the effective Kerr constant by measuring the power-dependence of the resonance line up to the regime where the nonlinearities kick in.
In this regime, the cavity lineshape begins to deform towards the famous shark-fin resonance of Duffing oscillators, i.e., it becomes asymmetric and the minimum moves towards lower frequencies, cf. Fig.~\ref{fig:Figure5}(b) and Supplementary Note VIII.
We also observe a small nonlinear damping effect, i.e., both the amplitude of $S_{11}$ in the minimum and the total linewidth increase with microwave power, most likely due to an increase of the time-averaged quasiparticle density.
We fit the experimentally obtained dataset based on the solution of the nonlinear cavity equation of motion \cite{Gely23}
\begin{equation}
	\dot{\alpha} = \left[i\left(\omega_0 + \mathcal{K}|\alpha|^2 \right) - \frac{\kappa_0 + \kappa_\mathrm{nl}|\alpha|^2}{2} \right]\alpha + i\sqrt{\kappa_\mathrm{e}}S_\mathrm{in}
\end{equation}
where $\alpha$ is the complex cavity field amplitude, which is normalized such that $|\alpha|^2 = n_\mathrm{c}$ is the intracavity photon number.
The microwave input field $S_\mathrm{in}$ is normalized such that $|S_\mathrm{in}|^2 = P_\mathrm{in}/(\hbar \omega)$ with the on-chip input power $P_\mathrm{in}$, the probe frequency $\omega$ and the reduced Planck constant $\hbar$.
As main fit parameter we obtain the Kerr constant $\mathcal{K}$ for each bias current $I_\mathrm{b}$.
More details on the fitting procedure, the free and fixed parameters during the fit, and on the error estimates can be found in Supplementary Note~VIII.
The obtained Kerr anharmonicity $\mathcal{K}$ increases strongly with increasing bias current $I_\mathrm{b}$, from $\sim 2\pi\cdot 18\,$Hz at $I_\mathrm{b} = 0$ to $\sim 2\pi\cdot 190\,$Hz at $I_\mathrm{b} \lesssim \pm I_\mathrm{sw}$, cf. Fig.~\ref{fig:Figure5}(c).
The theoretical relation between the CPR and the $\mathcal{K}$ is given by 
\begin{equation}
	\mathcal{K} = \frac{e^2}{2\hbar C_\mathrm{tot}}p_\mathrm{c}^3\left[\frac{g_4}{g_2} - \frac{3g_3^2}{g_2^2}\left(1-p_\mathrm{c} \right) - \frac{5}{3}\frac{g_3^2}{g_2^2}p_\mathrm{c} \right]
\end{equation}
where the coefficients
\begin{equation}
	g_{n} = \frac{\partial^{n-1} I}{\partial \delta^{n-1}}\bigg|_{\delta_0}
\end{equation}
encode the $(n-1)$-th derivatives of the CPR with respect to phase at the equilibrium phase $\delta_0$, and
\begin{equation}
	p_\mathrm{c} = \frac{L_\mathrm{c}(I_\mathrm{b})}{L_\mathrm{r} + L_\mathrm{c}(I_\mathrm{b})}
\end{equation}
is the constriction inductance participation ratio, cf. also Supplementary Note IV for the derivation.
Using the CPR, which we obtained from the linear-plus-sinusoidal model, we can easily calculate the theoretical Kerr constant now and the result is in acceptable agreement with the data. 
There are, however, also some deviations at the lowest and at the highest bias currents, the theoretical line is underestimating the Kerr constant at low currents and overestimating it at high currents.
Cetainly, there are some simple possible explanations for these deviations such as uncertainties in on-chip microwave power (error bars account for $\pm 1\,$dB), and fitting errors in external and internal cavity linewidths.
But there is also another, more interesting possibility: very small differences between the CPR fit-curve and the actual CPR.
To demonstrate how tiny differences in the CPR can have a large impact on the Kerr constant, we have therefore constructed an artificial CPR $I_\mathrm{ar}(\delta)$ based on an odd polynomial function (details in Supplementary Note VIII).
We then do a simultaneous fit of this new CPR to both the experimental CPR data and to the experimental Kerr data.
The result is shown as blue lines in Fig.~\ref{fig:Figure5}(c) and shows excellent agreement with both datasets.
The relative deviation $\left[I_\mathrm{ar}(\delta) - I(\delta)\right]/I(\delta)$ between this constructed CPR and the linear-plus-sinusoidal CPR $I(\delta)$ is smaller than $1\%$ in the complete range covered by our experiment ($I_\mathrm{b}\lesssim 40\,$\textmu A, $\delta \lesssim \pi/2$).
It also falls completely into the error range of the experimental CPR (gray-shaded area) for all phases except for a small region around $I_0$.
So even for nearly identical CPRs and very similar critical currents, the tiny details of the shape of the CPR, encoded in the $g_n$ coefficients, can have a strong impact on the actual experimental nonlinearities.
Most strikingly, and in stark contrast to perfectly sinusoidal CPR Josephson junctions, is there no unique relation between the junction inductance and the Kerr constant anymore, as can be seen from the values at $I_\mathrm{b} = 0$.
Both CPRs have identical first derivatives (=inductances) there, but Kerr constants that differ by $\gtrsim\sim 20\%$ due to the third derivatives being different.
Regarding possible applications, such small nonlinearities are well-suited for high dynamic range devices such as parametric amplifiers \cite{CastellanosBeltran08, Bergeal10, Xu23}, current detectors \cite{Schmidt20} or photon-pressure systems \cite{Bothner21, Rodrigues22}.
If desired, the Kerr constant could also be made much larger by simple adjustments, for instance by either making the constriction critical current smaller by milling it thinner \cite{Uhl23}, or by changing the circuit layout to a lumped element version with a very different $L_\mathrm{r}$.
One could even imagine to charaterize the cJJ just as we did here and then remove the cavity and circuitry around the cJJ again to replace it with a tailored target layout, optimized for exactly the existing junction.

\section*{Discussion}
\vspace{-2mm}

In conclusion, we have reported a superconducting half-wavelength microwave cavity with bias-current access, that has enabled us to comprehensively characterize the properties of a monolithically patterned 3D nano-constriction.
To this end we used a combination of microwave reflectometry and DC biasing of the device.
We demonstrated that we can tune both resonance frequency and cavity linewidth by biasing the nano-constriction with small currents in the \textmu A regime, a property potentially very useful for current detectors, parametric amplifiers or photon-pressure systems.
From the analysis of the bias-current- and temperature-dependent constriction-cavity properties, we were able to extract the constriction inductance and to reconstruct the constriction current-phase-relation for a wide range of temperatures.
We found a forward-skewed sinusoidal function, which is characteristic for nano-constrictions, and observed that the skewedness decreases with increasing temperature.
Furthermore, we found that the superconducting-resistive switching current of the constrictions is suppressed by about $20\%$ compared to the critical current and that the critical temperature of the constriction is considerably reduced compared to the bare niobium film.
On one hand our approach of measuring the CPR of a nonlinear superconducting inductance with DC-plus-microwave excitation has the disadvantage that we are limited to phases below the switching phase and cannot observe its behaviour in the regime of negative inductance (negative slope of the CPR).
This is possible in DC-SQUID experiments for instance \cite{Koops96, DellaRocca07, Nanda17, Wang23}.
On the other hand, however, can we directly compare and analyze switching current and critical current, and our approach allows us to characterize how the integration of the constriction into a microwave cavity impacts the circuit properties, in both cases biased and unbiased.
Our approach furthermore enables us to extract a value for the sub-switching constriction resistance, and by characterizing the Kerr constant we finally gain access to the derivatives of the CPR, something which is only possible in DC experiments by differentiating the resulting CPR dataset by hand.
Here, we could demonstrate by measuring and modeling the Kerr anharmonicity of the cavity, that very small changes of the absolute values of the CPR can lead to large differences in the higher order potential corrections though.
Interesting open questions are how exactly the constriction and cavity properties will behave at lower temperatures in the mK regime, and what the physical and/or technical origin is for the premature switching of the constriction.
The latter could be investigated by measuring switching statistics as a function of temperature, bias current sweep rates and in presence of various noise sources.
Further experiments could also be dedicated to investigate the origin of the suppression of $T_\mathrm{cc}$ compared to $T_\mathrm{c}$ and to the question if and how the CPR depends on constriction thickness and ion dose.

\subsection*{Acknowledgements}
\vspace{-2mm}
The authors thank Markus Turad, Ronny Löffler (both LISA$^+$) and Christoph Back for technical support.
This research was supported by the Deutsche Forschungsgemeinschaft (DFG) via grant numbers BO 6068/1-1, BO 6068/2-1 and KO 1303/13-2.
We also gratefully acknowledge support by the COST actions NANOCOHYBRI (CA16218) and SUPERQUMAP (CA21144).

\subsection*{Data and code availability}
\vspace{-2mm}
All data and processing scripts presented in this paper will be available via zenodo upon publication of this work.

\subsection*{Competing interest}
\vspace{-2mm}
The authors declare no competing interests.

\let\addcontentsline\oldaddcontentsline
\clearpage

\widetext
\begin{center}
	\noindent\textbf{\large Supplementary Material for:\\ \vspace{2mm} Extracting the current-phase-relation of a monolithic three-dimensional nano-constriction using a DC-current-tunable superconducting microwave cavity}
	
	\normalsize
	\vspace{.3cm}
	
	\noindent{K.~Uhl \textit{et al.}}
	\\
\end{center}
\vspace{0.2cm}

\tableofcontents

\newpage
\setcounter{figure}{0}
\setcounter{equation}{0}
\setcounter{table}{0}
\renewcommand{\figurename}{Supplementary Figure}
\renewcommand{\tablename}{Supplementary Table}
\renewcommand{\theequation}{S\arabic{equation}}
\renewcommand{\bibnumfmt}[1]{[S#1]}
\renewcommand{\citenumfont}[1]{S#1}

\section{Supplementary Note I: Device fabrication}
\label{Section:Fab}
\begin{itemize}
	\item \textbf{Step 1: Microwave cavity patterning.}
	\\
	The fabrication starts with sputtering $90\,$-nm-thick niobium (Nb) on top of a high-resistivity ($\rho > 10\,$k$\Omega\cdot$m) intrinsic two inch silicon wafer. 
	The thickness of the wafer is $\sim 500\,$\textmu m.
	Then, the complete wafer is covered with ma-P 1205 photoresist by spin-coating (resist thickness $\sim 600\,$nm) and structured by means of maskless scanning laser photolithography ($\lambda_\mathrm{Laser} = 365\,$nm).
	After development of the resist in ma-D 331/S for $25\,$s, the Nb film is dry etched by means of reactive ion etching using $\mathrm{SF}_6$.
	For cleaning, the wafer gets finally rinsed in multiple subsequent baths of acetone and isopropanol.
	\item \textbf{Step 2: Dielectric layer for the shunt capacitor.}
	\\
	As a second step, we again perform maskless scanning laser photolithography to define the areas on the chip, which will be covered with the dielectric for the parallel plate input shunt capacitor of the cavity. 
	After resist development identical to step 1, the wafer with the patterned resist structures is placed inside the vacuum chamber of a plasma-enhanced chemical vapour deposition (PECVD) system and is covered with $150\,$nm of silicon-nitride (Si$_3$N$_4$).
	Afterwards an ultrasonic-assisted lift-off procedure is performed in acetone, which removes the resist and all the Si$_3$N$_4$ except for the ellipsoidal plates on the parallel plate shunt capacitors, cf. Fig.~1 of the main paper.
	Finally, the wafer is rinsed in multiple baths of acetone and isopropanol again.
	\item \textbf{Step 3: Superconducting shunt capacitor top plate.}
	\\
	The third step is fully equivalent to step 2, but instead of PECVD-grown Si$_3$N$_4$, a $70\,$nm thick layer of niobium is deposited by magnetron sputtering.
	After liftoff in acetone, this second Nb layer is only covering the dielectric ellipse of the shunt capacitor and is removed at all other locations.
	To avoid getting shorts between the first and second niobium layers at the edge of the capacitor, the ellipsoid of the niobium top-layer is smaller than the corresponding Si$_3$N$_4$ ellipsoid by about $10\,$\textmu m along all edges.
	\item \textbf{Step 4: Dicing and mounting for pre-characterization.}
	\\
	At the end of the cavity fabrication, the wafer gets diced into individual $10\times10\,$mm$^2$ chips, and one chip at a time is mounted on a printed circuit board (PCB), where it is wirebonded to microwave feedlines and ground, and packaged in a radiation-tight copper housing.
	After mounting into the measurement setup, the pre-characterization of the device is performed.
	\item \textbf{Step 5: 3D constriction fabrication.}
	\\
	Each pre-characterized standing wave cavity contains a narrow part at its far end, where the constriction junction is placed after pre-characterization.
	To cut the constriction into the $\sim 3\,$\textmu m wide bridge, the sample is removed from the PCB and sample box again and mounted into a neon ion microscope (NIM).
	The NIM allows one to perform high-precision milling with a nano-scaled spot-size focused neon beam (Ne-FIB).   
	The 3D constriction patterning is performed by cutting two $\sim25\,$nm narrow slot-shaped rectangles from both sides into the bridge, and by additionally and simultaneously milling the constriction from the top with a third rectangle, but with a lower dose.
	The dose for the cut-through rectangles was chosen to be $18000\,$ions/nm$^2$ and on top of the constriction the dose was $6500\,$ions/nm$^2$.
	For this dose and an accelerating voltage of $20\,$kV, we expect a remaining constriction thickness of $\sim 40\,$nm.
	\item \textbf{Step 6: Dicing and mounting.}
	\\
	After the Ne-FIB cutting process the sample is mounted in the same way as in Step 4.
\end{itemize}

\section{Supplementary Note II: Measurement setup}
\label{Section:Setup}

\begin{figure*}
	\centerline{\includegraphics[width=0.75\textwidth]{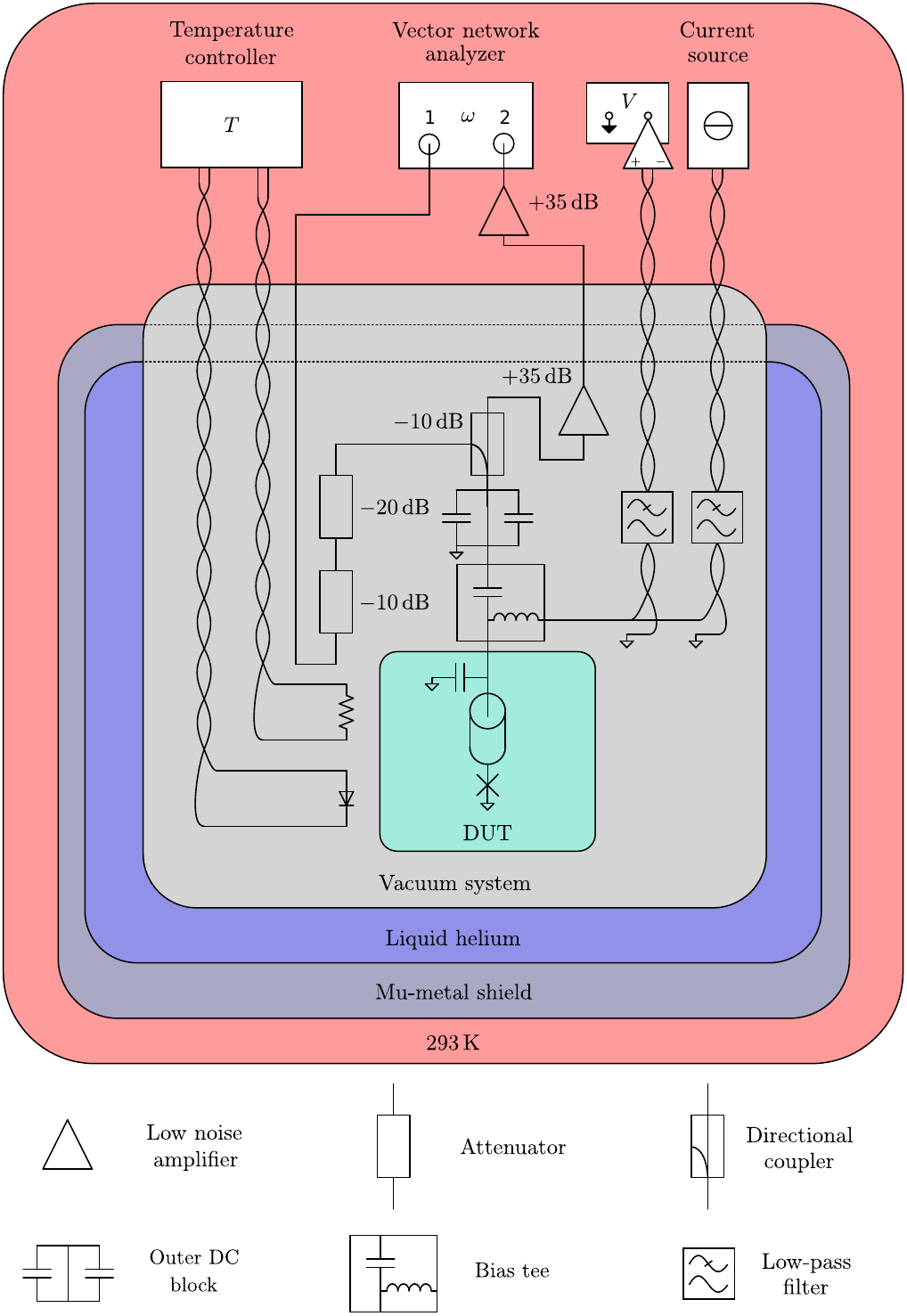}}
	\caption{\textsf{\textbf{Schematic of the measurement setup.} Detailed information is given in the text.}}
	\label{fig:FigureS1}
\end{figure*}
A schematic illustration of the measurement setup is shown in Supplementary Fig.~\ref{fig:FigureS1}.
Both the junction-less cavity and the cavity with constriction, here generically labeled as device under test (DUT), are characterized in an evacuated sample space located at the end of a cryostat dipstick, which is introduced into a liquid helium bath.
The cryostat is surrounded by a double-layer room-temperature mu-metal shield to provide magnetic shielding for the whole sample space. 
The DUT inside the copper housing is attached to an L-shaped copper mounting bracket, which is screwed to one of two circular mounting plates inside the evacuated sample volume.
The DUT is connected to a single coaxial cable for input and output of the microwave (measured in reflection) and the direct current (DC) signals (combined via bias-tee).
The microwave input and output signals, however, are sent and received through two separate coaxial lines, that are only combined via a directional coupler close to the device, in order to measure the reflection $S_{11}$ as transmission $S_\mathrm{21}$ by a vector network analyzer (VNA).
This way, the input signal can be strongly attenuated without any signal loss on the output line.
The input line is attenuated by $30\,$dB in order to balance the thermal radiation from room temperature to the cryostat temperature.
The directional coupler adds another $10\,$dB of input attenuation, as the input signal enters through the coupled port and the reflected signal propagates straight through it.
Including the cables, attenuators, couplers and all other components, we estimate the total attenuation between the VNA output and the device to be $\sim 53\pm 1\,$dB.
The attenuators are mounted in close proximity to the sample in the sample vacuum space and we assume them to have a temperature $T_\mathrm{att} \approx T_\mathrm{s}$, where $T_\mathrm{s}$ is the temperature of the sample.
For amplification of the weak microwave signal used here, a cryogenic high electron mobility transistor (HEMT) amplifier and a room temperature amplifier are mounted to the output line. 
The cryogenic HEMT is placed close to the DUT in order to minimize signal losses in between the sample and the amplifier chain.
In order to inject the DC current into the cavity and to measure the voltage in a near-ideal four-terminal configuration, a microwave bias-tee is added to the coaxial microwave lines just before the microwave signal reaches the device.
The DC port of the bias-tee and the ground are connected to two pairs of twisted copper wires by solder joints, one pair for the current and the other pair to measure the corresponding voltage.
Both twisted pairs are low pass filtered close to the microwave bias-tee with a cutoff frequency of $\sim 1.5\,$kHz in order to prevent noise in the kHz to MHz range from entering the device.
In order to minimize the low-frequency current and voltage noise entering the device through other channels, a ground-DC block is added in front of the bias-tee, which in combination with the bias-tee and the usage of non-conducting pieces and screws interrupts all galvanic connections between the cryostat/microwave lines and the sample box including the sample.
By these measures, we completely separate the cryostat/dipstick ground from the sample ground (the latter being highlighted with the empty triangular ground symbols in Supplementary Fig.~\ref{fig:FigureS1}).
The DC electronics are controlled via a National Instruments DAC/ADC measurement card, the current source has a floating ground and the voltage $V$ is preamplified by a room-temperature low-noise amplifier with a gain of $10^4$.
A temperature diode is attached to the sample housing/the mounting bracket in close proximity to the actual sample and both are coupled to the liquid helium bath via the copper mounting bracket and through a small amount of helium exchange gas. 
For controlling the sample temperature $T_\mathrm{s}$, the diode is glued with varnish to the DUT copper housing and a manganin heating resistor (made of a twisted pair wire to avoid stray magnetic fields) is placed nearby. 
Both the temperature diode (4 wires) and the heating resistor (2 wires) are also connected via twisted pairs of copper wire to a temperature controller. 
For cooling the device to temperatures below that of liquid helium ($4.2\,$K), we pump at the helium dewar of the cryostat and reach down to $T_\mathrm{s, min} \lesssim 2.4\,$K.
To achieve high-stability temperature control ($\Delta T_\mathrm{s} < 1 \,$mK) in the most relevant range for this work $2.4\,$K$ \lesssim T_\mathrm{s} \lesssim 6.5\,$K, we use the helium pumping and additionally heat the sample with the heating resistor, whose power is controlled via a PID feedback loop by the temperature controller.

\section{Supplementary Note III: The cavity model without constrictions}
\label{sec:NoteIII}
\subsection{Bare transmission line resonator}
Our device is based on a short-ended half wavelength coplanar waveguide cavity with a characteristic impedance $Z_1 = 32\,\Omega$ and a length of $l_1 = 7200\,$\textmu m.
The complex propagation constant along the transmission line resonator is given by $\gamma = \alpha + i\beta$ with the attenuation constant $\alpha$ and the phase constant $\beta = \omega/v_\phi$.
Here $\omega$ is the angular frequency of the propagating wave on the line and $v_\phi = 1/\sqrt{L'C'}$ is the phase velocity, where  $C'$ is the capacitance per unit length and $L'$ is the total inductance per unit length.
Note that $L'$ has both a geometric contribution $L_\mathrm{g}'$ and a kinetic contribution $L_\mathrm{k}'$ with $L' = L_\mathrm{g}' + L_\mathrm{k}'$.  
We can deduce the resonance frequency of the fundamental mode (without shunt capacitor and before junction cutting) as $\omega_1 = \pi v_\phi / l_1$.
For the input impedance of a short-ended and lossy transmission line at a distance $l_1$ from its shorted end, we have
\begin{equation}
	Z_\mathrm{in}^\mathrm{TL} = Z_1\frac{\tanh{\alpha l_1} + i\tan{\beta l_1}}{1 + i\tan{\beta l_1}\tanh{\alpha l_1}}
	\label{eqn:Z_input}
\end{equation}
which for small losses $\alpha l_1 \ll 1$ and close to its fundamental mode resonance $\omega \approx \omega_1$ can be Taylor-approximated as
\begin{equation}
	Z_\mathrm{in}^\mathrm{TL} \approx Z_1\alpha l_1 + iZ_1\pi\frac{\varDelta_1}{\omega_1}
\end{equation}
where $\varDelta_1 = \omega - \omega_1$.
When we compare this with the Taylor-approximated input impedance of a series RLC circuit
\begin{equation}
	Z_\mathrm{in}^\mathrm{RLC} \approx R_\mathrm{r} + 2iL_\mathrm{r}\varDelta_1
\end{equation}
we recognize that they actually look identical for
\begin{equation}
	R_\mathrm{r} = Z_1\alpha l_1, ~~~~~ L_\mathrm{r} = \frac{Z_1 \pi}{2\omega_1} = \frac{L'l_1}{2}, ~~~~~ C_\mathrm{r} = \frac{2}{\pi\omega_1 Z_1}=\frac{2C'l_1}{\pi^2}.
	\label{eqn:RLC_Circuit}
\end{equation}
From the lumped element equivalents, we can also now give expressions for the internal linewidth $\kappa_\mathrm{i, 1}$ and the internal quality factor $Q_\mathrm{i, 1} = \omega_1/\kappa_\mathrm{i, 1}$ of the resonator.
They are given as
\begin{equation}
	\kappa_\mathrm{i, 1} = \frac{R_\mathrm{r}}{L_\mathrm{r}} = \frac{2\omega_1 \alpha l_1}{\pi}, ~~~~~ Q_\mathrm{i, 1} = \frac{\pi}{2\alpha l_1}.
\end{equation}

\subsection{Shunt-coupled transmission line resonator}
When we couple the short-ended transmission line cavity to a microwave feedline with characteristic impedance $Z_0$ by means of a shunt capacitor to ground $C_\mathrm{s}$ as in our device, we need to consider a change in total capacitance as well as a splitting of the linewidth and quality factor into internal and external contributions.
To do this, we first consider the input impedance of the shunt-capacitor-and-feedline parallel combination as seen from the cavity, which is
\begin{equation}
	Z_\mathrm{e} = \left( \frac{1}{Z_0} + i\omega C_\mathrm{s} \right)^{-1} = \frac{Z_0}{1 + i\omega C_\mathrm{s}Z_0}.
\end{equation}
We can split this into its real and imaginary part and then write it as a combination of an effective frequency-dependent series combination of a resistor $R^*$ and capacitor $C^*$
\begin{equation}
	Z_\mathrm{e} = \frac{Z_0}{1 + \omega^2C_\mathrm{s}^2 Z_0^2} + \frac{1}{i\omega}\frac{\omega^2 C_\mathrm{s} Z_0^2}{1 + \omega^2C_\mathrm{s}^2 Z_0^2} = R^* + \frac{1}{i\omega C^*}.
\end{equation}
Next, we perform two approximations by using $\omega \approx \omega_\mathrm{b}$ and $\omega_\mathrm{b}^2 C_\mathrm{s}^2 Z_0^2 \gg 1$ and get
\begin{equation}
	R^* \approx \frac{1}{\omega_\mathrm{b}^2 C_\mathrm{s}^2 Z_0}, ~~~~~ C^* \approx C_\mathrm{s}
\end{equation}
which allows us to find the total capacitance
\begin{equation}
	C_\mathrm{tot} = \frac{C_\mathrm{r}C_\mathrm{s}}{C_\mathrm{r} + C_\mathrm{s}},
	\label{eqn:ctot}
\end{equation}
the new resonance frequency
\begin{equation}
	\omega_\mathrm{b} = \frac{1}{\sqrt{L_\mathrm{r}C_\mathrm{tot}}},
\end{equation}
the coupled resistance 
\begin{equation}
	R_\mathrm{b} = R_\mathrm{r} + R^*,
\end{equation}
and finally the total linewidth, which we can split into internal and external contibutions
\begin{eqnarray}
	\kappa_\mathrm{b} & = & \omega_\mathrm{b}^2 R_\mathrm{b} C_\mathrm{tot} \nonumber\\
	& = & \omega_\mathrm{b}^2 R_\mathrm{r} C_\mathrm{tot} + \omega_\mathrm{b}^2 R^* C_\mathrm{tot} \nonumber\\ 
	& = & \frac{\omega_\mathrm{b}^2 R_\mathrm{r} C_\mathrm{r} C_\mathrm{s}}{C_\mathrm{r} + C_\mathrm{s}} + \frac{C_\mathrm{r}}{Z_0 C_\mathrm{s}(C_\mathrm{r} + C_\mathrm{s})} \nonumber\\
	& = & \kappa_\mathrm{i, b} + \kappa_\mathrm{e, b}.
	\label{eqn:kappas}
\end{eqnarray}
The linewidths are related to the corresponding quality factors via $Q_\mathrm{i,b}=\omega_\mathrm{b}/\kappa_\mathrm{i,b}$ and $Q_\mathrm{e,b}=\omega_\mathrm{b}/\kappa_\mathrm{e,b}$.

\subsection{Circuit parameters and measurements}
\label{sec:NoteIIIC}

The parameters we need for our circuit are the total capacitance $C_\mathrm{tot}$ and the total inductance $L_\mathrm{r} = L_\mathrm{g} + L_\mathrm{k}$, which has a geometric and a kinetic contribution. 
What makes things more complicated on one hand but also experimentally accessible on the other hand is that the kinetic contributions are dependent on the niobium London penetration depth $\lambda_\mathrm{L}$, which is a function of sample temperature $T_\mathrm{s}$.
We start our parameter extraction procedure by calculating the geometric inductance per unit length 
\begin{equation}
	L_\mathrm{g}' = \frac{\mu_0}{4}\frac{K(k_1')}{K(k_1)},
\end{equation}
where $\mu_0$ is the vacuum permittivity, $K(k)$ is the complete elliptic integral of the first kind and 
\begin{equation}
	k_1 = \frac{S_1}{S_1 + 2W_1}, ~~~~~ k_1' = 1 - k_1^2.
\end{equation}
here, the width of the coplanar waveguide center conductor is $S_1 = 50\,$\textmu m and the gap between ground and center conductor is $W_1 = 5\,$\textmu m.
As result we obtain $L_\mathrm{g}' = 261\,$nH/m, a value that we also obtain with less than $1\%$ deviation by numerical simulations.
In addition to the geometric inductance we need to take the kinetic inductance into account. 
The relation between $\lambda_\mathrm{L}$ and the kinetic inductance per unit length is given by \cite{Watanabe94_SI}
\begin{equation}
	L_\mathrm{k}' = \mu_0 g \frac{\lambda_\mathrm{eff}}{S_1},
\end{equation}
where $d_\mathrm{Nb}$ is the film thickness, $g$ is a geometrical, dimensionless factor taking into account the details of the transmission line cavity by \cite{Watanabe94_SI, Clem13_SI}
\begin{equation}
	g = \frac{1}{2\left[k_1'K(k_1)\right]^2}\left[-\ln(\frac{d_\mathrm{Nb}}{4S_1})-k_1\ln(\frac{d_\mathrm{Nb}}{4(S_1+2W_1)})+2\frac{S_1+W_1}{S_1+2W_1}\ln(\frac{W_1}{S_1+W_1}) \right]
\end{equation}
and the effective penetration depth $\lambda_\mathrm{eff}$ of a thin film $d_\mathrm{Nb}\lesssim \lambda_\mathrm{L}$ is \cite{Klein90_SI}
\begin{equation}
	\lambda_\mathrm{eff} = \lambda_\mathrm{L}\coth{\frac{d_\mathrm{Nb}}{\lambda_\mathrm{L}}}.
\end{equation}
Then the total inductance per unit length is
\begin{equation}
	L' = L_\mathrm{g}' + \mu_0 g \frac{\lambda_\mathrm{L}\coth{\frac{d_\mathrm{Nb}}{\lambda_\mathrm{L}}}}{S_1}.
	\label{eqn:l_lambda}
\end{equation}
In the experiment, we do not vary directly $\lambda_\mathrm{L}$ but the sample temperatur $T_\mathrm{s}$, the relation between the two is given by
\begin{equation}
	\lambda_\mathrm{L}(T_\mathrm{s}) = \frac{\lambda_0}{\sqrt{1 - \left(\frac{T_\mathrm{s}}{T_\mathrm{c}}\right)^4}}
\end{equation}
where $\lambda_0$ is the zero-temperature penetration depth and $T_\mathrm{c}$ is the critical temperature.
We measure the cavity resonance frequency $\omega_\mathrm{b}(T_\mathrm{s})$ and fit the experimentally obtained data with
\begin{eqnarray}
	\omega_\mathrm{b}(T_\mathrm{s}) & = & \frac{1}{\sqrt{C_\mathrm{tot}L_\mathrm{r}(T_\mathrm{s})}} \nonumber \\
	& = & \frac{1}{\sqrt{C_\mathrm{tot}\frac{l_1}{2} \left[ L_\mathrm{g}' + \frac{\mu_0 g}{S_1} \frac{\lambda_0}{\sqrt{1 - \left(\frac{T_\mathrm{s}}{T_\mathrm{c}} \right)^4}}\coth{\left[ \frac{d_\mathrm{Nb}}{\lambda_0}\sqrt{1 - \left(\frac{T_\mathrm{s}}{T_\mathrm{c}}\right)^4} \right]} \right]}}
	\label{eqn:omega_t}
\end{eqnarray}
with the geometrical parameters $S_1 = 50\,$\textmu m, $W_1 = 5\,$\textmu m, $d_\mathrm{Nb} = 90\,$nm, $g = 3.8$, $l_1 = 7200\,$nm and $L_\mathrm{g}' = 261\,$nH$\cdot$m$^{-1}$ as constants and with $C_\mathrm{tot} = 367\,$fF, $\lambda_0 = 141\,$nm and $T_\mathrm{c} = 9.0\,$K as fit parameters.
The result is shown in Supplementary Fig.~\ref{fig:FigureS2}(a).
\begin{figure*}
	\centerline{\includegraphics[width=0.85\textwidth]{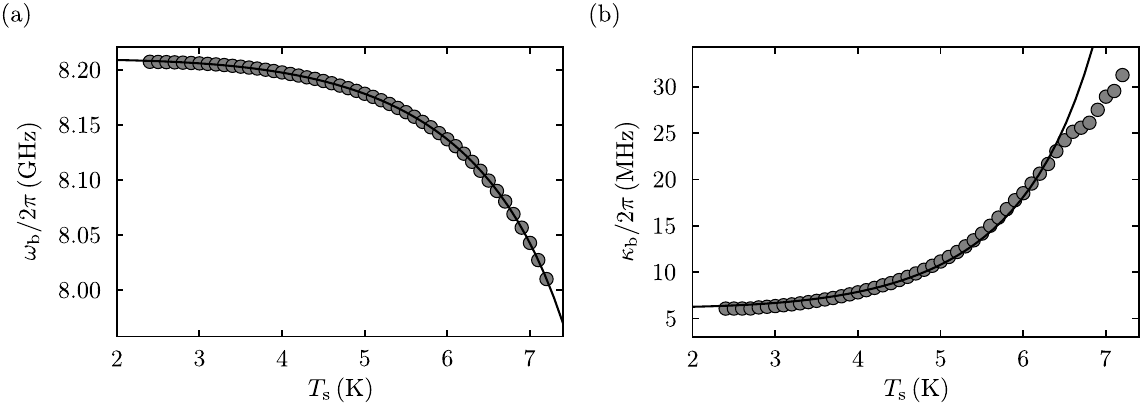}}
	\caption{\textsf{\textbf{Temperature dependence of cavity parameters before junction cutting.} In (a) we show the cavity resonance frequency $\omega_\mathrm{b}$ vs sample temperature $T_\mathrm{s}$. The shift in frequency with increasing temperature occurs due to a change of the total circuit inductance $L_\mathrm{r}(T_\mathrm{s}) = L_\mathrm{g} + L_\mathrm{k}(T_\mathrm{s})$. Circles are data, line is a fit using Eq.~(\ref{eqn:omega_t}) and with $T_\mathrm{c} = 9.0\,$K, $C_\mathrm{tot} = 367\,$fF and $\lambda_0 = 141\,$nm as fit parameters. In panel (b), the total circuit linewidth $\kappa_\mathrm{b} / 2 \pi$ vs $T_\mathrm{s}$ is shown. The linewidth increases with increasing temperature, indicating growing losses by thermal quasiparticles in the superconductor. Circles are data, line is a fit using Eq.~(\ref{eqn:kappa_T}) with $\kappa_\mathrm{0,b} = 2\pi \cdot 6.2\,$MHz and $A_\kappa = 4.1 \cdot 10^\mathrm{-15}\,$$\mathrm{s}^3 / \mathrm{m}^3$ as fit parameter.}}
	\label{fig:FigureS2}
\end{figure*}
The next relevant parameter is the effective shunt capacitance $C_\mathrm{s}$, which we obtain from measurement of the external linewidth $\kappa_\mathrm{e, b} = 2\pi\cdot 6.3\,$MHz and the knowledge of $Z_0 = 50\,\Omega$ and $C_\mathrm{tot}$.
By using Eqs.~(\ref{eqn:ctot}) and (\ref{eqn:kappas}) we obtain for the coupling shunt capacitance $C_\mathrm{s} = 17.5\,$pF and $C_\mathrm{r} = 375\,$fF for the cavity capacitance at $T_\mathrm{s} = 3.9\,$K.
We can compare this with the theoretical value for $C_\mathrm{r}^\mathrm{theo} = 388\,$fF, where we used
\begin{equation}
	C' = 4\epsilon_0\epsilon_\mathrm{eff}\frac{K(k_1)}{K(k_1')}
\end{equation}
with $\epsilon_\mathrm{eff} = (\epsilon_\mathrm{Si} + 1)/2$ and $\epsilon_\mathrm{Si} = 11.5$, and find reasonable agreement.
Similarly, we get for the shunt capacitance $C_\mathrm{s}^\mathrm{theo} = 12.4\,$pF, which we obtain by calculating $C_\mathrm{s1}^\mathrm{theo} = \epsilon_0\epsilon_\mathrm{SiN}A_1/d_\mathrm{SiN} = 24.9 \,$pF and $C_\mathrm{s2}^\mathrm{theo} = \epsilon_0\epsilon_\mathrm{SiN}A_2/d_\mathrm{SiN} = 24.9\,$pF.
Here, we used $\epsilon_\mathrm{SiN} = 7$, $A_1 = 0.06\,$mm$^2$, $A_2 = 0.06\,$mm$^2$ and $d_\mathrm{SiN} = 150\,$nm.
The two capacitances $C_\mathrm{s1}^\mathrm{theo}$ and $C_\mathrm{s2}^\mathrm{theo}$ are the capacitances between center conductor and top shunt electrode and between top shunt electrode and the ground planes, respectively, which in series result in $C_\mathrm{s}^\mathrm{theo} = C_\mathrm{s1}^\mathrm{theo}C_\mathrm{s2}^\mathrm{theo}/(C_\mathrm{s1}^\mathrm{theo}+C_\mathrm{s2}^\mathrm{theo})$.
Again, we find acceptable agreement, and possible deviations originate most likely from variations in $d_\mathrm{SiN}$ and $\epsilon_\mathrm{SiN}$, but possibly also from inductive contributions in the shunt capacitor plates or from the feedline input impedance deviating from $Z_0$ due to cable resonances and parasitic reflections.
In addition to the resonance frequency, we also extract the total resonance linewidth $\kappa_\mathrm{b}$ as a function of temperature, data are shown in Supplementary Fig.~\ref{fig:FigureS2}(b). 
At the elevated temperatures we are operating here $T_\mathrm{s} \gtrsim T_\mathrm{c}/4$, the internal decay rate will be dominated by quasiparticle losses.
From the two-fluid model, the effective surface resistance of a superconductor with the corresponding correction factor for thin films and around the cavity resonance frequency is given by \cite{Klein90_SI}
\begin{equation}
	R_\mathrm{s, eff} = \frac{1}{2}\omega_\mathrm{b}^2 \mu_0^2 \lambda_\mathrm{L}^3 \sigma_\mathrm{n} \frac{n_\mathrm{n}}{n}\left[\coth{\left( \frac{d_\mathrm{Nb}}{\lambda_\mathrm{L}}\right)} + \frac{d_\mathrm{Nb}/\lambda_\mathrm{L}}{\sinh^2{\left( \frac{d_\mathrm{Nb}}{\lambda_\mathrm{L}}\right)}} \right],
\end{equation}
where $\sigma_n$ is the normal state conductivity, $n_\mathrm{n}$ is the quasiparticle density and $n = n_\mathrm{n} + n_\mathrm{s}$ is the total electron density with $n_\mathrm{s}$ being the superconducting charge carrier density (twice the Cooper pair density).
The temperature dependence of the quasiparticle density is given by
\begin{equation}
	\frac{n_\mathrm{n}(T_\mathrm{s})}{n} = \left(\frac{T_\mathrm{s}}{T_\mathrm{c}}\right)^4.
\end{equation}
Since the quasiparticle loss channel is equivalent to the kinetic inductance channel in terms of current density distribution, the resulting circuit model lumped element resistance $R_\mathrm{r} \propto R_\mathrm{s, eff}$ is expected to be in series with $L_\mathrm{r}$.
Combining this result with Eq.~(\ref{eqn:kappas}) we get
\begin{eqnarray}
	\kappa_\mathrm{i, b}(T_\mathrm{s}) & = & \omega_\mathrm{b}^2(T_\mathrm{s}) R_\mathrm{r}(T_\mathrm{s}) C_\mathrm{tot} \nonumber \\
	& = & A_\kappa\omega_\mathrm{b}^4(T_\mathrm{s})\lambda_\mathrm{L}^3(T_\mathrm{s})\left( \frac{T_\mathrm{s}}{T_\mathrm{c}}\right)^4\left[\coth{\left( \frac{d_\mathrm{Nb}}{\lambda_\mathrm{L}(T_\mathrm{s})}\right)} + \frac{d_\mathrm{Nb}/\lambda_\mathrm{L}(T_\mathrm{s})}{\sinh^2{\left( \frac{d_\mathrm{Nb}}{\lambda_\mathrm{L}(T_\mathrm{s})}\right)}} \right]
	\label{eqn:kappa_qp}
\end{eqnarray}
with the fit parameter $A_\kappa$ that contains geometry, material properties and other temperature-independent contributions.
Since we are not certain that we can reliably discriminate between $\kappa_\mathrm{i, b}$ and $\kappa_\mathrm{e, b}$ due to cable resonances and impedance mismatches in the setup leading to Fano interferences, we fit the temperature dependence of the total linewidth using
\begin{equation}
	\kappa_\mathrm{b}(T_\mathrm{s}) = \kappa_\mathrm{0, b} + \kappa_\mathrm{i, b}(T_\mathrm{s})
	\label{eqn:kappa_T}
\end{equation}
with $\kappa_\mathrm{0, b}$ as second fit parameter.
The agreement between the experimental data and the fit line is very good, cf. Supplementary Fig.~\ref{fig:FigureS2}(b), with considerable deviations only appearing for $T_\mathrm{s} \gtrsim 6.5\,$K.
Note also that the fit value $\kappa_\mathrm{0, b} = 2\pi\cdot 6.2\,$MHz is very close to $\kappa_\mathrm{e, b} = 2\pi\cdot 6.3\,$MHz obtained at $T_\mathrm{s} = 3.9\,$K.
Supplementary Table~\ref{tab:Table1} summarizes all relevant cavity parameters before constriction cutting again in a single spot.
\begin{center}
	\begin{table}[h]
		\caption{\label{tab:params}\textsf{\textbf{Circuit parameters before cutting the nanobridge junction}. The geometric inductance $L_\mathrm{g}$ is obtained using $L_\mathrm{g} = L_\mathrm{g}'l_1/2$. From a fit to the temperature-dependence of $\omega_\mathrm{b}$ we obtain the zero-temperature penetration depth $\lambda_0$, the critical temperature $T_\mathrm{c}$ and the total capacitance $C_\mathrm{tot}$. Additionally we get the kinetic inductance $L_\mathrm{k} = L_\mathrm{k}'l_1/2$ and therefore the circuit inductance $L_\mathrm{r} = L_\mathrm{g} + L_\mathrm{k}$. From the measured external linewidth $\kappa_\mathrm{e, b}$ we subsequently find the coupling capacitance $C_\mathrm{s}$ and the circuit capacitance $C_\mathrm{r}$. For completeness we also give $\kappa_\mathrm{i, b}$. All experimental values are given for $T_\mathrm{s} = 3.9\,$K. \\}}
		\begin{tabular}{  l | l | l | l | l | l | l | l | l | l | l | l }
			$l_1\,$(nm) & $L_\mathrm{g}\,$(pH) & $L_\mathrm{k}\,$(pH) & $L_\mathrm{r}\,$(pH) & $C_\mathrm{r}\,$(fF) & $C_\mathrm{s}\,$(pF) & $C_\mathrm{tot}\,$(fF) & $\lambda_\mathrm{0}\,$(nm) & $T_\mathrm{c}\,$(K) & $\frac{\omega_\mathrm{b}}{2\pi}\,$ (GHz) &  $\frac{\kappa_\mathrm{e, b}}{2\pi}\,$(MHz) & $\frac{\kappa_\mathrm{i, b}}{2\pi}\,$(MHz) \\ \hline
			7200 & 939.4 & 87.4 & 1026.8 & 375 & 17.5 & 367 & 141 & 9 & 8.199 & 6.3 & 1.3 \\ \hline
		\end{tabular}
		\label{tab:Table1}
	\end{table}
\end{center}
\section{Supplementary Note IV: The cavity model with constrictions}
\label{Sec:NoteIV}
\subsection{The nano-constriction: Potential, current-phase-relation and inductance}
We model the constriction similar to an ideal Josephson junction by assuming a $2\pi$-periodic current-phase-relation (CPR) $I(\delta)$ and a $2\pi$-periodic corresponding potential energy $E_\mathrm{c}(\delta)$.
The relation between the two can be established by
\begin{equation}
	E_\mathrm{c}(\delta) = \int_0^t I(\delta)V(t')dt'.
\end{equation}
With the voltage
\begin{equation}
	V(t) = \frac{\Phi_0}{2\pi}\dot{\delta}
\end{equation}
this can be written as
\begin{equation}
	E_\mathrm{c}(\delta) = \frac{\Phi_0}{2\pi}\int I(\delta)d\delta,
\end{equation}
i.e., the CPR is essentially the derivative of the potential energy
\begin{equation}
	I(\delta) = \frac{2\pi}{\Phi_0}\frac{\partial E_\mathrm{c}}{\partial\delta}.
\end{equation}
We can Taylor-expand the total potential up to fourth order around the equilibrium phase $\delta_0$ and get
\begin{equation}
	E(\delta) = E_\mathrm{c}(\delta_0) + \frac{\partial E_\mathrm{c}}{\partial\delta}\bigg|_{\delta_0} \delta + \frac{1}{2}\frac{\partial^2 E_\mathrm{c}}{\partial\delta^2}\bigg|_{\delta_0} \delta^2 + \frac{1}{6}\frac{\partial^3 E_\mathrm{c}}{\partial\delta^3}\bigg|_{\delta_0} \delta^3 + \frac{1}{24}\frac{\partial^4 E_\mathrm{c}}{\partial\delta^4}\bigg|_{\delta_0} \delta^4 - \frac{\Phi_0 I_\mathrm{b}}{2\pi}\delta
\end{equation}
where we also included a tilt of the potential $E_\mathrm{c}$ by the bias current $I_\mathrm{b}$ and where $\delta$ (kept the same for simplicity) is the new dynamical variable around the equilibrium phase $\delta_0$.
For very small $\delta$, we can drop the third and fourth order terms as well as the constant offset and get
\begin{equation}
	E(\delta) \approx \frac{\partial E_\mathrm{c}}{\partial\delta}\bigg|_{\delta_0} \delta + \frac{1}{2}\frac{\partial^2 E_\mathrm{c}}{\partial\delta^2}\bigg|_{\delta_0} \delta^2  - \frac{\Phi_0 I_\mathrm{b}}{2\pi}\delta.
\end{equation}
From here we find the equilibrium position $\delta_0$ from the condition $\partial E/\partial\delta = 0$, i.e., by
\begin{equation}
	\frac{\partial E_\mathrm{c}}{\partial\delta}\bigg|_{\delta_0} = \frac{\Phi_0 I_\mathrm{b}}{2\pi}
\end{equation}
or
\begin{equation}
	I(\delta_0) = I_\mathrm{b}.
\end{equation}
Using the generalized flux $\varPhi = \frac{\Phi_0}{2\pi}\delta$, we can write the last remaining term in the potential as an inductive energy
\begin{equation}
	E_\mathrm{c}(\Phi) \approx  \frac{\varPhi^2}{2L_\mathrm{c}}
\end{equation}
with the constriction inductance
\begin{equation}
	L_\mathrm{c} = \frac{\Phi_0^2}{4\pi^2}\left( \frac{\partial^2 E_\mathrm{c}}{\partial \delta^2} \bigg|_{\delta_0}\right)^{-1}.
\end{equation}
In the next subsection, we will discuss how to model the overall cavity when including this inductance as well as a corresponding resistor, then how to reconstruct the CPR from the meaurement of $L_\mathrm{c}$ for different $I_\mathrm{b}$, and finally we will discuss the consequences of the higher order terms to the high-power dynamics of the resonator.

\subsection{Shunt-coupled transmission line resonator with nano-constriction}
We observe that cutting the constrictions into the circuit leads to a shift of the resonance frequency and to a broadening of the resonance linewidth.
Similar to the two-fluid model and following our considerations in the previous subsection, we therefore model the circuit elements introduced by the junction for low powers as a constriction inductance $L_\mathrm{c}$ in parallel with a constriction resistance $R_\mathrm{c}$.
We note that we omit any additional capacitance, as according to our simulations the impedance of a possible constriction capacitance is negligible compared to its inductance impedance.
For the parallel combination of $R_\mathrm{c}$ and $L_\mathrm{c}$ near the relevant frequency $\omega_0$, we get the input impedance
\begin{eqnarray}
	Z_\mathrm{c} & = & \frac{i\omega L_\mathrm{c} R_\mathrm{c}}{R_\mathrm{c} + i\omega L_\mathrm{c}} \nonumber \\
	& \approx & \frac{R_c\omega_0^2L_\mathrm{c}^2}{R_\mathrm{c}^2 + \omega_0^2 L_\mathrm{c}^2} + i\omega \frac{L_\mathrm{c}R_\mathrm{c}^2}{R_\mathrm{c}^2 + \omega_0^2 L_\mathrm{c}^2} \nonumber \\
	& = & R_\mathrm{c}^* + i\omega L_\mathrm{c}^*.
\end{eqnarray}
Unfortunately, we cannot do any approximation here, as a priori we cannot assume anything for the ratio $\omega_0 L_\mathrm{c}/R_\mathrm{c}$.
However, adding these new series elements to the circuit, we get the final resonator elements
\begin{eqnarray}
	R_\mathrm{tot} & = & R_\mathrm{r} + R^* + R_\mathrm{c}^* \\
	L_\mathrm{tot} & = & L_\mathrm{r} + L_\mathrm{c}^* \\
	C_\mathrm{tot} & = & \frac{C_\mathrm{r}C_\mathrm{s}}{C_\mathrm{r} + C_\mathrm{s}}.
\end{eqnarray}
and we can express the final resonance frequency as
\begin{eqnarray}
	\omega_\mathrm{0} & = & \frac{1}{\sqrt{C_\mathrm{tot}\left(L_\mathrm{r} + L_\mathrm{c}^* \right)}} \nonumber \\
	& \approx & \frac{\omega_\mathrm{b}}{ 1 + \frac{L_\mathrm{c}^*}{2L_\mathrm{r}}}
	\label{eqn:w0_JJ}
\end{eqnarray}
where the approximation is valid for $L_\mathrm{c}^* \ll L_\mathrm{r}$, a condition safely fulfilled in our device.
For the linewidth we find
\begin{eqnarray}
	\kappa_0 & = & \omega_\mathrm{0}^2 R_\mathrm{tot} C_\mathrm{tot} \nonumber\\
	& = & \omega_\mathrm{0}^2 R_\mathrm{r} C_\mathrm{tot} + \omega_\mathrm{0}^2 R_\mathrm{c}^* C_\mathrm{tot} + \omega_\mathrm{0}^2 R^* C_\mathrm{tot} \nonumber\\
	& = & \kappa_\mathrm{i} + \kappa_\mathrm{e}.
	\label{eqn:kappa_JJ}
\end{eqnarray}
where
\begin{equation}
	\kappa_\mathrm{i} = \omega_\mathrm{0}^2 R_\mathrm{r} C_\mathrm{tot} + \omega_\mathrm{0}^2 R_\mathrm{c}^* C_\mathrm{tot}
\end{equation}
contains also the junction contribution from $R_\mathrm{c}$.
An alternative route to finding expressions for $\omega_0$ and $\kappa_\mathrm{i}$ starts with equalizing the input impedances of the transmission line resonator at the point of the constriction and the input impedance of the cJJ
\begin{eqnarray}
	Z_1\alpha l_1 + iZ_1\tan{\beta l_1} = - R_\mathrm{c}^* - i\omega L_\mathrm{c}^*
\end{eqnarray}
where the minus signs on the right hand side stem from the opposite current directions in the two parts.
When we use our expressions for $R_\mathrm{r}$ and $L_\mathrm{r}$ and Taylor-expand the tangent function around the first mode, we obtain
\begin{eqnarray}
	R_\mathrm{r} + i\frac{2L_\mathrm{r}\omega_\mathrm{1}}{\pi}\left(\beta l_1 - \pi  \right) = - R_\mathrm{c}^* - i\omega L_\mathrm{c}^*.
\end{eqnarray}
Now we use $\beta = \omega/v_\phi$ and look for a complex-valued solution for $\omega$, that we call $\tilde{\omega}_0 = \omega_0 + i\frac{\kappa_\mathrm{i}}{2}$.
We get
\begin{equation}
	\tilde{\omega}_0 = \frac{\omega_1}{1 + \frac{L_\mathrm{c}^*}{2L_\mathrm{r}}} + i\frac{R_\mathrm{r} + R_\mathrm{c}^*}{2L_\mathrm{r} + L_\mathrm{c}^*}.
\end{equation}
This is equivalent to
\begin{eqnarray}
	\omega_0 & = & \frac{\omega_1}{1 + \frac{L_\mathrm{c}^*}{2L_\mathrm{r}}} \\
	\kappa_\mathrm{i} & = & \frac{R_\mathrm{r} + R_\mathrm{c}^*}{L_\mathrm{r} + L_\mathrm{c}^*/2}
\end{eqnarray}
which is slightly different from our earlier obtained lumped element expression, since we did not take the shunt capacitance into account and that it leads to a shift from $\omega_1$ to $\omega_\mathrm{b}$ already.
If we replace the constriction-less resonance frequency $\omega_1$ by $\omega_\mathrm{b}$ though, we get for the resonance frequency the above result
\begin{equation}
	\omega_0 = \frac{\omega_\mathrm{b}}{1 + \frac{L_\mathrm{c}^*}{2L_\mathrm{r}}}.
\end{equation}
The deviation in $\kappa_\mathrm{i}$ is given by the ratio $(L_\mathrm{r} + L_\mathrm{c}^*)/(L_\mathrm{r} + L_\mathrm{c}^*/2)$, which is one the order of $\omega_\mathrm{b}/\omega_0 \lesssim 1.01$.

\subsection{Getting the CPR from the constriction inductance}
Based on our above considerations, we can use that the second derivative of the potential is the first derivative of the CPR
\begin{equation}
	\frac{\partial^2 E_\mathrm{c}}{\partial\delta^2} = \frac{\Phi_0}{2\pi}\frac{\partial I}{\partial \delta}
\end{equation}
and insert this into our equation for the constriction inductance
\begin{equation}
	L_\mathrm{c} = \frac{\Phi_0}{2\pi}\left( \frac{\partial I}{\partial \delta} \bigg|_{\delta_0} \right)^{-1}.
\end{equation}
So once we know the constriction inductance $L_\mathrm{c}$ for each bias current and corresponding equilibrium phase, we can invert this relation to get
\begin{equation}
	\delta = \frac{2\pi}{\Phi_0}\int_0^{I_\mathrm{b}}L_\mathrm{c} dI + \delta_\mathrm{off}.
\end{equation}
We assume the offset phase $\delta_\mathrm{off} = 0$.
In practice, we have a finite number of points and do 
\begin{equation}
	\delta_j = \frac{2\pi}{\Phi_0}\sum_j^N L_{\mathrm{c}, j} \Delta I.
\end{equation}
instead with $j > 0$, $N$ the total number of bias-current values and $\Delta I = I_{\mathrm{b}, j} - I_{\mathrm{b}, j-1}$.
Finally, we can plot pairwise values for $(\delta_j, I_{\mathrm{b}, j})$, i.e., the experimentally determined CPR.

\subsection{Kerr anharmonicity of the circuit}
So far, we have neglected the third and forth order terms of the Taylor series of $E_\mathrm{c}$.
We will now derive the resulting Kerr constant $\mathcal{K}$ in two different ways.
In one derivation, we will not make any assumptions about the shape of the CPR, except that it is $2\pi$-periodic and single-valued.
In the second derivation further below, we will assume the constriction to be a series combination of a linear inductance and an ideal Josephson inductance with a sinusoidal CPR.
Both derivations lead to the same $\mathcal{K}$ here.
In the first derivation, we start by assuming that the potential can be described by a prefactor $E_\mathrm{J}$ multiplied by a $2\pi$-periodic function.
Then, we can write the Taylor series as
\begin{equation}
	\frac{E(\delta)}{E_\mathrm{J}} = \tilde{g}_0 + \tilde{g}_1 \delta + \frac{1}{2}\tilde{g}_2 \delta^2 + \frac{1}{6}\tilde{g}_3 \delta^3 + \frac{1}{24}\tilde{g}_4 \delta^4 - \frac{\Phi_0 I_\mathrm{b}}{2\pi E_\mathrm{J}}\delta
\end{equation}
where the coefficients are given by
\begin{equation}
	\tilde{g}_n = \frac{1}{E_\mathrm{J}}\frac{\partial^n E_\mathrm{c}}{\partial \delta^n}\bigg|_\mathrm{\delta_0}.
\end{equation}
Of course, we can also write the coefficients for $n>0$ as
\begin{equation}
	\tilde{g}_n = \frac{\Phi_0}{2\pi E_\mathrm{J}}\frac{\partial^{n-1} I}{\partial \delta^{n-1}}\bigg|_\mathrm{\delta_0},
\end{equation}
which gives a much more direct relation to the CPR.
The minimum phase is determined by $\tilde{g}_1 = \frac{\Phi_0 I_\mathrm{b}}{2\pi E_\mathrm{J}}$ or just by $I(\delta_0) = I_\mathrm{b}$.
We can also obtain $\delta_0$ experimentally as described above for the reconstruction of the CPR, when we just integrate up to $I_\mathrm{b}$.
Taking also into account the additional linear inductance of the cavity $L_\mathrm{r}$ and following Ref.~\cite{Frattini18_SI} now, we find that we need to define a total phase variable $\delta_\mathrm{tot}$, that describes the phase across the series combination of $L_\mathrm{r}$ and the constriction.
With the linear inductive energy $E_L = \frac{\Phi_0^2}{4\pi^2 L_\mathrm{r}}$, the total inductive energy is given by
\begin{equation}
	E_\mathrm{tot}(\delta_\mathrm{tot}, \delta) = \frac{1}{2}E_L\left(\delta_\mathrm{tot} - \delta \right)^2 + E_\mathrm{c}(\delta) - \frac{\Phi_0 I_\mathrm{b}}{2\pi}\delta_\mathrm{tot}.
\end{equation}
From the condition $\partial E_\mathrm{tot}/\partial\delta = 0$ we find the current-conservation condition \cite{Frattini18_SI}
\begin{equation}
	0 = E_L\left(\delta - \delta_\mathrm{tot}\right) + \frac{\partial E_\mathrm{c}}{\partial\delta}
\end{equation}
which defines $\delta$ as a function of $\delta_\mathrm{tot}$, i.e., $\delta[\delta_\mathrm{tot}]$.
This function allows us to calculate the derivative now
\begin{equation}
	\frac{\partial\delta}{\partial\delta_\mathrm{tot}} = \frac{E_L}{E_L + \frac{\partial^2 E_\mathrm{c}}{\partial\delta^2}}
\end{equation}
as well as the corresponding higher order derivatives $\partial^2\delta/\partial\delta_\mathrm{tot}^2$ and $\partial^3\delta/\partial\delta_\mathrm{tot}^3$.
Similarly to above, we now Taylor-expand the total potential as
\begin{equation}
	\frac{E_\mathrm{tot}(\delta_\mathrm{tot})}{E_\mathrm{J}} = \tilde{G}_0 + \tilde{G}_1 \delta_\mathrm{tot} + \frac{1}{2}\tilde{G}_2 \delta_\mathrm{tot}^2 + \frac{1}{6}\tilde{G}_3 \delta_\mathrm{tot}^3 + \frac{1}{24}\tilde{G}_4 \delta_\mathrm{tot}^4 - \frac{\Phi_0 I_\mathrm{b}}{2\pi E_\mathrm{J}}\delta_\mathrm{tot}
\end{equation}
where the new coefficients (for $n\geq 2$) are given by
\begin{eqnarray}
	\tilde{G}_2 & = & \frac{E_L}{E_\mathrm{J}}\left( 1 - \frac{\partial\delta}{\partial\delta_\mathrm{tot}} \right) \\
	\tilde{G}_3 & = & -\frac{E_L}{E_\mathrm{J}}\frac{\partial^2\delta}{\partial\delta_\mathrm{tot}^2} \\
	\tilde{G}_4 & = & -\frac{E_L}{E_\mathrm{J}}\frac{\partial^3\delta}{\partial\delta_\mathrm{tot}^3}
\end{eqnarray}
After some algebra and using
\begin{equation}
	p_\mathrm{c} = \frac{L_\mathrm{c}}{L_\mathrm{r} + L_\mathrm{c}}
\end{equation}
we find that the coefficients can be expressed as
\begin{eqnarray}
	\tilde{G}_2 & = & p_\mathrm{c}\tilde{g}_2 \\
	\tilde{G}_3 & = & p_\mathrm{c}^3 \tilde{g}_3 \\
	\tilde{G}_4 & = & p_\mathrm{c}^4\left[\tilde{g}_4 - \frac{3\tilde{g}_3^2}{\tilde{g}_2}\left(1-p_\mathrm{c}\right) \right]
\end{eqnarray}
and that the Kerr constant in the end is given by \cite{Frattini18_SI}
\begin{equation}
	\mathcal{K} = \frac{e^2}{2\hbar C_\mathrm{tot}}p_\mathrm{c}^3\left[\frac{\tilde{g}_4}{\tilde{g}_2} - \frac{3\tilde{g}_3^2}{\tilde{g}_2^2}\left(1 - p_\mathrm{c} \right) - \frac{5}{3} \frac{\tilde{g}_3^2}{\tilde{g}_2^2}p_\mathrm{c}\right].
\end{equation}
In order to make the very close connection to the system CPR obvious again, we define new coefficients in terms of the CPR $I(\delta)$ as
\begin{eqnarray}
	g_2 & = & \frac{\partial I}{\partial \delta} \\
	g_3 & = & \frac{\partial^2 I}{\partial \delta^2} \\
	g_4 & = & \frac{\partial^3 I}{\partial \delta^3}.
\end{eqnarray}
These are connected to the previous coefficients by $g_n = 2\pi E_\mathrm{J}\tilde{g}_n /\Phi_0$ and lead to the Kerr constant
\begin{equation}
	\mathcal{K} = \frac{e^2}{2\hbar C_\mathrm{tot}}p_\mathrm{c}^3\left[\frac{g_4}{g_2} - \frac{3g_3^2}{g_2^2}\left(1 - p_\mathrm{c} \right) - \frac{5}{3} \frac{g_3^2}{g_2^2}p_\mathrm{c}\right].
\end{equation}
In the second variant we split the constriction inductance into a linear part $L_\mathrm{lin}$ and the Josephson part $L_\mathrm{J}$.
As phase variables we consider the phase $\delta_\mathrm{J}$ across the Josephson part $L_\mathrm{J}$ with a sinusoidal CPR $I = I_0\sin{\delta_\mathrm{J}}$ and the total phase $\delta_\mathrm{tot}$.
The total linear inductance is the given by $L_\mathrm{r}' = L_\mathrm{r} + L_\mathrm{lin}$.
The equilibrium phase is given by
\begin{equation}
	\delta_\mathrm{J, 0} = \arcsin{\frac{I_\mathrm{b}}{I_0}}
\end{equation}
and the Josephson energy by $E_\mathrm{J} = \frac{\Phi_0 I_0}{2\pi}$.
With the modified coefficients
\begin{equation}
	c_n = \frac{1}{E_\mathrm{J}}\frac{\partial^n E_\mathrm{cos}}{\partial \delta_\mathrm{J}^n}\bigg|_{\delta_\mathrm{J, 0}}
\end{equation}
and $E_\mathrm{cos} = E_\mathrm{J}(1-\cos{\delta_\mathrm{J}})$ we follow the same derivation steps as above and find the Kerr constant as
\begin{equation}
	\mathcal{K} = \frac{e^2}{2\hbar C_\mathrm{tot}}p_\mathrm{J}^3\left[\frac{c_4}{c_2} - \frac{3c_3^2}{c_2^2}\left( 1 - p_\mathrm{J}\right) - \frac{5}{3} \frac{c_3^2}{c_2^2}p_\mathrm{J}\right]
\end{equation}
where 
\begin{equation}
	p_\mathrm{J} = \frac{L_\mathrm{J}}{L_\mathrm{r}' + L_\mathrm{J}}, ~~~ L_\mathrm{J} = \frac{\Phi_0}{2\pi I_0 \cos{\delta_\mathrm{J, 0}}}.
\end{equation}
\section{Supplementary Note V: Circuit response model}
\label{sec:Note6}
\subsection{Equation of motion and general considerations}
We model the classical intracavity field $\alpha$ of the constriction cavity with effective Kerr nonlinearity and nonlinear damping using the equation of motion \cite{Gely23_SI}
\begin{equation}
	\dot{\alpha} = \left[i(\omega_\mathrm{c} + \mathcal{K}|\alpha|^2) - \frac{\kappa + \kappa_\mathrm{nl}|\alpha|^2}{2}\right]\alpha + i\sqrt{\kappa_\mathrm{e}}S_\mathrm{in}.
\end{equation}
Here, $\omega_\mathrm{c}$ is the cavity resonance frequency ($=\omega_\mathrm{b}$ before cutting and $=\omega_0$ after), $\mathcal{K}$ is the Kerr nonlinearity (frequency shift per photon), $\kappa$ is the bare total linewidth ($=\kappa_\mathrm{b}$ before cutting and $=\kappa_0$ after), $\kappa_\mathrm{nl}$ is the nonlinear damping constant, $\kappa_\mathrm{e}$ is the external linewidth ($=\kappa_\mathrm{e, b}$ before cutting) and $S_\mathrm{in}$ is the input field.
The intracavity field is normalized such that $|\alpha|^2 = n_\mathrm{c}$ corresponds to the intracavity photon number $n_\mathrm{c}$ and $|S_\mathrm{in}|^2$ to the input photon flux (photons per second) on the coplanar waveguide feedline.
We do not explicitly take into account the third order nonlinearity here, but the nonlinear frequency shift that results from its existence is contained in $\mathcal{K}$ as described above.
The solution of this equation of motion depends significantly on the input power.
The ideal reflection response function, however, will always be of the form
\begin{equation}
	S_{11}^\mathrm{ideal} = - 1 - i\sqrt{\kappa_\mathrm{e}}\frac{\alpha}{S_\mathrm{in}}
\end{equation}
with the solution of interest $\alpha$.

\subsection{The linear single-tone regime}
In the linear single-tone regime, valid for low microwave-probing powers, we set $\mathcal{K} = \kappa_\mathrm{nl} = 0$.
Then, we can solve the remaining equation by Fourier transform and obtain
\begin{equation}
	\alpha = \frac{i\sqrt{\kappa_\mathrm{e}}}{\frac{\kappa}{2} + i(\omega - \omega_\mathrm{c})}S_\mathrm{in}.
\end{equation}
The ideal reflection response is then given by
\begin{equation}
	S_{11}^\mathrm{ideal} = - 1 + \frac{2\kappa_\mathrm{e}}{\kappa + 2i(\omega - \omega_\mathrm{c})}.
	\label{eqn:S21ideal}
\end{equation}

\subsection{The nonlinear single-tone regime}
In the nonlinear single-tone regime, we have to solve the full equation of motion and start by setting the input field to $S_\mathrm{in} = S_0e^{i\phi}e^{i\omega t}$ with real-valued $S_0$.
For the intracavity field, we make the Ansatz $\alpha(t) = \alpha_0e^{i\omega t}$ with real-valued $\alpha_0$.
The phase delay between input and response is fully encoded in $\phi$.
Then the equation of motion reads
\begin{equation}
	i\omega \alpha_0 = \left[i\left(\omega_\mathrm{c} + \mathcal{K}\alpha_0^2\right) - \frac{\kappa + \kappa_\mathrm{nl}\alpha_0^2}{2}\right]\alpha_0 + i\sqrt{\kappa_\mathrm{e}}S_0e^{i\phi}
	\label{eqn:NL_EOM}
\end{equation}
which after multiplication with its complex conjugate yields the characteristic polynomial for the intracircuit photon number $n_\mathrm{c} = \alpha_0^2$
\begin{equation}
	n_\mathrm{c}^3\left[\mathcal{K}^2 + \frac{\kappa_\mathrm{nl}^2}{4}\right] + n_\mathrm{c}^2\left[ \frac{\kappa \kappa_\mathrm{nl}}{2} - 2\mathcal{K}\varDelta \right] + n_\mathrm{c}\left[\varDelta^2 + \frac{\kappa^2}{4}\right] - \kappa_\mathrm{e}S_0^2 = 0.
	\label{eqn:ploy_st}
\end{equation}
Here $\varDelta = \omega - \omega_\mathrm{c}$ is the detuning between the microwave input tone and the bare cavity resonance.
The real-valued roots of this polynomial correspond to the physical solutions for the amplitude $\alpha_0$, the highest and lowest amplitudes are the stable states in the case of three real-valued roots.
For the complete complex reflection, we also need the phase $\phi$, which we obtain via
\begin{equation}
	\phi = \atan2\left(-\frac{\kappa + \kappa_\mathrm{nl}n_\mathrm{c}}{2}, \varDelta - \mathcal{K}n_\mathrm{c}\right)
\end{equation}
Having both parts of the complex field solution at hand, we can also calculate the reflection
\begin{eqnarray}
	S_{11, \mathrm{nl}}^\mathrm{ideal} & = & - 1 - i\sqrt{\kappa_\mathrm{e}}\frac{\alpha}{S_\mathrm{in}} \nonumber \\
	& = & - 1 - i\sqrt{\kappa_\mathrm{e}}\frac{\alpha_0}{S_\mathrm{0}}e^{-i\phi}.
	\label{eqn:single_tone}
\end{eqnarray}
We use this equation to fit the nonlinear response curves in the higher-power regime, from which we determine the Kerr nonlinearity, cf. Fig.~5 of the main paper and Supplementary Note~\ref{sec:Note8}.

\section{Supplementary Note VI: $S$-parameter background correction and fitting}
\label{sec:NoteVI}
\begin{figure*}
	\centerline{\includegraphics[width=0.85\textwidth]{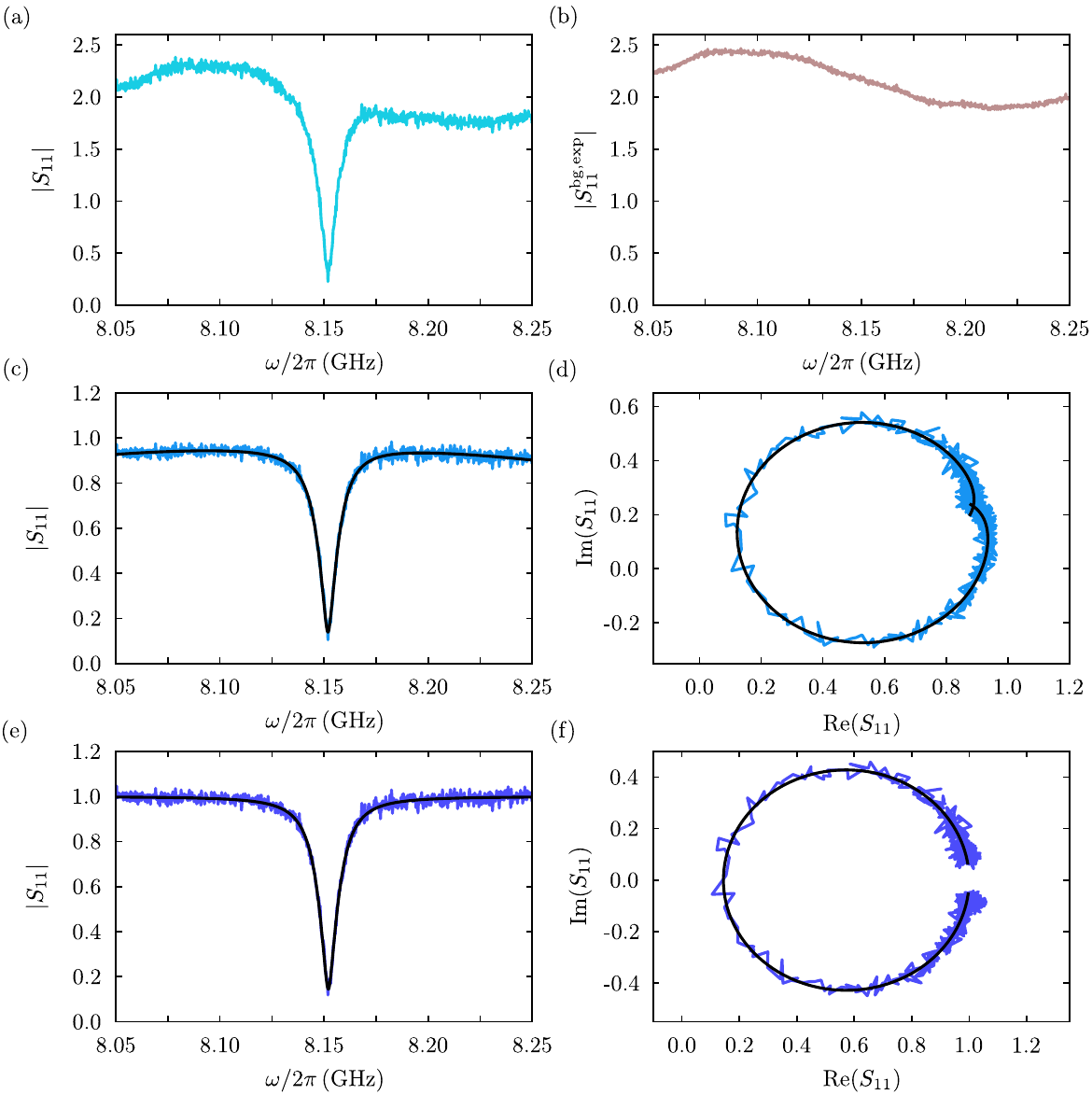}}
	\caption{\textsf{\textbf{Background correction and fitting routine.} (a) Reflection $|S_{11}|$ vs probe frequency of the constriction cavity for a current bias value $I_\mathrm{b} = 0$. The absorption resonance dip around $8.15\,$GHz is clearly visible, the measurement temperature is $T_\mathrm{s} = 3.9\,$K. (b) Identical to (a), but at an elevated temperature $T_\mathrm{s} = 8.1\,$K. What we detect here is the experimental background $S_{11}^\mathrm{bg, exp}$, slightly modified by temperature-dependent signal propagation on the chip and in the coldest parts of the microwave cables. We measure not only the amplitude, but also the phase of $S_{11}$ and $S_{11}^\mathrm{bg, exp}$. (c) shows the magnitude of $S_{11}/S_{11}^\mathrm{bg, exp}$, the background is nearly a flat line, but not yet at $|S_{11}| = 1$ as expected for an ideal reflection. (d) shows the imaginary part of the background-divided reflection vs the real part. Noisy light blue lines in (c) and (d) are data, black smooth lines are a fit with Eq.~(\ref{eqn:S21real}). (e) and (f) show the final background-corrected data, where also the remaining background from the fit is divided off and the resonance circle is corrected by the Fano rotation $\theta$. Noisy blue lines in (e) and (f) are data, black smooth lines are the fits.}}
	\label{fig:FigureS3}
\end{figure*}
\subsection{The real-world reflection function and fit-based background correction}
Due to impedance imperfections in both, the input and output lines, the ideal response is modified by cable resonances and interferences within the setup \cite{Wenner11_SI, Rieger23_SI}.
Origin of these imperfections are connectors, attenuators, wirebonds, transitions to or from the PCB etc. in the signal lines.
In addition, the cabling has a frequency-dependent attenuation.
To take all these modifications into account, we model the final reflection parameter $S_{11}^\mathrm{real}$ by
\begin{equation}
	S_{11}^\mathrm{real} = \left(a_0 + a_1\omega + a_2\omega^2  \right)\left[- 1 + f(\omega)e^{i\theta}\right]e^{i\left(\phi_0  + \phi_1\omega \right)}
	\label{eqn:S21real}
\end{equation}
when the ideal response would be given by
\begin{equation}
	S_{11}^\mathrm{ideal} = - 1 + f(\omega).
\end{equation}
The real-valued numbers $a_0, a_1, a_2, \phi_0, \phi_1$ describe a frequency dependent modification of the background reflection, and the phase factor $\theta$ takes into account possible interferences such as parasitic reflection just before the device or interferences from e.g. imperfect isolation in the directional coupler.
Our standard fitting routine begins with removing the actual resonance signal from the measured $S_{11}$, leaving us with a gapped background reflection, which we fit using
\begin{equation}
	S_{11}^\mathrm{bg} = \left(a_0 + a_1\omega + a_2\omega^2  \right)e^{i\left(\phi_0  + \phi_1\omega \right)}.
\end{equation}
Subsequently, we remove this background function from all measurement traces by complex division.
The resonance circle rotation angle $\theta$ is then rotated off additionally.
The result of both corrections is what we present as background-corrected data or reflection/response data in all figures.
For the power dependence measurements, we determine the background from the measurement in the linear regime and perform a background correction based on that single linear response line for all powers.

\subsection{Data-based background correction}
\label{Section:raw_data_processing}
As the cavity in our experiment has a rather large linewidth of tens of MHz and as the background reflection often cannot be described over such a large frequency span with a simple second order polynomial as suggested by Eq.~(\ref{eqn:S21real}), we perform a two-step background correction to obtain as clean $S$-parameters as possible.
The procedure is exemplarily shown for one resonance of the constriction cavity in Supplementary Fig.~\ref{fig:FigureS3}.
In the first step, we record for each measurement (e.g. the one in panel (a)) also the resonance-less reflection function as shown in panel (b).
The resonance-less $S_{11}$ is obtained by increasing the sample temperature to about $T_\mathrm{s} = 8.1\,$K, where the resonance frequency is out of the measurement window and $\kappa_\mathrm{i}$ is so large that the resonance is not impacting the data anymore.
The elevated temperature leads to a slight upshift of the overall background, but since its frequency-dependence is not modified and since we perform a second-step-correction, this is not impacting the final result.
Then we perform a complex division of the full $S_{11}$ signal by the bare background signal $S_{11}^\mathrm{bg, exp}$, the result is a resonance with a nearly flat background as shown in (c), the complex-valued version can be seen in (d).
Subsequently, we perform a fit using Eq.~(\ref{eqn:S21real}) from which we obtain a second background function as well as a Fano rotation angle $\theta$.
We divide off the fit-background, again by complex division, and finally rotate the resonance circle by $\theta$ around its anchor point.
The final result including the corresponding fits can be seen in panels (e) and (f).
For the circuits with constrictions, we perform this data processing with all $S_{11}$ spectra used for the data analysis and all shown resonances have been treated this way.
For the data before constriction cutting, we do only the fit-based background-correction.
\section{Supplementary Note VII: Additional data and analyses}
\label{sec:Note7}
\subsection{Properties of the superconducting niobium film}

For analyzing the characteristics of the niobium film additionally from transport data, the complete feedline-plus-cavity resistance $R_\mathrm{s}$ is tracked via a DC 4-point measurement, while slowly changing the sample temperature $T_\mathrm{s}$ between room temperature and $\sim 5\,$K.
The resistance $R_\mathrm{s} = V_\mathrm{s}/I_\mathrm{s}$ is obtained by sending a current $I_\mathrm{s} = 1\,$\textmu A and by measuring the corresponding voltage $V_\mathrm{s}$.
For the 4-point measurement, two wirebonds are attached to the microwave launcher of the cavity feedline and two are attached to the ground plane near the end of the cavity, where also the constriction is placed. 
One pair is sending the current and the second pair is detecting the voltage.
The experiment is conducted in a separate dipstick with a stepper motor, that is slowly immersing the sample into liquid helium.
\begin{figure*}[h!]
	\centerline{\includegraphics[width=0.95\textwidth]{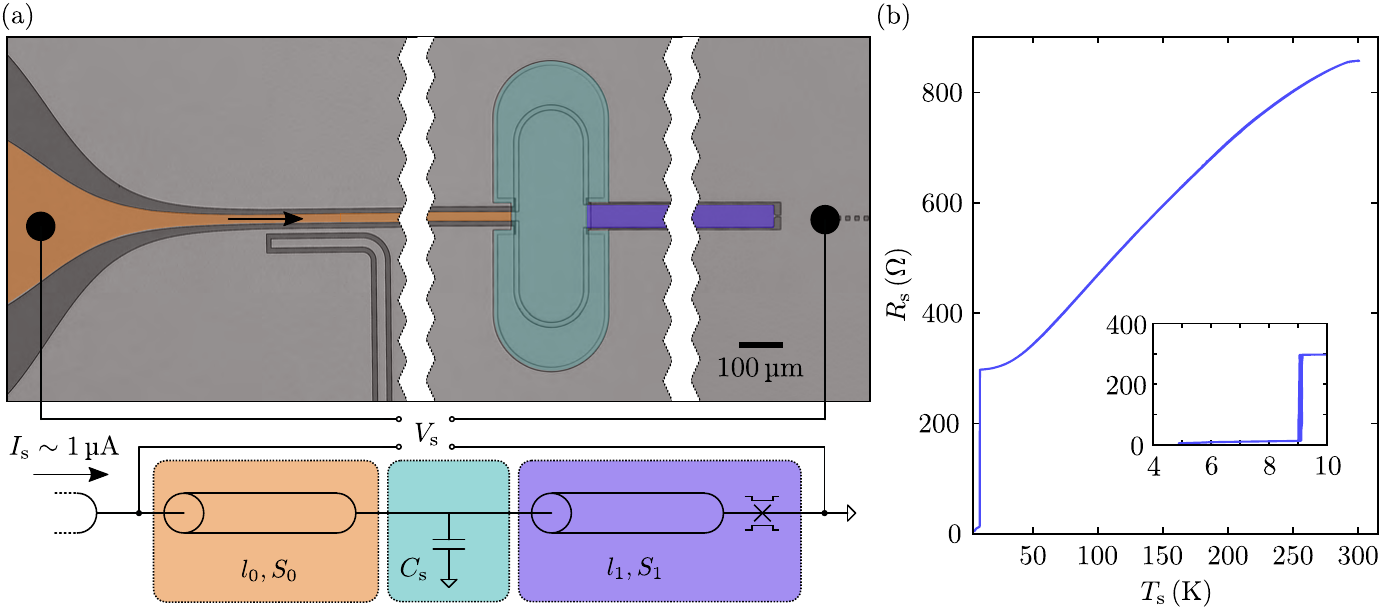}}
	\caption{\textsf{\textbf{Temperature dependence of film resistance.} (a) Schematic of the experiment; top: false-color optical images, bottom: circuit equivalent. By implementing a four-probe measurement we measure the voltage $V_\mathrm{s}$ across the whole feedline-cavity center conductor when sending a current $I_\mathrm{s} = 1\,$\textmu A for varying sample temperature $T_\mathrm{s}$. The feedline has a length $l_0$ and a cross-section $A_0 = S_0d_\mathrm{Nb}$; the cavity has the length $l_1$ and the cross-section $A_1 = S_1d_\mathrm{Nb}$. The circuit is slowly moving into a helium bath for changing the temperature and measuring the voltage after every 0.005$\,$s. (b) Resistance $R_\mathrm{s} = V_\mathrm{s}/I_\mathrm{s}$ vs temperature $T_\mathrm{s}$. We obtain a critical temperature $T_\mathrm{c,4p}=9.08\pm0.04\,$K, a resistance $R_\mathrm{s}(10\,\mathrm{K}) = 297.6\,\Omega$ at 10$\,$K, a resistance $R_\mathrm{c, 4p}(8.9\,\mathrm{K}) = 12.8\,\Omega$ and a residual film resistivity $\rho= 7.3\,$\textmu $\Omega\cdot$cm, see main text for more details. Inset shows a zoom-in to temperatures around and below the superconducting transition.}}
	\label{fig:FigureS4}
\end{figure*}
Supplementary Fig.~\ref{fig:FigureS4} shows a schematic of the DC measurement and the resulting resistance $R_\mathrm{s}(T_\mathrm{s})$.
We find a sudden and large drop of $R_\mathrm{s}$ at the critical temperature of the niobium film $T_\mathrm{c,4p}=9.08\pm0.04\,$K, which is in a good agreement with the result obtained from the microwave data in Supplementary Note~\ref{sec:NoteIIIC}.
The remaining resistance for $T_\mathrm{s} < T_\mathrm{c, 4p}$ is attributed to the residual resistance of the constriction, which has a considerably reduced transition temperature $T_\mathrm{cc}$.
The fact that there is not a second sharp step in the resistance at $T_\mathrm{s} \approx T_\mathrm{cc}$ indicates that there might be a wider $T_\mathrm{cc}$-distribution in and around the constriction, possibly induced by ion implantation or surface damage.
To evaluate the film properties further, we calculate the resistance drop at $T_\mathrm{c, 4p}$ as $R_\mathrm{Nb} = R_\mathrm{s}(T_\mathrm{s} = 10\,\mathrm{K}) - R_\mathrm{s}(T_\mathrm{s} = 8.9\,\mathrm{K}) \approx 284.8\,\Omega$.
The residual resistivity of the film at $T_\mathrm{s} = 10\,$K (without constriction) is then determined by
\begin{equation}
	\rho = \frac{d_\mathrm{Nb}R_\mathrm{Nb}}{ \left(\frac{l_1}{S_1} + \frac{l_0}{S_0} \right)},
\end{equation}
where $l_1 = 7200\,$\textmu m and $S_1 = 50\,$\textmu m are length and center conductor width, respectively, of the cavity, and where $l_0 = 4416\,$\textmu m and $S_0 = 20\,$\textmu m are the corresponding values of the feedline.
We obtain $\rho = 7.3\,$\textmu$\Omega\cdot$cm.
In addition, we estimate the electron mean free path $l_\mathrm{e} = 5.7\,$nm using the material constant $\rho l_\mathrm{e} = 3.72\cdot 10^{-6}\,$\textmu $\Omega\cdot$cm$^2$ \cite{Mayadas72_SI}.
This shows, that our films are well-described by the so-called dirty limit, where $l_\mathrm{e} \ll \xi_0$ with the BCS coherence length $\xi_0 = 38\,$nm.

\subsection{Impact of junction cutting to a reference cavity and constriction parameter uncertainty}
\label{sec:NoteVIIB}

\begin{figure*}
	\centerline{\includegraphics[width=0.85\textwidth]{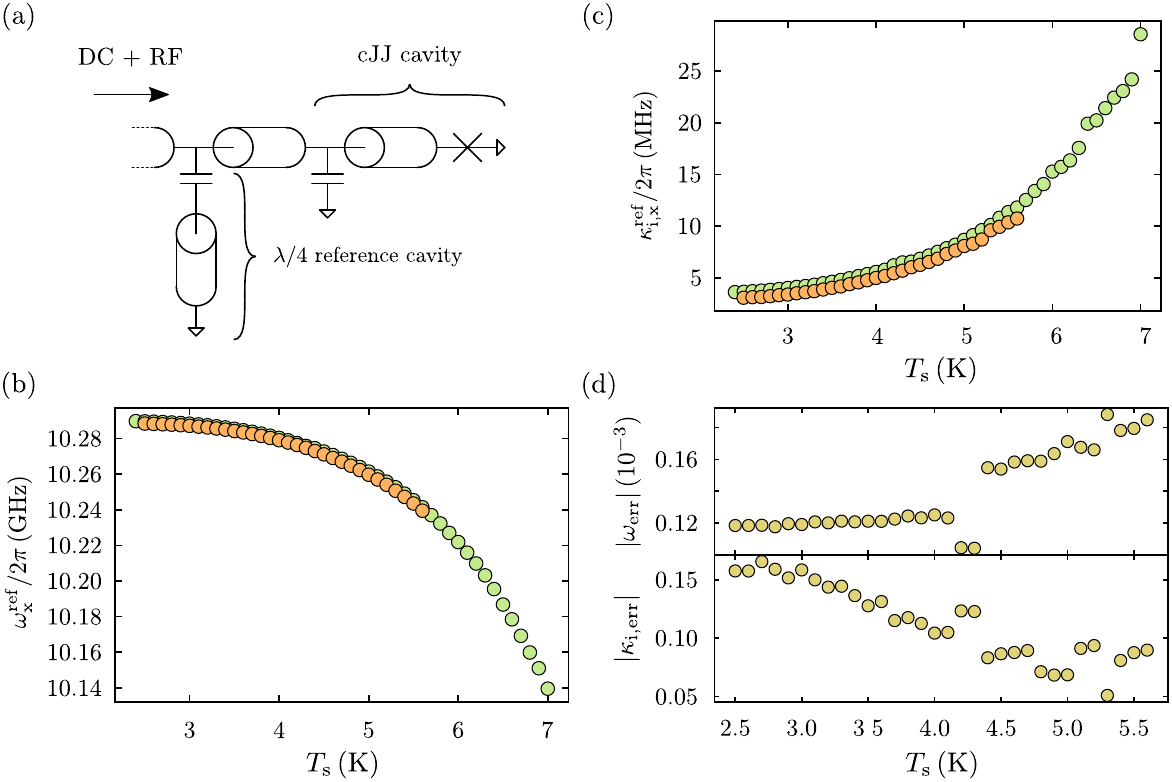}}
	\caption{\textsf{\textbf{Side-coupled $\lambda/4$ reference-resonator properties before and after the neon-ion-beam process.} (a) Circuit equivalent of the complete feedline and cavities, including the side-coupled $\lambda/4$ reference cavity and the transmission line cavity with cJJ at the end. The reflection of both cavities is measured with a vector network analyzer. (b) shows the resonance frequency $\omega_\mathrm{x}^\mathrm{ref}$ of the $\lambda/4$ reference cavity before and after cJJ cutting (into the cJJ cavity only) vs temperature. Green circles are before cutting and orange after. In panel (c) the internal linewidth of the $\lambda/4$ reference cavity $\kappa_\mathrm{i, x}^\mathrm{ref}/2\pi$ vs temperature is shown. Color of datapoints is equivalent to (b). In (d) we show the normalized change of the resonance frequency $|\omega_\mathrm{err}|=|(\omega_0^\mathrm{ref} - \omega_\mathrm{b}^\mathrm{ref}) / \omega_\mathrm{b}^\mathrm{ref}|$ and of the internal linewidth $|\kappa_\mathrm{i, err}|=|(\kappa_\mathrm{i}^\mathrm{ref} - \kappa_\mathrm{i, b}^\mathrm{ref}) / \kappa_\mathrm{i, b}^\mathrm{ref}|$ vs temperature in the range of $2.5\,$K$ \leq T_\mathrm{s} \leq 5.6\,$K.}}
	\label{fig:FigureS5}
\end{figure*}
In main paper Fig.~1(e), we present the resonances of the transmission line cavity before and after cutting the nano-constrictions and discuss their impact on the cavity at $T_\mathrm{s}=3.9\,$K.
In fact, for all our data regarding the constriction inductance and current-phase-relation, the knowledge of the cavity parameters before cutting $\omega_\mathrm{b}, \kappa_\mathrm{b}, L_\mathrm{r}, C_\mathrm{r}$ is crucial.
The reliable determination of the junction properties therefore requires that the niobium film itself or the bare cavity are not impacted by the ion-beam-cutting procedure and by exposing the chip to air, nitrogen atmosphere and room-temperature vacuum for some time (on the order of hours to days).
Otherwise for instance the kinetic inductance might change between the two measurements, which would impact the apparent junction properties extracted from the comparison of $\omega_\mathrm{b}, \kappa_\mathrm{b}$ and $\omega_0, \kappa_0$.
In order to estimate the impact of the additional nano-patterning step to the unirradiated parts of the niobium film as precisely as possible, we add a $\lambda/4$ reference resonator with higher resonance frequency than the constriction-cavity to the same feedline, cf. Supplementary Fig~\ref{fig:FigureS5}(a).
The coupler of the hanger-type $\lambda/4$-cavity can also be seen in the optical image in the top left part of Supplementary Fig.~\ref{fig:FigureS4}.
The reference cavity is capacitively side-coupled with $\kappa_\mathrm{e, x}^\mathrm{ref} \lesssim 2 \pi \cdot 0.9\,$MHz (x stands for 'b' or '0', i.e., indicates whether before and after cutting) to the transmission line and it is shorted to ground at the opposite end, forming a $\lambda/4$ transmission line cavity. 
Similar to the constriction-cavity, we characterize the reference cavity before and after constriction cutting, although it does not contain any constrictions.
Except for the Ne ion irradiation, it has experienced the same environments as the constriction-cavity and we can estimate the uncertainty in constriction-cavity parameters from its property changes.
In Supplementary Fig.~\ref{fig:FigureS5}(b)-(d) we discuss how resonance frequency and internal linewidth of the reference cavity changed between the two measurements for several temperatures. 
The resonance frequency has shifted by $\Delta\omega^\mathrm{ref}=\omega_\mathrm{0}^\mathrm{ref} - \omega_\mathrm{b}^\mathrm{ref} \lesssim -2\pi \cdot 1.1\,$MHz to lower frequencies. 
The total linewidth and therefore the internal linewidth has decreased by $\Delta\kappa_\mathrm{i}^\mathrm{ref}=\kappa_\mathrm{i}^\mathrm{ref} - \kappa_\mathrm{b,i}^\mathrm{ref} \lesssim -2\pi \cdot 0.6\,$MHz. 
The knowledge of these slight changes allows us to estimate an analog inaccuracy for the resonance frequency and linewidth of the cJJ cavity before cutting.    
To do so, we introduce the normalized deviation of the frequency as $|\omega_\mathrm{err}|=|(\omega_0^\mathrm{ref} - \omega_\mathrm{b}^\mathrm{ref}) / \omega_\mathrm{b}^\mathrm{ref}| \leq 0.19\cdot10^\mathrm{-3}$ and of the internal linewidth as $|\kappa_\mathrm{i, err}|=|(\kappa_\mathrm{i}^\mathrm{ref} - \kappa_\mathrm{i,b}^\mathrm{ref}) / \kappa_\mathrm{i, b}^\mathrm{ref}| \leq 0.17$.
Although the exact numbers depend on the temperature, we take the maximum value as upper threshold for all temperatures and integrate it into our analysis of the constriction properties.
To be more precise, we assume, that the parameters before cutting are given by
\begin{eqnarray}
	\omega_\mathrm{b, err} & = & \omega_\mathrm{b}\left(1 \pm |\omega_\mathrm{err}|\right) \\
	\kappa_\mathrm{i, b, err} & = & \kappa_\mathrm{i, b}\left(1 \pm |\kappa_\mathrm{i, err}|\right)
\end{eqnarray}
and carry these errors through the data analysis and parameter extraction.
This leads to the shaded areas around the constriction inductance shown in main paper Fig.~2(e) and main paper Fig.~4(a), and to the areas around the resistances shown in Supplementary Fig.~\ref{fig:FigureS7}.
It also leads to the error bars in the inductancess and resistances in Supplementary Figs.~\ref{fig:FigureS6} and \ref{fig:FigureS7}.
The error is furthermore included in all experimental CPRs, but too small to be visible.
\subsection{Bias current-tuning curves of $\omega_0$ and constriction inductance $L_\mathrm{c}$ for all sample temperatures}
In main paper Fig.~2 we show the bias current-tuning curve of the resonance frequency $\omega_\mathrm{0}(I_\mathrm{b})$ for the sample temperature $T_\mathrm{s}=3.9\,$K.
\begin{figure*}
	\centerline{\includegraphics[width=0.85\textwidth]{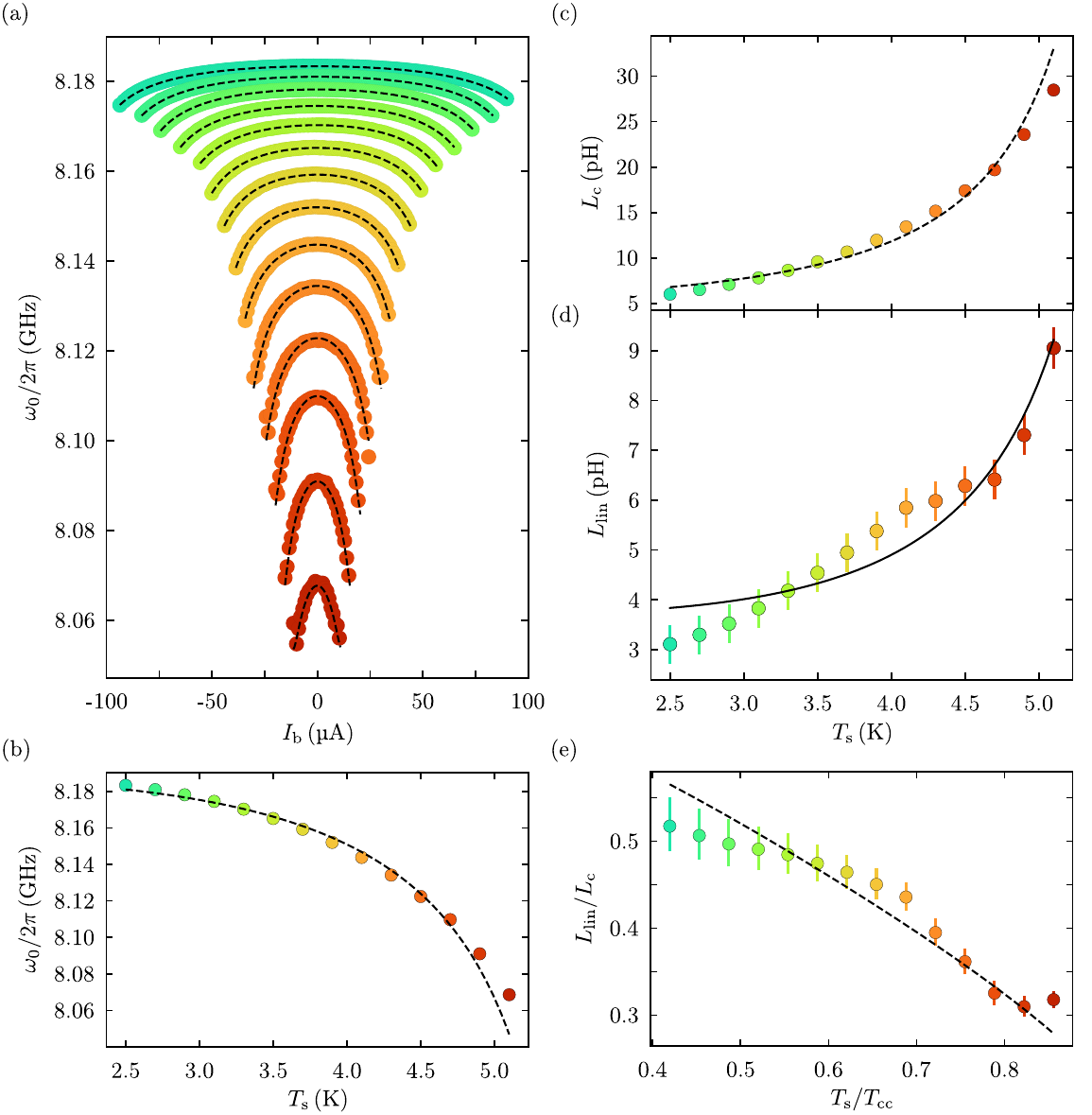}}
	\caption{\textsf{\textbf{Bias-current tunability of resonance frequency $\omega_0$ and analysis of constriction inductances vs $T_\mathrm{s}$.} (a) shows the cavity resonance frequency vs bias current $\omega_\mathrm{0}(I_\mathrm{b})$ for temperatures $2.5-5.1\,$K. Temperature increases from top curve to bottom curve in steps of $0.2\,$K. In (b) we show the resonance frequency at zero bias current vs temperature $T_\mathrm{s}$. For (a) and (b), symbols are data and dashed lines are calculated using Eq.~(\ref{eqn:w0(Ib_Ts)}) in combination with fit curves of the constriction inductances $L_\mathrm{c}(I_\mathrm{b})$, $L_\mathrm{J}(T_\mathrm{s})$ and $L_\mathrm{lin}(T_\mathrm{s})$. In panel (c) we show $L_\mathrm{c}$ for zero bias current ($I_\mathrm{b} = 0$) vs sample temperature. Symbols are data, the dashed theory line is calculated from the individual fits of $L_\mathrm{lin}(T_\mathrm{s})$ and $I_\mathrm{0}(T_\mathrm{s})$. (d) Linear constriction inductance contribution $L_\mathrm{lin}$ as obtained from the $L_\mathrm{c}(I_\mathrm{b})$-tuning curve fits vs sample temperature $T_\mathrm{s}$. Symbols are data, the solid black line is a fit using Eq.~(\ref{eqn:Llin_Ts}) (e) shows the participation ratio of the linear inductance $L_\mathrm{lin}$ to the total constriction inductance $L_\mathrm{c}$ vs reduced temperature $T_\mathrm{s}/T_\mathrm{cc}$, demonstrating that with increasing temperature the linear contribution gets less significant, i.e., that the CPR gets less skewed. Symbols are data, the dashed line is calculated from the individual fits of $L_\mathrm{lin}(T_\mathrm{s})$ and $I_\mathrm{0}(T_\mathrm{s})$.}}
	\label{fig:FigureS6}
\end{figure*}
For completeness and for additional analyses of the constriction temperature-dependence, we present in Supplementary Fig.~\ref{fig:FigureS6}(a) and (b) the bias-current tuning-curves for all measured temperatures and the temperature-dependence of the resonance frequency at zero bias current, respectively.
The resonance frequency at $I_\mathrm{b}=0$ decreases with increasing $T_\mathrm{s}$, most dominantly due to the increasing constriction inductance. 
At the same time, the bias-current tuning range $\omega_0^\mathrm{max}-\omega_0^\mathrm{min}$ increases with $T_\mathrm{s}$ until $4.7\,$K is reached, and decreases again for even higher $T_\mathrm{s}$.
The trend for increasing tuning range with increasing $T_\mathrm{s}$ originates from a larger participation ratio of the constriction inductance with increasing temperature, which is closely related to $T_\mathrm{cc} < T_\mathrm{c}$.
In other words, the constriction inductance grows faster with absolute temperature than the total inductance of the remaining cavity.
The observation that for the highest temperatures the tuning range decreases again is related to the constriction switching current being more suppressed compared to $I_0$ when its temperature approaches $T_\mathrm{cc}$, possibly induced by thermal current noise.
For a more quantitative discussion of the constriction inductance, we extract $L_\mathrm{c}$ and the constriction resistance $R_\mathrm{c}$ (which will be discussed in the next subsection) by using the coupled Eqs.~(\ref{eqn:w0_JJ}) and (\ref{eqn:kappa_JJ}) and as also described in the main paper.
We model the bias-current-dependence by
\begin{eqnarray}
	L_\mathrm{c}(I_\mathrm{b}) & = & L_\mathrm{lin} + L_\mathrm{J} \nonumber\\ 
	& = & L_\mathrm{lin} + \frac{L_\mathrm{J0}}{ \cos\delta_\mathrm{J}} \nonumber\\ 
	& = & L_\mathrm{lin} + \frac{\Phi_0}{2 \pi \sqrt{I_\mathrm{0}^2 - I_\mathrm{b}^2}},
	\label{eqn:LJ(Ib)}
\end{eqnarray}
i.e., by the total inductance being a series combination of an ideal Josephson inductance $L_\mathrm{J}$ with a sinusoidal current-phase-relation (CPR) and a linear inductance $L_\mathrm{lin}$.
Here, $\delta_\mathrm{J}$ is the phase difference across the Josephson element, $L_\mathrm{J0} = \Phi_0 \cdot (2\pi I_\mathrm{0})^{-1}$ with $\Phi_0 \approx 2.068 \cdot 10^{-15}\,$Tm$^2$ the flux quantum and $I_\mathrm{0}$ the critical current of the junction.
For all temperatures the experimental data can be fitted with high reliability using that simple model, cf. panel (a) in main paper Fig.~4.
From the fits to the $L_\mathrm{c}(I_\mathrm{b})$ data for all temperatures we obtain as fit parameters the values for $L_\mathrm{lin}(T_\mathrm{s})$ as well as $I_0(T_\mathrm{s})$. 
The latter is already discussed in the main paper (cf. main paper Fig.~4(c)) and can be well described by
\begin{equation}
	I_0(T_\mathrm{s}) = I_\mathrm{c, 0}\left( 1 - \frac{T_\mathrm{s}}{T_\mathrm{cc}} \right)^{3/2}
\end{equation}
in the temperature range relevant here, despite the fact that most likely there is a $T_\mathrm{cc}$-distribution present instead of a single sharp value.
The fit parameters are the critical current at zero temperature $I_{c, 0} = 252\,$\textmu A and the constriction critical temperature $T_\mathrm{cc} = 6.0\,$K.
The values for $L_\mathrm{lin}(T_\mathrm{s})$ increase with temperature from about $3\,$pH at the lowest $T_\mathrm{s}$ to around $9\,$pH at the highest $T_\mathrm{s}$.
Since the linear contribution is likely a predominantly kinetic inductance (the geometric inductance of a constriction is very small), we model its temperature dependence by
\begin{equation}
	L_\mathrm{lin}(T_\mathrm{s}) = L_\mathrm{off} + \frac{L_\mathrm{lin, 0}}{1 - \left(\frac{T_\mathrm{s}}{T_\mathrm{cc}}\right)^4},
	\label{eqn:Llin_Ts}
\end{equation}
where $L_\mathrm{lin,0}$ is the inductance at zero temperature, $T_\mathrm{cc}$ is the constriction critical temperature and $L_\mathrm{off}$ a possible temperature-independent offset.
Although the data points suggest a more complicated temperature dependence, possibly due to a $T_\mathrm{cc}$ distribution in the constriction or to cavity-linewidth fitting errors, the overall trend is described acceptably by this simple model with the fit parameters $L_\mathrm{off} = -1.0\,$pH and $L_\mathrm{lin, 0} = 4.8\,$pH.
The negative offset is small and most likely does not have a physical meaning.
Using the two fits for $I_0(T_\mathrm{s})$ and $L_\mathrm{lin}(T_\mathrm{s})$, we can also calculate the expected curves for $\omega_0(T_\mathrm{s})$, $L_\mathrm{c}(T_\mathrm{s})$, and $L_\mathrm{lin}(T_\mathrm{s})/L_\mathrm{c}(T_\mathrm{s})$, cf. dashed lines in panels (b), (c), and (e), respectively.
The latter, $L_\mathrm{lin}/L_\mathrm{c}$, shows a clear trend for a decrease with increasing temperature, indicating again that the CPR gets less skewed with increasing $T_\mathrm{s}$, cf. also main paper Fig.~4.
The calculation of $L_\mathrm{c}(I_\mathrm{b}, T_\mathrm{s})$ and $\omega_0(I_\mathrm{b}, T_\mathrm{s})$ is done via
\begin{equation}
	L_\mathrm{c}(I_\mathrm{b}, T_\mathrm{s}) = L_\mathrm{lin}(T_\mathrm{s}) + \frac{\Phi_0}{2 \pi \sqrt{I_\mathrm{0}(T_\mathrm{s})^2 - I_\mathrm{b}^2}}
	\label{eqn:LJ(Ib_Ts)}
\end{equation}
and
\begin{eqnarray}
	\omega_\mathrm{0}(I_\mathrm{b}, T_\mathrm{s}) & = & \frac{\omega_\mathrm{b}(T_\mathrm{s})}{\sqrt{1 + \frac{L_\mathrm{c}(I_\mathrm{b}, T_\mathrm{s})}{2L_\mathrm{r}(T_\mathrm{s})}}} \nonumber\\ 
	& = & \frac{1}{C_\mathrm{tot}L_\mathrm{r}(T_\mathrm{s})}\frac{1}{\sqrt{1 + L_\mathrm{lin}(T_\mathrm{s}) + \frac{L_\mathrm{J0}}{ \sqrt{I_\mathrm{0}(T_\mathrm{s})^2 - I_\mathrm{b}^2}} \frac{1}{2L_\mathrm{r}(T_\mathrm{s})}}}
	\label{eqn:w0(Ib_Ts)}
\end{eqnarray}
where in the latter we used the approximation $R_\mathrm{J} \gg \omega_\mathrm{b}L_\mathrm{J}$, i.e., $L_\mathrm{c} \approx L_\mathrm{c}^*$.
\subsection{Internal linewidth $\kappa_\mathrm{i}$ and resistance $R_\mathrm{c}$ vs bias current and temperature}

In main paper Fig.~4(a) we show the bias current-tuning curves of the cJJ inductance $L_\mathrm{c}$ for varying sample temperature $T_\mathrm{s}$. 
For the determination of $L_\mathrm{c}$ we use the coupled Eqs.~(\ref{eqn:w0_JJ}) and (\ref{eqn:kappa_JJ}).
Therefore, we obtain also the constriction resistance $R_\mathrm{c}$ from the microwave behaviour by considering the internal linewidth $\kappa_\mathrm{i}$.
The internal linewidth for all $I_\mathrm{b}$ and $T_\mathrm{s}$ is shown in Supplementary Fig.~\ref{fig:FigureS7}(a). 
The device shows a strong dependence of $\kappa_\mathrm{i}$ on both $I_\mathrm{b}$ and $T_\mathrm{s}$, and increases with both.
The linewidth tuning range follows similar trends as the resonance frequency equivalent and grows with increasing temperature.
At the lowest temperature, the circuit has a linewidth tuning range of $\sim 3\,$MHz, which increases up to $\sim 65\,$MHz for highest $T_\mathrm{s}$.
We believe the increase of internal loss rate with both bias current and temperature is related to a locally reduced superconducting energy gap and therefore to an increased quasiparticle density in the constriction, mainly due to a reduced critical temperature compared to the rest of the niobium film.
It will be interesting to analyze the losses at much lower temperatures in future experiments, since currently we do not have a solid model to understand the temperature and bias current dependence of $\kappa_\mathrm{i}$.
How does this dependence translate to the constriction resistance $R_\mathrm{c}(I_\mathrm{b}, T_\mathrm{s})$ now?
Some examples for its bias-current-dependence $R_\mathrm{c}(I_\mathrm{b})$ are shown in Supplementary Fig.~\ref{fig:FigureS7}(b), the values change by about $3-4\,\Omega$, interestingly nearly independent of temperature.
The temperature only modifies moderately the zero-bias-current value of $R_\mathrm{c}$ and the bias current required for a certain change.
In panel (c) we show the the slight variations of $R_\mathrm{c}$ at zero bias current as a function of temperature, all values are between $10\,\Omega$ and $7.5\,\Omega$ with more of an oscillating behaviour than a strong trend towards decreasing or increasing. 
The values are close to the resistances $R_\mathrm{dc}$ measured in the IV-characteristics and have a deviation of $(R_\mathrm{dc} - R_\mathrm{c}) / R_\mathrm{dc} \sim 0.02-0.38$.
As an easy approach to obtain a function for $\kappa_\mathrm{i}(I_\mathrm{b})$, we fit the resistance with a polynomial
\begin{equation}
	R_\mathrm{c}(I_\mathrm{b}) = R_0 + R_1 I_\mathrm{b}^2 + R_2 I_\mathrm{b}^4
	\label{eqn:RJ_Ib}
\end{equation} 
with $R_0$, $R_1$ and $R_2$ as fit parameter. 
\begin{figure*}
	\centerline{\includegraphics[width=0.85\textwidth]{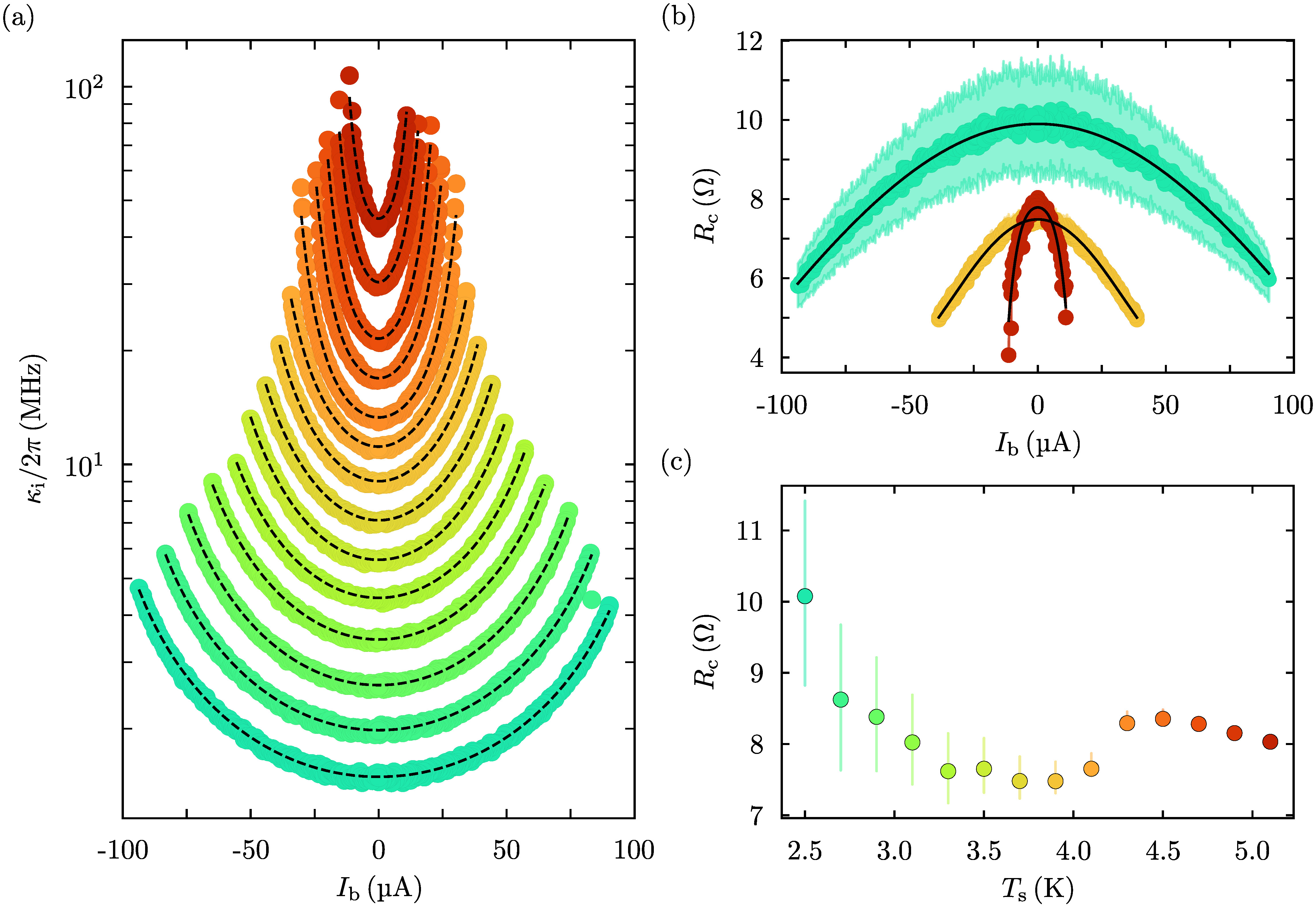}}
	\caption{\textsf{\textbf{Internal linewidth $\kappa_\mathrm{i}$ and resistance $R_\mathrm{c}$ for varying bias current and sample temperatures.} (a) Internal linewidth $\kappa_\mathrm{i}$ vs bias current $I_\mathrm{b}$ for all measurement temperatures $T_\mathrm{s}$ in the range from $2.5\,$K to $5.1\,$K. Points are data and dashed lines are calculated theory lines using Eq.~(\ref{eqn:kappa_ib}). In panel (b) the constriction resistance $R_\mathrm{c}$ vs bias current for three different temperatures $T_\mathrm{s}=(2.5,3.9,5.1)\,$K is shown. Points are data and the solid black lines are fits using Eq.~(\ref{eqn:RJ_Ib}). In (c) we show $R_\mathrm{c}$ at zero bias current vs temperature $T_\mathrm{s}$. Points are data. The unvertainties in (a)-(c), shown as shaded areas and error bars, respectively, are related to the uncertainties in $\kappa_\mathrm{i, b}$ and $\omega_\mathrm{b}$, cf. Supplementary Note~\ref{sec:NoteVIIB}.}}
	\label{fig:FigureS7}
\end{figure*}
The bias-current-dependent internal linewidth can then be written as
\begin{equation}
	\kappa_\mathrm{i}(I_\mathrm{b}) = \kappa_\mathrm{b,i} + \frac{R_\mathrm{c}(I_\mathrm{b}) \omega_\mathrm{b}^2 L_\mathrm{c}(I_\mathrm{b})^2}{R_\mathrm{c}(I_\mathrm{b})^2 + \omega_\mathrm{b}^2 L_\mathrm{c}(I_\mathrm{b})^2} C_\mathrm{tot} \left(\frac{\omega_\mathrm{b}}{1 + \frac{L_\mathrm{c}(I_\mathrm{b}) R_\mathrm{c}(I_\mathrm{b})^2}{R_\mathrm{c}(I_\mathrm{b})^2 + \omega_\mathrm{b}^2 L_\mathrm{c}(I_\mathrm{b})^2}\frac{1}{2L_\mathrm{r}}} \right)^2,
	\label{eqn:kappa_ib} 
\end{equation} 
where $L_\mathrm{c}(I_\mathrm{b})$ is the constriction inductance, $\omega_\mathrm{b}$ is the frequency before cutting, $C_\mathrm{tot}$ is the total cavity capacitance and $L_\mathrm{r}$ is the linear inductance of the cavity without the constriction. 
The curves we obtain from this agree very well with the experimental data, cf. dashed lines in Supplementary Fig.~\ref{fig:FigureS7}.
The inaccuracy in $R_\mathrm{c}$ is due to the inaccuracy of $\kappa_\mathrm{i, b}$ obtained from analyzing the reference cavity before and after cutting with NIM as described in Supplementary Note~\ref{sec:NoteVIIB}.
\subsection{The ratio $I_\mathrm{sw}/I_0$ and comparison with SQUID circuits}
\label{Section:CPR}
To (piecewise) calculate the theoretical curves for the linear-plus-sinusoidal CPRs as shown in main paper Figs.~3 and 4, we use the total phase $\delta$ as function of the current $I \leq I_0$
\begin{eqnarray}
	\delta & = & \delta_\mathrm{J} + \delta_\mathrm{lin} \nonumber \\
	& = & (-1)^n \arcsin{\left( \frac{I}{I_\mathrm{0}}\right)} + \frac{2\pi}{\Phi_0}L_\mathrm{lin}I + n\pi.
\end{eqnarray}
and plot the result inverted as $I(\delta)$.
The corresponding curves for all measurement temperatures relevant in this paper are shown in main paper Fig.~4 and again in Supplementary Fig.~\ref{fig:FigureS8}(a) for further discussions.
Note that there are also other ways for plotting the CPR, for instance as numerical derivative of the potential $E_\mathrm{c}(\delta)$, which leads to exactly the same curves.
\begin{figure*}
	\centerline{\includegraphics[width=0.85\textwidth]{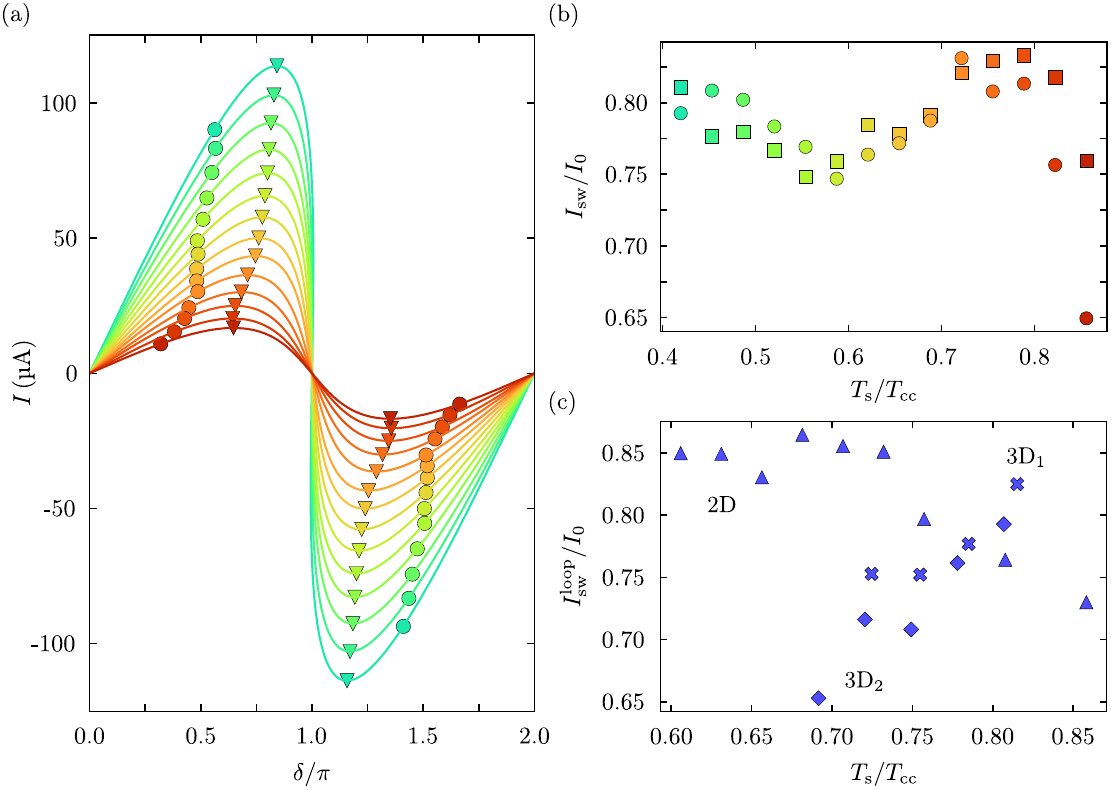}}
	\caption{\textsf{\textbf{Discussion of $I_\mathrm{sw}/I_\mathrm{0}$ and a comparison with the same quantity from SQUID circuits.} (a) CPRs for all measurement temperatures $T_\mathrm{s}$, calculated via $L_\mathrm{lin}$ and expected critical current $I_0$, for details cf. text. Increasing $T_\mathrm{s}$ corresponds to decreasing $I_0$ and decreasing forward-skewedness of the CPR, which can be seen from the maximum of the curves $I_0$ (triangular symbols) shifting to smaller phases. Circles are the switching current $I_\mathrm{sw}$ measured by using microwave reflectometry. (b) Ratio of measured switching current to CPR critical current $I_\mathrm{sw}/I_0$ using the data from $\omega_\mathrm{0}(I_\mathrm{b})$ (circles) and from IV-characteristics (squares). In panel (c) we show data of $I^\mathrm{loop}_\mathrm{sw}/I_0$, obtained from experiments with niobium SQUID microwave circuits \cite{Uhl23_SI}, where $I^\mathrm{loop}_\mathrm{sw}$ is the maximum circulating ring current in the SQUID loop. The ring current is induced by applying external magnetic flux. Switching loop currents are observed through flux hysteresis jumps of the circuit resonance frequencies, and the $I_0$ are obtained similarly to here by a linear-plus-sinusoidal inductance model and by fitting the measured resonance-frequency shifts with magnetic field. There are data for three different types of cJJs. The cJJ types are a 2D (thickness $\sim 90\,$nm equal for the leads and cJJ) and two 3D (constriction thinner than the superconducting leads) versions (thicknesses $\sim 30\,$nm for 3D$_1$ and $\sim20\,$nm for 3D$_2$). Interestingly, $I^\mathrm{loop}_\mathrm{sw}/I_0$ for all three devices lies in the same range as for the device of the present work, although the trends with reduced temperature slightly differ.}}
	\label{fig:FigureS8}
\end{figure*}
The measured switching current $I_\mathrm{sw}$ from the IV-characteristics and from the corresponding bias-current-tuning of the resonance frequency deviate both considerably from the (expected) critical current $I_0$, that we obtain from the fit of $L_\mathrm{c}(I_\mathrm{b})$.
In Supplementary Fig.~\ref{fig:FigureS8}(a), we added as circles the switching current measured in the microwave experiment and as tringles we highlight the points of $I = I_0$.
In panel (b) we plot the ratio $I_\mathrm{sw}/I_\mathrm{0}$ for both IV-switching-currents and microwave-switching-currents.
The two values are very similar, although for the highest temperature there seems to be a deviation and the microwave-switching currents are somewhat lower, possibly due to microwave-activated escape of the phase particle from the potential minimum.
For the ratio $I_\mathrm{sw}/I_\mathrm{0}$ we find values in the range $\sim 0.65-0.84$ in the reduced-temperature range $0.42 < T_\mathrm{s} / T_\mathrm{cc} < 0.86$.
Except for the value obtained at the highest temperature in the microwave experiment, there is not a clear trend towards increasing or decreasing values and the ratio is nearly constant and oscillating around $\sim 0.8$.
This indicates that most likely constant-amplitude current noise or a thermal current noise (whose amplitude increases with $T_\mathrm{s}$) cannot be responsible for the observation of premature switching.
Due to this observation we believe that noise by the HEMT, thermal cavity noise, and environmental noise (e.g. $50\,$Hz noise) are not the culprits.
Interestingly, however, we realized that an almost identical deviation is observable when analyzing the data in Ref. \cite{Uhl23_SI}.
In that work, we have investigated microwave circuits with integrated superconducting quantum interference devices (SQUIDs) and with no DC bias-current access.
The SQUIDs there are also based on niobium and on neon-ion-beam-patterned nano-constrictions, they even have the same film thickness of $\sim 90\,$nm.
Regarding the constrictions, we have investigated three different types in Ref.~\cite{Uhl23_SI}.
One of the three is a 2D cJJ with thickness equal to the leads ($90\,$nm), and the other two are 3D versions, where the constrictions are thinner than the superconducting leads, similar to the constrictions studied in the present work.
The 3D cJJs only differ in the thickness $\sim 30\,$nm (3D$_1$) and $\sim 20\,$nm (3D$_2$).
All three types had a length of $\sim 20\,$nm and a width of $\sim 40\,$nm, i.e., they are very comparable to the one described here.
By applying an external magnetic field to the circuits, that introduces magnetic flux into the SQUIDs, the flux-tunability of the resonance frequencies was studied. 
From a careful analysis of the resulting data, we determined the circulating ring current $I_\mathrm{loop}$ in the SQUID, the switching ring-current $I_\mathrm{loop}^\mathrm{sw}$, where the number of flux quanta in the loop is jumping, and the theoretical critical current $I_0$ of a single cJJ in the SQUID.
We find that $I_\mathrm{loop}^\mathrm{sw}/I_0$ for all three SQUID devices lie in the same range as for the single-cJJ device of the current paper.
For the 2D-SQUID device, the values seem rather constant at lower $T_\mathrm{s}$ with a tendency to decrease for the higher temperatures.
For the 3D-SQUIDs the trend suggests that the ratio is actually increasing with temperature, which would in agreement with the data from the current system.
In any case does this similarity in values, but also in trends for the 3D-cJJs, support the idea that the premature switching has an origin in the intrisic constriction properties rather in external source.
It would be highly unlikely that an external noise source would couple equally into both systems.
We believe that the suppressed critical current of the constrictions and therefore the premature switching into the normal state is related to phase slips, and very similar $I_\mathrm{sw}I_0$ values have been reported by both experiments and theoretical works in the past \cite{Pekker09_SI, Li11_SI, Aref12_SI, Baumans17_SI, Friedrich19_SI}.
To illuminate the current switching in detail, however, further experiments, e.g. measuring switching statistics for varying current sweep rates and experiments at lower temperatures, will be necessary.

\section{Supplementary Note VIII: Additional data and theory for the extraction of the Kerr constant}
\label{sec:Note8}
\subsection{Observation of cavity response nonlinearities and fits in the weakly nonlinear regime}
In Supplementary Fig.~\ref{fig:FigureS9}, we show reflection data and corresponding fits of the cavity response around the onset of nonlinearities in (a) and data only for even higher powers in (b).
For the determination of the Kerr anharmonicity $\mathcal{K}$, we use the reflection data in the weakly nonlinear regime shown in (a), since the data for the higher powers shown in (b) are beyond the validity of our simple nonlinear model.
They show features that cannot be reproduced by our theory.
\begin{figure*}
	\centerline{\includegraphics[width=0.85\textwidth]{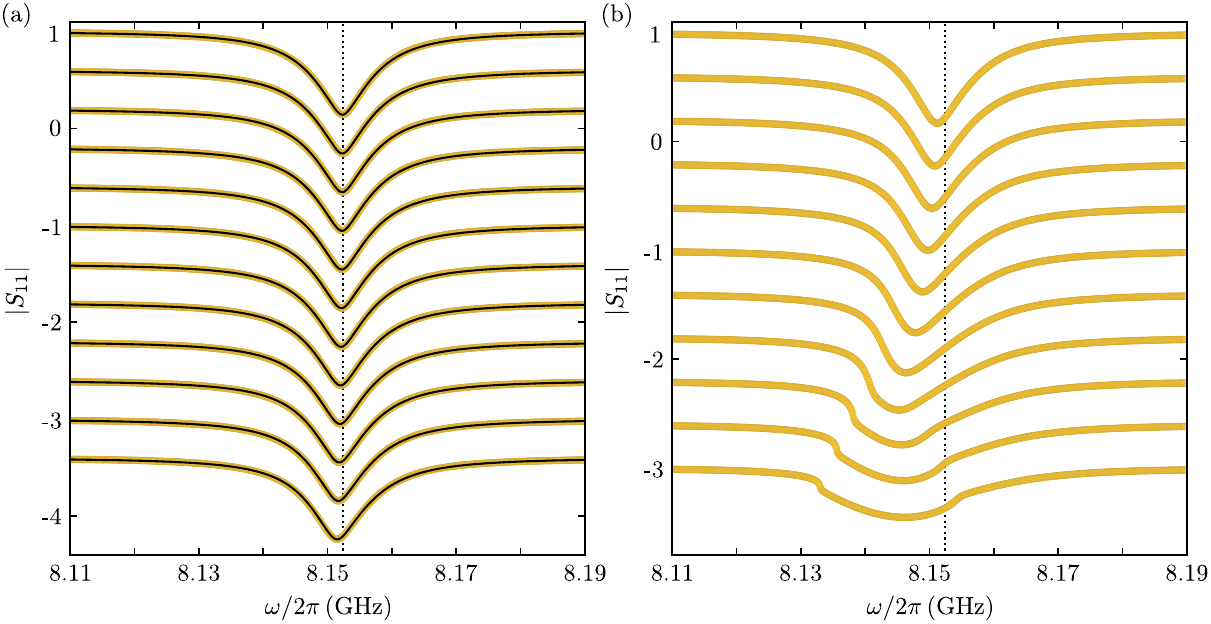}}
	\caption{\textsf{\textbf{Cavity resonances in the nonlinear regime.} (a) Cavity reflection $|S_{11}|$ in the weakly nonlinear regime. We define the weakly nonlinear regime as the power range in which the data can be well-described by our model. Input power increases in steps of $1\,$dB; top curve: lowest power, bottom curve: highest power. Subsequent datasets are offset by $-0.4\,$dB each for clarity, colored points are data, black smooth lines are fits. (b) Cavity reflection $|S_{11}|$ for input powers beyond the weakly nonlinear regime. Data only (no fits), since our model is not able to capture the resonance shapes anymore. Temperature for both panels is $T_\mathrm{s} = 3.9\,$K and bias current is $I_\mathrm{b} = 0$. In (a) and (b) the vertical dotted line shows the resonance frequency in the linear regime}}
	\label{fig:FigureS9}
\end{figure*}
In the weakly nonlinear regime, the resonance minimum shifts slightly to lower values with increasing microwave input power, the total linewidth increases and the resonance shape becomes slightly asymmetric towards the shape of a Duffing resonance.
After a background correction, we use Eqs.~(\ref{eqn:NL_EOM})-(\ref{eqn:single_tone}) to fit the data in two steps.
In the first step, we fit all curves simultaneously with a single $\mathcal{K}$, a single $\kappa_0$ and a single $\kappa_\mathrm{nl}$ as fit parameters.
For $\kappa_\mathrm{e}$ and $\omega_0$ we use a constant average value, which we obtained from multiple resonances below the nonlinear regime, and for $P_\mathrm{in}^\mathrm{av}$ we use the on-chip input power calculated from the VNA output power and the cable attenuation.
This first fit gives already a very good agreement with the data.
We use a second round of fitting afterwards to improve the fits of panel (a), in which we allow each individual resonance dataset to have its own $\kappa_\mathrm{nl}$.
The result of this second round is shown as lines overlaid to the data in Supplementary Fig.~\ref{fig:FigureS9}(a).
Strictly speaking this approach is somehow in contradiction to the model itself, since if the model was completely valid, we would only get a single $\kappa_\mathrm{nl}$ for all curves.
However, due to the large total cavity linewidth we believe that the background correction is prone to having small frequency-dependent uncertainties, the external linewidth is also not a constant over the complete resonance line since it is frequency-dependent, and there might be other contributions to the nonlinearity of $\kappa$ and $\kappa_\mathrm{e}$ present at low powers, e.g. due to nonlinear dielectric losses and two-level systems in the shunt capacitor.
The resulting $\kappa_\mathrm{nl}$ and the injected $\kappa_\mathrm{e}$ are shown in Supplementary Fig.~\ref{fig:FigureS10}.
We emphasize, that this second round of fitting is not really necessary and not changing the value of $\mathcal{K}$, its purpose is merely to demonstrate that the resonances can be well-explained by a single Kerr constant and power-dependent decay rates, although the exact power-dependence of the decay rates is somewhat obscured when each line gets its own $\kappa_\mathrm{nl}$.
To calculate the possible deviation of the fit parameters due to uncertainties in the on-chip input power, we repeat this procedure for $P_\mathrm{in}^\mathrm{min} = P_\mathrm{in}^\mathrm{av} - 1\,\mathrm{dB}$ and $P_\mathrm{in}^\mathrm{max} = P_\mathrm{in}^\mathrm{av} + 1\,\mathrm{dB}$.
The resulting values for $\mathcal{K}_+$ and $\mathcal{K}_-$ represent the tips of the error bars in main paper Fig.~5.
We repeat the same routine for each bias current and obtain the values for $\mathcal{K}$ (as well as $\mathcal{K}_+$ and $\mathcal{K}_-$) plotted in main paper Fig.~5.
For higher input powers, cf. data in panel (b), we observe that the resonances deviate strongly from Duffing resonances.
For this reason we exclude the higher powers from the extraction of $\mathcal{K}$.
\begin{figure*}
	\centerline{\includegraphics[width=0.85\textwidth]{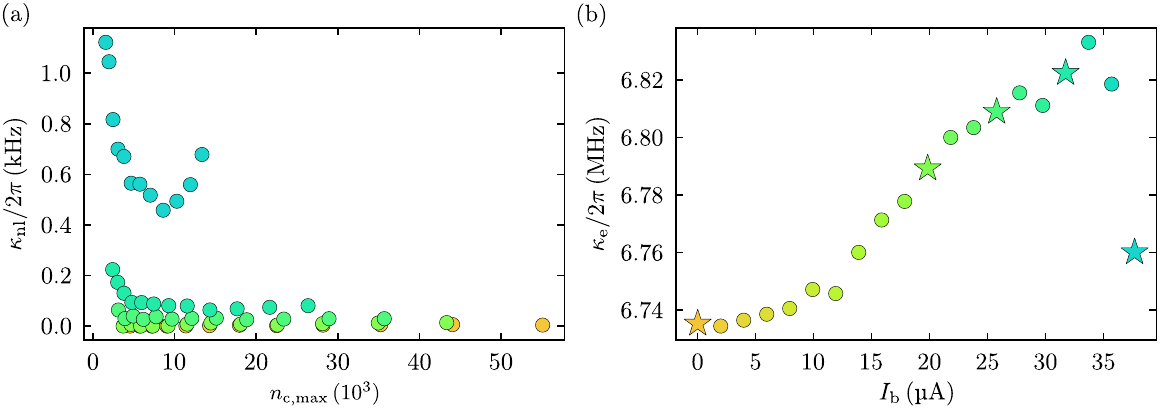}}
	\caption{\textsf{\textbf{The nonlinear damping and external linewidth of the Kerr fitting.} (a) Nonlinear damping parameter $\kappa_\mathrm{nl}$ for 5 different bias-currents (color-coded) and plotted vs the maximum intracavity photon number on resonance, as obtained from the second round of fitting, cf. text. (b) the external linewidth $\kappa_\mathrm{e}$ vs bias-current $I_\mathrm{b}$ used for the nonlinear fits and obtained as average value from multiple resonances in the linear regime. The bias-current-dependence originates most likely from a frequency-dependence $\kappa_\mathrm{w}(\omega)$ and could be explained by parasitic reflections in the setup or by fitting errors due to Fano resonances. The star symbols correspond to the 5 bias currents, that are also displayed in (a).}}
	\label{fig:FigureS10}
\end{figure*}
To double-check the results for $\mathcal{K}$, that we obtained by this full-model fitting, we also implement a more common and simple routine for finding $\mathcal{K}$.
We know that within our model the microwave probe tone with frequency $\omega$ is on the shifted resonance frequency $\omega_0'$ of the cavity when $\varDelta = \mathcal{K}n_\mathrm{c, max}$ with $\varDelta = \omega - \omega_0$.
The resonance intracavity photon number $n_\mathrm{c, max}$ is the maximum photon number achieved with a constant input power during the frequency sweep at $\omega_0'$.
Equation~(\ref{eqn:NL_EOM}) then becomes
\begin{equation}
	\frac{\kappa_\mathrm{eff}}{2}\alpha_\mathrm{0, max} = i\sqrt{\kappa_\mathrm{e}}S_0e^{i\phi}
\end{equation}
with $\kappa_\mathrm{eff} = \kappa_0 + \kappa_\mathrm{nl}n_\mathrm{c, max}$.
To calculate $n_\mathrm{c, max}$, we then get the magnitude squared of this equation, re-sort some factors and arrive at
\begin{equation}
	n_\mathrm{c, max} = \frac{4P_\mathrm{in}^\mathrm{av}}{\hbar\omega}\frac{\kappa_\mathrm{e}}{\kappa_\mathrm{eff}^2}.
\end{equation}
The only unknown in this relation is then the effective decay rate $\kappa_\mathrm{eff}$, but we can easily find its value by the reflection response $S_{11}$ at this particular frequency.
It can be calculated from (valid in the effective undercoupled regime $\kappa_\mathrm{e}< \kappa_\mathrm{eff}/2$)
\begin{equation}
	|S_{11}| = 1 - \frac{2\kappa_\mathrm{e}}{\kappa_\mathrm{eff}}.
\end{equation}
without the knowledge of $\kappa_\mathrm{nl}$ or $n_\mathrm{c, max}$.
We obtain both $\delta\omega_0 = \omega_0' - \omega_0$ and $\kappa_\mathrm{e}/\kappa_\mathrm{eff}$ from the data by fitting the resonance minimum with a simple parabola as shown exemplarily in the inset of Supplementary Fig.~\ref{fig:FigureS11}(a).
The shifted resonance frequency $\omega_0'$ is just given by the frequency coordinate of the parabola minimum and the corresponding $|S_{11}|$ by the corresponding $y$-axis value of the minimum.
Finally we calculate $n_\mathrm{c, max}$.
We repeat this procedure for all powers at a fixed bias current, and then perform a linear fit of $\delta\omega_0 = \mathcal{K}n_\mathrm{c, max}$ with $\mathcal{K}$ as fit parameter.
Doing the same for all bias-currents then gives us the anharmonicity as a function of current.
To consider a $\pm1\,$dB uncertainty in input power, which then translates to an error in the intracavity photon number and finally to an error in anharmonicity, we repeat the same procedure for $P_\mathrm{in}^\mathrm{min/max} = P_\mathrm{in}^\mathrm{av} \pm 1\,$dB.
In Supplementary Fig.~\ref{fig:FigureS11} we show the result of these fits in comparison with the results obtained from the full response fit explained above.
The short conclusion is that the two approaches lead to nearly identical values for $\mathcal{K}$.
\begin{figure*}
	\centerline{\includegraphics[width=0.85\textwidth]{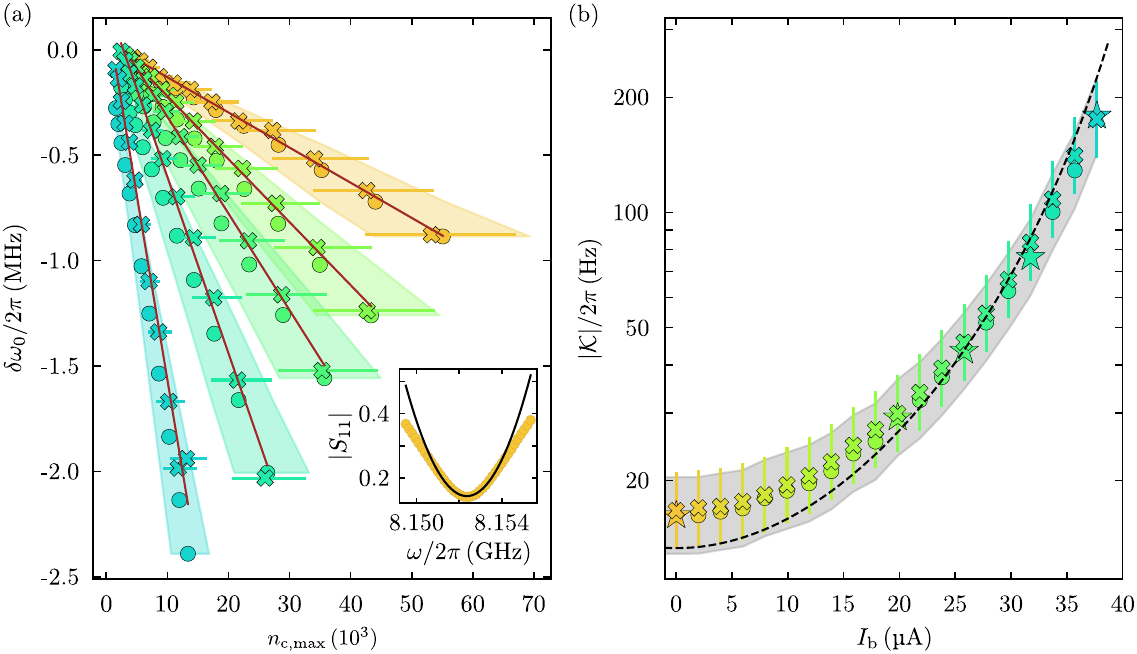}}
	\caption{\textsf{\textbf{Comparing full-model Kerr anharmonicity with the one obtained by a linear-fit approach.} (a) Frequency shift $\delta\omega_0 = \omega_0' - \omega_0$ of the cavity as function of intracavity photon number at the power-shifted resonance frequency $\omega_0'$. Data are shown for five values of the bias current (color-coded). Circles are obtained from fitting the nonlinear resonances with the full response model for $S_{11}(\omega)$ and naturally lie on a straight line due to the underlying model. Crosses are the result of measuring just the position of the nonlinear resonance minima and calculating from that the corresponding intracavity photon number $n_\mathrm{c, max}$ by hand, cf. text for details and inset for a parabolic fit around one of the resonance minima. Error bars to the crosses consider $\pm 1\,$dB uncertainty in input power, error shades are the anaologous quantity for the full response model. Brown solid lines are linear fits $\delta\omega_0 = \mathcal{K}n_\mathrm{c, max}$ with $\mathcal{K}$ as single fit parameter. The values for the extracted $\mathcal{K}$ are plotted in (b) as $|\mathcal{K}| = -\mathcal{K}$. Circles are the values of the full response model, crosses from the linear fit, star shaped symbols in the full response data correspond to th ebias currents discussed in (a). Error bars belong to the circles and correspond to the color shades in (a), error shade in gray belongs to the crosses and corresponds to the error bars in (a). The dashed line is the theoretical curve based on the linear-plus-sinusoidal model for the CPR. All values lie within the error range of the experiment. }}
	\label{fig:FigureS11}
\end{figure*}
\subsection{Construction of an artificial CPR}
To construct the artificial CPR used in main paper Fig.~5 for the calculation of the alternative Kerr constant and shown in the inset of panel (c), we use the Ansatz of an odd polynomial function
\begin{equation}
	I_\mathrm{ar}(\delta) = \sum_{n > 0} \alpha_{2n-1} \delta^{2n-1}
\end{equation}
and the validity of the function is $\delta \in [-\pi, \pi[$.
Outside of the intervall, the function will then be just set to be periodically repeating.
After trying for a bit by hand, we found that restricting ourselves to a few term is sufficient to model CPR and Kerr simultaneously and we use
\begin{equation}
	I_\mathrm{ar}(\delta) = \alpha_1\delta + \alpha_3 \delta^3 + \alpha_7\delta^7 + \alpha_{15}\delta^{15}
\end{equation}
For the first coefficient, we use the actual slope
\begin{equation}
	\alpha_1 = \frac{\Phi_0}{2\pi (L_\mathrm{J0} + L_\mathrm{lin})}
\end{equation}
to ensure that the zero bias-current inductance is not changed by a modified CPR.
The second and third coefficients $\alpha_3$ and $\alpha_7$ are used as fit parameters, and the final coefficient $\alpha_{15}$ is set to be
\begin{equation}
	\alpha_{15} = -\frac{\alpha_1\pi + \alpha_3\pi^3 + \alpha_7\pi^7}{\pi^{15}}
\end{equation}
to ensure that $I_\mathrm{ar}(\pi) = 0$, i.e., that the CPR goes to zero at $\delta = \pi$
We use this function to fit simultaneously the CPR and the Kerr data.
Most likely there are even better fitting polynomial curves, when allowing more or different terms, but this one is sufficient to demonstrate the principle.
\let\oldaddcontentsline\addcontentsline
\renewcommand{\addcontentsline}[3]{}

\end{document}